**Structured Quantum Baths with Memory: A QuTiP Framework for Spectral Diagnostics and Machine Learning Inference**

Ridwan Sakidja
Dept. of Physics, Astronomy and Materials Science
Missouri State University

**Abstract**

We introduce a compact simulation framework for modeling open quantum systems coupled to structured, memory-retaining baths using QuTiP. Our method models the bath as a finite set of layered qubits with adjustable connections, interpreted either as a physical realization or as a conceptual representation, rather than as a continuum. This explicit modeling enables direct control over non-Markovian dynamics and allows spectral diagnostics via Fast Fourier Transform (FFT) of system observables. Using a triangle-based bath motif and its extension to a six-qubit anisotropic fractal-like architecture, we demonstrate how spectral fingerprints encode bath topology and memory depth. Standard machine learning tools such as Principal Component Analysis (PCA) and gradient boosting can then be employed to infer bath parameters and estimate proximity to exceptional points (EPs). The results suggest that spectral analysis can serve as a unifying, quantum-platform agnostic tool across theory, simulation, and experiment, offering both a student-accessible and experimentally relevant approach to understanding coherence loss and memory flow in quantum hardware. Rather than treating noise as an adversary to be eliminated, our approach views structured baths as collaborative partners, enabling controlled memory and delocalized memory and information flow for engineered quantum dynamics. In addition to its diagnostic power, the framework offers a modular and reproducible platform for teaching open quantum systems. Ultimately, we frame this as a pedagogical tool: students can pair FFT-based spectral features with lightweight ML (e.g., PCA and gradient boosting) to extract data-rich, interpretable signatures of open-system and non-Hermitian dynamics.

## 1. Introduction

Open quantum systems are often modeled using Markovian Lindblad dynamics, where the environment is treated as an infinite, memoryless sink. In reality, engineered quantum hardware is almost always embedded within finite, structured environments. These may be intentional, such as resonators and couplers in superconducting qubits, or unavoidable, such as spectator modes in trapped ions or neighboring sites in neutral atom arrays. In all these cases, the immediate environment has a limited and defined size, specific connectivity, and thus, has the potential capacity to not only store, but also to return information, making it inherently non-Markovian.

Throughout this work, the term bath will refer specifically to this finite, structured near-field bath that we explicitly model as qubit nodes. This near-field bath is distinct from the global Markovian environment, which is still assumed to act as an infinite reservoir in the background. In other words, the system is always globally connected to a Markovian bath, but our focus is on the immediate, structured layers where memory effects originate. For brevity, we will simply use bath to mean this finite near-field component unless otherwise noted.

To address this, we consider a simpler and discretized framework that explicitly models the bath using a potentially scalable, anisotropic topology, which we identify as the minimal structural module that can capture nonlinear and non-chain system-bath connections. This enables us to investigate non-Markovian effects through spectral diagnostics that extend beyond conventional temporal characterizations such as $T_1$/$T_2$ measurements.

Recent work by Prof. Jens Eisert's research group [1] demonstrated that thermalization does not require strong entanglement, opening the door to reconsidering the role of bath structure. In our framework, this distinction is explicit: the system qubit undergoes only dephasing, while genuine thermalization occurs indirectly through coherent exchange with bath qubits that are themselves coupled to thermal reservoirs. Although Eisert's study did not address bath memory or non-Markovian effects, their result motivates a natural question: what changes when the immediate environment is modeled as a finite, structured, memory-retaining system? To explore this, we represent baths as small dissipative subsystems composed of qubit nodes, showing that such structures can generate non-Markovian dynamics even in weakly entangled regimes. These bath constructs may be interpreted either as physical components or as conceptual models, but in both cases they can be tuned to control memory flow. The notion of an "immediate environment" is therefore not tied to a single hardware realization but serves as a modeling tool that isolates the degrees of freedom most responsible for short-time memory and coherence effects.

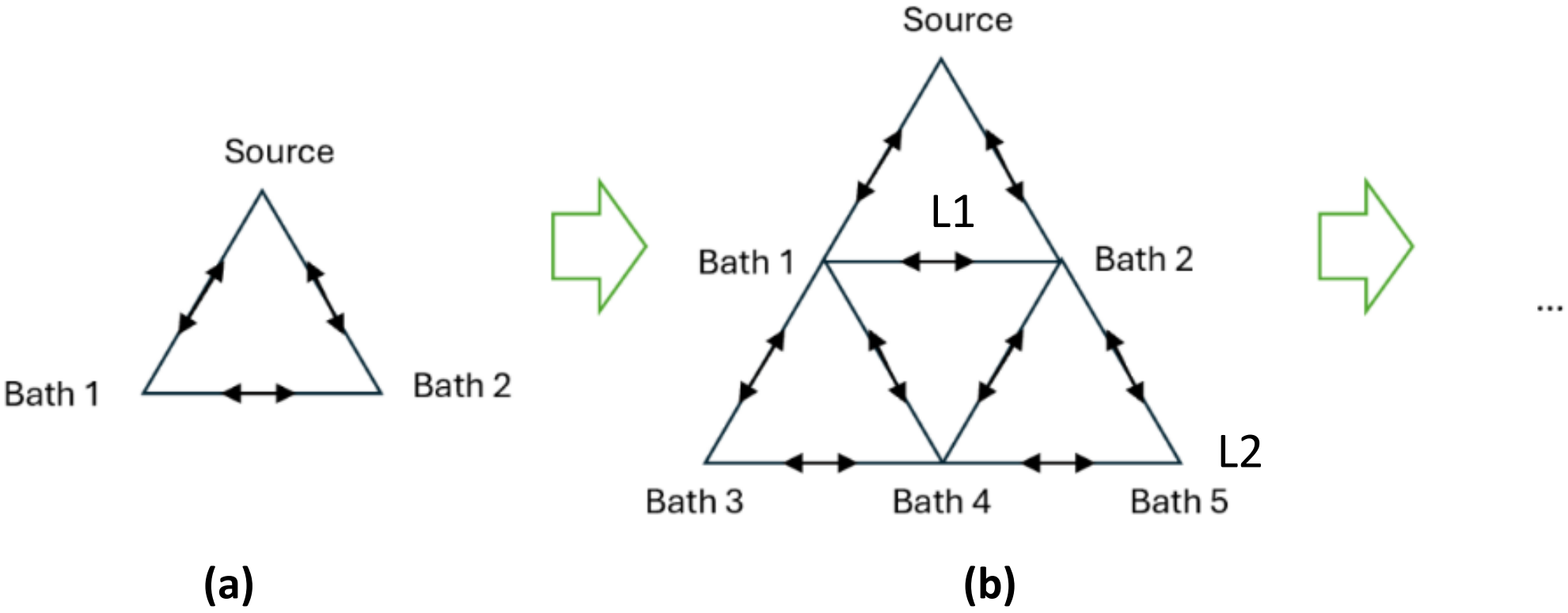

**Figure 1**. Schematic of triangle-based bath geometries. (a) Three-qubit configuration consisting of a system qubit (Source) coupled to two bath qubits (Bath 1 and Bath 2) that are also mutually coupled as "$L_1$" environment. (b) Extended six-qubit configuration with a second bath layer (Bath 3–Bath 5) as "$L_2$" environment that is coupled to the first layer. The design can be generalized to larger, more complex environments.

Throughout this work, the term “bath qubit” refers to such a modeling construct; while some platforms may realize similar configurations physically, the framework itself is general and platform independent.

As a whole, the three-qubit system remains Markovian at the global level, i.e. connected to an infinite bath, but we hypothesize that the internal connectivity from the system to the two connecting baths as well as the connections amongst two bath qubits, labeled as $L_1$ environment, acts as a memory channel, enabling partial information backflow from the baths to the system (Figure 1a). We then extend this analysis to a second layer of memory $L_2$, forming a six-qubit system (Figure 1b) where the second-layer baths interact both with the first-layer baths and with each other. This six-qubit construction can be further extended for more complex systems. Such a simple design is not only useful for practical research into non-Markovian dynamics but also serves as an educational tool for learning about open quantum systems beyond the standard Markovian approximation. Within the context of the foundational work by Breuer et al. [2] who have long established the theoretical formalism to measure for non-Markovian behavior, our work extends the concept by explicitly modeling the physical mechanisms, namely, a structured, hierarchical bath, that implements these memory effects into practice. Hence, our objectives go beyond Markovian reservoir engineering by demonstrating how the non-Markovian nature of a structured bath can now be leveraged as a resource for an additional quantum control. Further, our method may offer a physically motivated framework for exploring concepts from non-Hermitian physics, such as the emergence of exceptional points (EPs)[3]. By connecting the study of EPs into a realistic, discrete bath model, our work may provide a pathway for their experimental realization and application within a current quantum architecture.

It is worth pointing out that many theoretical models of non-Markovian dynamics have indeed been advanced for some time including the Nakajima–Zwanzig (NZ) memory kernel [4, 5], time convolutionless (TCL) master equations [6], and the Hierarchical Equations of Motion (HEOM) framework [7, 8]. These approaches, however, share a common feature: the environment is treated as an infinite continuum of harmonic oscillators with a prescribed spectral density. This is understandable since the main objective has been to extract the critical information about the bath correlation functions rather than to explicitly model the bath’s discretized internal structure. This goal has been accomplished either by integrating out the environment entirely in NZ and TCL or by encoding it indirectly through auxiliary density operators in HEOM.

In contrast, our work employs an explicit finite bath construction in which the environmental degrees of freedom are represented as physical qubits in the immediate vicinity, with network connectivity between these bath qubits defined by a tunable topology. This design choice is deliberate. It connects the bath’s layered, fractal-like structure to measurable quantities such as entropy growth, population dynamics, energy flow, and trace distance backflow. By modeling the bath structure directly, we can track how memory moves and how coupling paths compete.

Even though our approach models only a finite bath for the near field environment, it still supports spectral analysis of system and bath dynamics. The key distinction is that the extracted spectral features can be mapped directly to structural parameters such as layer couplings, competing channels, and topology related signatures rather than being averaged over an abstract spectral density. This enables a more granular and interpretable view of how the bath's internal structure governs non-Markovian effects, while retaining the benefits of frequency domain diagnostics in experimentally realizable architectures.

We also would like to note the related and recent work of Brand et al. [9], who developed a hardware-agnostic Markovian noise-modelling and parameter-extraction framework for quantum devices, focusing on time-domain analysis of relaxation and dephasing channels. This approach differs with our nodal-based approach as our method embeds finite, structured baths directly into the Hilbert space and extract the full frequency-domain spectra of, $\langle \sigma_x \rangle, \langle \sigma_y \rangle, \langle \sigma_z \rangle$, preserving oscillatory and phase information to enable a more comprehensive inference of bath topology and memory effects.

The remainder of this paper is organized as follows.

In **Section 2**, we introduce the finite-bath model and Hamiltonian, implemented as a Heisenberg-type network in QuTiP [10, 11], incorporating both coherent and dissipative components. **Section 3** presents the simulation protocol and initial results, demonstrating thermal equilibration in a minimal three-qubit configuration as a baseline for weak-coupling thermalization. **Section 4** extends the framework to a six-qubit system to classify bath memory categories through FFT-based diagnostics, while **Section 5** leverages these observables within a machine learning pipeline to infer bath parameters. Building on this, **Section 6** connects the finite-bath nodal framework to continuum models of non-Markovian dynamics, such as the Nakajima–Zwanzig formalism. **Section 7** moves beyond bath topology to examine the role of system–bath coupling strength. **Section 8** provides an application example by probing proximity to exceptional points using dynamical Dissipative Exceptional Point Spacing (or we call it "DEPS"). **Section 9** explores the generalization of this approach to many-body logical qubit systems. **Section 10** offers perspectives on broader implications, including coarse-graining strategies and experimental relevance. **Section 11** outlines limitations and future directions, and **Section 12** concludes with final remarks.

## 2. Model and Hamiltonian (Heisenberg network in QuTiP)

We simulate a small open quantum network in QuTiP that consists of a single system qubit $Q_0$ coupled to an environment represented using qubit like nodes which serve as conceptual nodes though in some cases they may map to physical degrees of freedom. These nodes can correspond to physical elements or conceptual degrees of freedom. The minimal motif is a triangle (three qubits) as shown in Figure 1. The extended case has two layers of baths for a total of six qubits as depicted also in Figure 1. The goal is to show how bath connections support memory and backflow while overall dynamics remain Markovian at the master equation level.

**2.1 Hilbert space and initial state**

The system qubit $Q_0$ starts with the equal superposition:

$$|+\rangle = \frac{|0\rangle+|1\rangle}{\sqrt{2}} \qquad (1)$$

with a density matrix:

$$\rho_{sys} = |\psi_{sys}\rangle\langle\psi_{sys}| \qquad (2)$$

All bath qubits $Q_1 \dots Q_{N-1}$ begin in identical thermal states:

$$\rho_{th}(\beta) = \frac{e^{-\beta H_q}}{Tr\left[e^{-\beta H_q}\right]}\ , H_q = \frac{\omega}{2}\sigma_z \qquad (3)$$

where $\beta = \frac{1}{k_B T}$ and Tr denotes the matrix trace, i.e., the sum of the diagonal elements. The denominator $Tr\left[e^{-\beta H_q}\right]$ is the partition function *Z* which ensures a proper normalization $Tr[\rho_{th}(\beta)] = 1$. In our simulation, the initial state of the full N-qubit system is the tensor product:

$$\rho_0 = \rho_{sys} \otimes {\rho_{th}}^{\otimes(N-1)} \qquad (4)$$

So, for the three-qubit case (system $Q_0$ qubit and two baths $Q_1$ and $Q_2$)

$$\rho_0 = \rho_{sys} \otimes {\rho_{th}}^{(1)} \otimes {\rho_{th}}^{(2)} \qquad (5)$$

where $\rho_{sys}$is the initial system state and ${\rho_{th}}^{(i)}$ is the thermal state of bath qubit $Q_i$. For the six-qubit case including a system qubit , first-layer ($L_1$) baths, and second layer ($L_2$) baths :

$$\rho_0 = \rho_{sys} \otimes {\rho_{th}}^{(1)} \otimes {\rho_{th}}^{(2)} \otimes {\rho_{th}}^{(3)} \otimes {\rho_{th}}^{(4)} \otimes {\rho_{th}}^{(5)} \qquad (6)$$

Using this set-up, each bath qubit is initialized in the same thermal state but remains uncorrelated with the system and with the other baths at *t=0*.

**2.2 Local Operators**

For a single qubit, the standard Pauli operators are:

$$\sigma_x = \begin{pmatrix} 0 & 1 \\ 1 & 0 \end{pmatrix}, \sigma_y = \begin{pmatrix} 0 & -i \\ i & 0 \end{pmatrix}, \sigma_z = \begin{pmatrix} 1 & 0 \\ 0 & -1 \end{pmatrix} \qquad (7)$$

In an N-qubit system, the operator $\sigma_\alpha$ acting on qubit $Q_i$ is lifted to the full Hilbert space as:

$$\sigma_\alpha^{(i)} = I^{\otimes i} \otimes \sigma_\alpha \otimes I^{\otimes(N-i-1)}, \alpha \in \{x, y, z\} \qquad (8)$$

where matrix *I* is the 2x2 identity matrix.

In the case of 3-qubit system for example, we can define the nine operators acting on a multi-qubit Hilbert space as:

$$\sigma_x^{(0)} = \sigma_x \otimes I \otimes I\ , \sigma_y^{(0)} = \sigma_y \otimes I \otimes I\ , \sigma_z^{(0)} = \sigma_z \otimes I \otimes I \qquad (9)$$

$$\sigma_x^{(1)} = I \otimes \sigma_x \otimes I\ , \sigma_y^{(1)} = I \otimes \sigma_y \otimes I\ , \sigma_z^{(1)} = I \otimes \sigma_z \otimes I \qquad (10)$$

$$\sigma_x^{(2)} = I \otimes I \otimes \sigma_x\,, \sigma_y^{(2)} = I \otimes I \otimes \sigma_y\,, \sigma_z^{(2)} = I \otimes I \otimes \sigma_z \qquad (11)$$

With on-site frequency $\omega$, the local Hamiltonian can be expressed as (N=3):

$$H_{local} = \sum_{i=0}^{N-1} \frac{\omega}{2}\sigma_z^{(i)} = \frac{\omega}{2}\left(\sigma_z^{(0)} + \sigma_z^{(1)} + \sigma_z^{(2)}\right)$$
$$= \frac{\omega}{2}(\sigma_z \otimes I \otimes I + I \otimes \sigma_z \otimes I + I \otimes I \otimes \sigma_z) \qquad (12)$$

**2.3 Coherent portions of the Hamiltonian**

For simplicity, we model the qubit–qubit interactions using an isotropic Heisenberg-type coupling that includes the $XX, YY, and\ ZZ$ terms with a coupling constant $J$. In this framework, each pair of coupled qubits interacts via the Hamiltonian:

$$H_{ij} = J\left(\sigma_x^{(i)}\sigma_x^{(j)} + \sigma_y^{(i)}\sigma_y^{(j)} + \sigma_z^{(i)}\sigma_z^{(j)}\right) \qquad (13)$$

Here, $J_{sb}$ is to denote the coupling between the source and Bath 1 or Bath 2 , and $J_{L1}$ is to mark the coupling strength between Bath1 and Bath2.

$Q_0$ and $Q_1$:

$$H_{01} = J_{sb}\left(\sigma_x^{(0)}\sigma_x^{(1)} + \sigma_y^{(0)}\sigma_y^{(1)} + \sigma_z^{(0)}\sigma_z^{(1)}\right) = J_{sb}\left(\sigma_x \otimes \sigma_x \otimes I + \sigma_y \otimes \sigma_y \otimes I + \sigma_z \otimes \sigma_z \otimes I\right) \ (14)$$

$Q_0$ and $Q_2$:

$$H_{02} = J_{sb}\left(\sigma_x^{(0)}\sigma_x^{(2)} + \sigma_y^{(0)}\sigma_y^{(2)} + \sigma_z^{(0)}\sigma_z^{(2)}\right) = J_{sb}\left(\sigma_x \otimes I \otimes \sigma_x + \sigma_y \otimes I \otimes \sigma_y + \sigma_z \otimes I \otimes \sigma_z\right) \ (15)$$

And between $Q_1$ and $Q_2$:

$$H_{12} = J_{L1}\left(\sigma_x^{(1)}\sigma_x^{(2)} + \sigma_y^{(1)}\sigma_y^{(2)} + \sigma_z^{(1)}\sigma_z^{(2)}\right) = J_{L1}\left(I \otimes \sigma_x \otimes \sigma_x + I \otimes \sigma_y \otimes \sigma_y + I \otimes \sigma_z \otimes \sigma_z\right) \ (16)$$

Thus, the 3-qubit Hamiltonian (triangle) is made of the system qubit $Q_0$ and the "bath triangle" $Q_1 - Q_2$ with a system–bath edge coupling $J_{sb}$ and a mutual coupling $J_{L1}$:

$$H = H_{local} + J_{sb}\left(\sigma_x^{(0)}\sigma_x^{(1)} + \sigma_y^{(0)}\sigma_y^{(1)} + \sigma_z^{(0)}\sigma_z^{(1)}\right) + J_{sb}\left(\sigma_x^{(0)}\sigma_x^{(2)} + \sigma_y^{(0)}\sigma_y^{(2)} + \sigma_z^{(0)}\sigma_z^{(2)}\right)$$
$$+ J_{L1}\left(\sigma_x^{(1)}\sigma_x^{(2)} + \sigma_y^{(1)}\sigma_y^{(2)} + \sigma_z^{(1)}\sigma_z^{(2)}\right) \qquad (17)$$

Using the same argument, we can expand the whole coherent portion of the Hamiltonian for a 6-qubit system (see Fig. 1b) including the local Hamiltonian plus the nine combined terms of interacting coupling:

$$H = H_{local} + J_{sb}\sum_{\alpha\in\{x,y,z\}}\left(\sigma_\alpha^{(0)}\sigma_\alpha^{(1)} + \sigma_\alpha^{(0)}\sigma_\alpha^{(2)}\right) + J_{L1}\sum_{\alpha\in\{x,y,z\}}\sigma_\alpha^{(1)}\sigma_\alpha^{(2)} + J_{L12}\sum_{\alpha\in\{x,y,z\}}\left(\sigma_\alpha^{(1)}\sigma_\alpha^{(3)} + \sigma_\alpha^{(1)}\sigma_\alpha^{(4)} + \sigma_\alpha^{(2)}\sigma_\alpha^{(4)} + \sigma_\alpha^{(2)}\sigma_\alpha^{(5)}\right) + J_{L2}\sum_{\alpha\in\{x,y,z\}}\left(\sigma_\alpha^{(3)}\sigma_\alpha^{(4)} + \sigma_\alpha^{(4)}\sigma_\alpha^{(5)}\right) \qquad (18)$$

$$H_{local} = \sum_{i=0}^{5}\frac{\omega}{2}\sigma_z^{(i)} = \frac{\omega}{2}\left(\sigma_z^{(0)} + \sigma_z^{(1)} + \sigma_z^{(2)} + \sigma_z^{(3)} + \sigma_z^{(4)} + \sigma_z^{(5)}\right) \qquad (19)$$

It bears repeating here that the definition of the interaction terms is consistent with the 2-layered bath structure as depicted in Figure 1b:

$$J_{01} = J_{02} = J_{sb},\ J_{12} = J_{L1},\ J_{13} = J_{14} = J_{24} = J_{25} = J_{L12}\ and\ J_{34} = J_{45} = J_{L2} \qquad (20)$$

While the present model adopts a Heisenberg-type Hamiltonian for clarity and tractability, the framework is readily extensible to more complex bath–system interactions. This includes incorporation into higher-dimensional or composite quantum circuit designs, where structured bath elements could potentially be directly embedded and tuned within hardware-specific architectures.

**2.4 Dissipative Dynamics in Structured Baths**

It is important to reemphasize that in this framework, every qubit (system and bath nodes alike) remains coupled to an infinite Markovian reservoir via Lindblad dissipators. Memory effects arise solely from the explicit qubit–qubit couplings within the finite bath structure, not from the dissipators themselves. Once we establish the coherent portion of the Hamiltonian, we introduce its dissipative segment using the Lindblad master equation [12, 13]:

$$\frac{\mathrm{d}\rho}{\mathrm{dt}} = -i[H,\rho] + \sum_k \left( L_k \rho L_k^\dagger - \frac{1}{2}\left\{ L_k^\dagger L_k, \rho \right\} \right) \qquad (21)$$

where $L_k$'s are the collapse operators that represent dissipative components including thermal relaxation and dephasing, each representing an irreversible process. These operators are strictly Markovian: thermal relaxation, thermal excitation, and pure dephasing processes are modeled with constant rates $\gamma$, corresponding to time-independent exponential decay or excitation. These terms assume the Born–Markov approximation, in which bath correlation times are negligibly short and no explicit memory kernel appears.

However, by embedding the bath degrees of freedom directly into the Hamiltonian as explicit qubits, we depart from the conventional picture. These immediate bath qubits are now part of the full system Hilbert space and can exchange energy and coherence with each other and with the central system qubit. As a result, information can flow back from the bath into the system through coherent interactions, even though the local dissipators remain purely Markovian. This explicit modeling of bath structure allows us to investigate and quantify information backflow and memory effects arising from unitary bath–bath and bath–system couplings.

There are three types of collapse operators employed:

1. Pure dephasing of the system qubit $(Q_0)$ to model phase randomization without energy exchange:

   $$L_{deph} = \sqrt{\gamma_{sys}}\sigma_z^{(0)} \qquad (22)$$

   Referring back to Lindblad master equation, this contributes to the dissipator as:

   $$\mathcal{D}_{deph}^{(0)}[\rho] = \left( \gamma_{sys}\sigma_z^{(0)}\rho\sigma_z^{(0)} - \frac{1}{2}\left\{ \gamma_{sys} I, \rho \right\} \right) = \gamma_{sys}\left( \sigma_z^{(0)}\rho\sigma_z^{(0)} - \rho \right) \qquad (23)$$

   $$\text{since } \left( \sigma_z^{(0)} \right)^2 = I$$

2. Thermal relaxation for Layer 1 Baths

   These are applicable to the two qubits forming Layer 1 bath. Following the standard treatment of thermal relaxation in open quantum systems [14, 15], we set each bath qubit to be coupled to an external thermal reservoir at inverse temperature $\beta$ wherein each has two Lindblad operators:

   $$L_\downarrow^{(i)} = \sqrt{\gamma_i(1+n_{th})}\sigma_i^- \ (emision) \qquad (24)$$

$$L_{\uparrow}^{(i)} = \sqrt{\gamma_i n_{th}}\sigma_i^{+} \ (absorption) \qquad (25)$$

These Lindblad operators describe energy exchange with a thermal bath. The emission term quantifies the way the system loses an excitation to the bath whereas the absorption term marks the way the system gains an excitation from the bath, due to non-zero temperature. For $L_1$, $\gamma_i = \gamma_{L1}$ which is the damping rate for both upward and downward processes for each qubit $Q_1$ and $Q_2$. It controls how fast Layer 1 exchanges energy with its environment. The thermal occupation number $n_{th} = \frac{1}{e^{\beta\omega}-1}$ is directly linked to the preservation of the detailed balance so as to maintain the ratio of the upward and downward transition rates ($\Gamma_{\uparrow}$ *and* $\Gamma_{\downarrow}$ respectively) obey:

$$\frac{\Gamma_{\uparrow}}{\Gamma_{\downarrow}} = \frac{\gamma_i n_{th}}{\gamma_i(n_{th}+1)} = e^{-\beta\omega} \quad (26)$$

which aims at maintaining the correct thermal equilibrium state at temperature T, consistent with the Boltzmann distribution.

3. Thermal relaxation for Layer 2 Baths

   In principle, this has a similar two-operator structure as Layer 1, but with different thermal damping rate for $\gamma_{L2}$ applied to $Q_3 - Q_5$. By differentiating the magnitude of $\gamma$, we can model slower or faster equilibration in the outer bath layer compared to Layer 1.

Putting these together, the Lindblad master equation contribution for each bath qubit becomes:

$$\mathcal{D}_{thermal}^{(i)}[\rho] = L_{\downarrow}^{(i)}\rho L_{\downarrow}^{(i)\dagger} - \frac{1}{2}\left\{L_{\downarrow}^{(i)\dagger}L_{\downarrow}^{(i)}, \rho\right\} + L_{\uparrow}^{(i)}\rho L_{\uparrow}^{(i)\dagger} - \frac{1}{2}\left\{L_{\uparrow}^{(i)\dagger}L_{\uparrow}^{(i)}, \rho\right\} \qquad (27)$$

(For more details in Supplementary Section S1)

Additional note on lifetimes**.** In this work we specify decoherence directly through Lindblad rates, since they provide the most transparent link between the master equation and our simulations. Nevertheless, many experimental reports use the language of relaxation and coherence times, $T_1$ , the time constant of exponential relaxation of the excited-state population, and $T_2$ , the time constant of the exponential decay of coherence. For thermal relaxation channels [24-27], the effective equilibrium time is $T_1 = 1/\gamma_i(1 + 2n_{th})$ ,which reduces to $T_1 = 1/\gamma_i$ in the zero-temperature limit ($n_{th} = 0$). For the system pure-dephasing channel with collapse operator $L_{deph} = \sqrt{\gamma_{sys}}\sigma_z^{(0)}$, the off-diagonal elements in the density matrix decay as $e^{-2\gamma_{sys}t}$, corresponding to $T_2$ as $1/2\gamma_i$. These relations are provided solely for comparison with experimental conventions; all analysis in this work is carried out directly in terms of the Lindblad rates.

**3. Simulation protocol and first results: thermal equilibration**

We first simulate the master equation with the Hamiltonians and collapse operators defined above using *mesolve* in QuTiP by only using the 3-qubit system. The parameters are

$$\omega = 1, \beta = 1, J_x = 0.1, J_y = 0.1, J_z = 0.05$$

$$J_{sb} = 0.2, J_{L1} = 0.2, \gamma_{sys} = 0.005 \ and \ \gamma_{L1} = 0.005$$

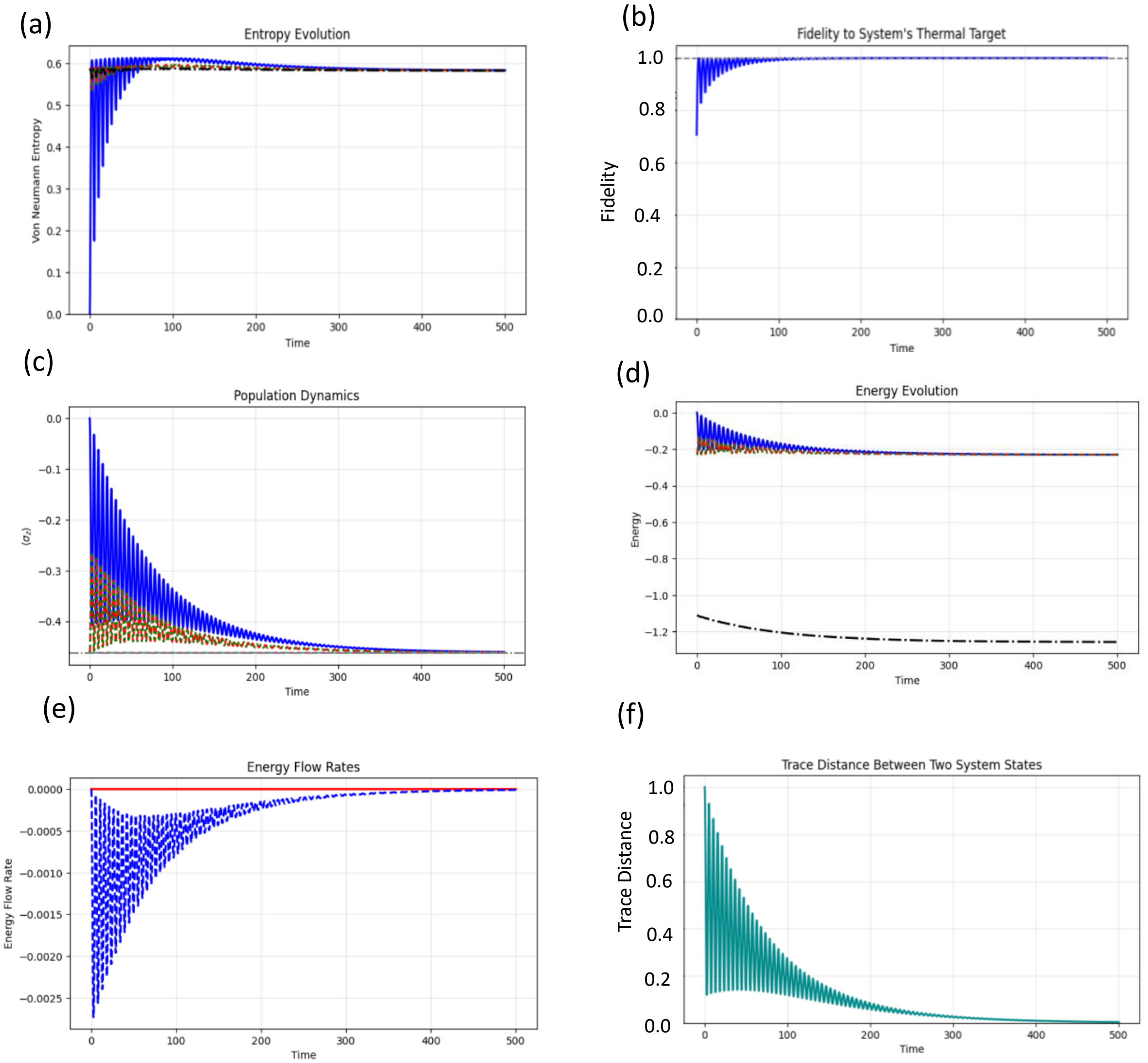

**Figure 2a-f:** Results from the minimal 3-qubit simulation under weak coupling conditions (parameters listed above), showing: (a) von Neumann entropy of the system and baths converging to a common equilibrium value; (b) fidelity of the system state to its thermal target rapidly approaching unity; (c) population dynamics of the system and baths relaxing toward thermal values; (d) local and total energy evolution, with a small initial drop in total energy before stabilization; (e) coherent and dissipative energy flow rates during equilibration; and (f) trace-distance dynamics between orthogonal initial states, yielding a significant backflow score to mark a measure of non-Markovianity.

To connect with the criterion outlined by Eisert and co-workers that even weak coupling to a thermal bath can suffice for full thermalization, we first verify that our parameter choices place the minimal 3 qubit setup firmly in the weak coupling regime. All coupling strengths satisfy $J_x, J_y, J_z \ll \omega$ and the dephasing rate is small relative to $\omega$, ensuring the system operates in a weak-coupling Lindblad regime consistent with the framework “thermalization without large entanglement” analysis.

Below are some of the results from the minimal 3-qubit system using the parameters specified above. Our results follow this line of reasoning by showing that, despite the Markovian nature of the Lindblad terms, the inclusion of explicit bath qubits and their interactions allows the system to approach a thermal state under weak coupling conditions. Specifically:

Figure 2a shows the von Neumann entropy of the system and baths. The system entropy rises rapidly from zero to its equilibrium value, matching the bath entropies and remaining stable thereafter, indicative of successful thermalization.

The von Neumann entropy is defined as [16]:

$$S(t) = -Tr\left[\rho_{sys}(t)\log\rho_{sys}(t)\right] \tag{28}$$

Figure 2b presents the fidelity of the system state to its thermal target:

$$F(t) = \left[Tr\left(\sqrt{\sqrt{\rho_{th}}\rho_{sys}(t)\sqrt{\rho_{th}}}\right)\right]^2 \tag{29}$$

which is the standard Uhlmann fidelity for mixed quantum states [17] (see also the formulation in [18]). In our simulation, the fidelity reaches unity within the first ~50-time units and remains constant. This further confirms convergence to the thermal state.

Figure 2c depicts the population dynamics $\langle\sigma_z\rangle$ for the system and baths:

$$\langle\sigma_z^{(i)}(t)\rangle = Tr\left[\rho(t)\sigma_z^{(i)}\right] \tag{30}$$

wherein the system population relaxes toward the thermal target value, with bath populations approaching complementary values as expected for a thermal exchange process [19].

Figure 2d presents the local energy evolution for the system and baths, along with the total Hamiltonian expectation value. The local energy of each subsystem is computed from:

$$E_i(t) = Tr[\rho(t)H_i] \tag{31}$$

where $H_i$is the single-qubit Hamiltonian. The total energy:

$$E_{tot}(t) = \sum_i E_i(t) \tag{32}$$

shows an initial transient drop before stabilizing, indicating early-time energy leakage into dissipative channels before reaching a steady state. Once past the transient, $E_{\text{tot}}$ remains constant within numerical tolerance, confirming energy conservation in the asymptotic regime.

Figure 2e reports the instantaneous coherent and dissipative energy flow rates, defined following Spohn's treatment of entropy production [20]:

$$\Phi_{\text{coh}}(\text{t}) = \frac{dE_{coh}}{dt}, \Phi_{\text{diss}}(\text{t}) = \frac{dE_{diss}}{dt} \tag{33}$$

demonstrating how thermalization proceeds through both coherent exchange and irreversible dissipation. Finally, the trace-distance backflow plot was evaluated over time using orthogonal $|0\rangle$ and $|1\rangle$ initial system states as shown in Figure 2f. The trace distance between these initially distinct system states was calculated:

$$D(t) = \frac{1}{2}Tr|\rho_1(t) - \rho_2(t)| \tag{34}$$

where $|\rho_1(t) - \rho_2(t)|$ denotes the trace norm. Following Breuer, Laine, and Piilo (BLP) [2], non-Markovianity is then characterized by the total amount of increase in this quantity,

$$\mathcal{N} = \int_{\dot{D}(t)>0} \dot{D}(t) dt \qquad (35)$$

In practice, we compute $D(t)$ for the chosen orthogonal pair $|0\rangle$ and $|1\rangle$. This yields a lower bound on the full BLP measure, which is defined via maximization over all initial state pairs. The resulting backflow score $\mathcal{N}$ for this run thus provides quantitative evidence of information return and non-Markovian behavior, even in the weak-coupling regime. For comparability across simulations with different total durations, the script also reports a value $\frac{\mathcal{N}}{T}$, where $T = t_{final} - t_{initial}$, representing the average backflow per unit time. Further details of the numerical procedure are provided in Supplementary Section S2.

These results show that even a minimal three qubit setup made of a central system qubit connected to two bath qubits can reach a thermal state with only weak coupling between them. The entropy, fidelity, and population figures all point to a quick approach to steady values that match the target thermal distribution. The total energy levels off after an early drop which means that energy is conserved once the system settles. The trace distance backflow shows that memory effects are still present even at weak coupling. This agrees with the work of Eisert and colleagues who found that small baths can still cause thermalization without the need for very strong coupling or very high entanglement. Our bath qubit model gives a clear and direct example of this idea and sets the stage for the more complex bath designs studied in the next sections.

**4. Bath memory categories using the 6-qubit system and FFT extraction**

Building on the three-qubit demonstrations of thermalization from Segment 3, we extend the model to a six-qubit structured bath system arranged in two hierarchical layers. Each logical system qubit is coupled to its immediate bath qubits (Layer 1), which are in turn connected to secondary bath qubits (Layer 2). This configuration enables explicit control over memory retention, transfer, and dissipation pathways, allowing us to systematically investigate how bath topology affects observable spectral features.

The simulation employs a similarly Heisenberg-type coupling $(XX + YY + ZZ)$ with tunable parameters for system–bath and intra-bath interactions. Thermal Lindblad dissipators are applied to selected bath qubits $(Q_1 - Q_5)$ to mimic finite-temperature environments. The QuTiP-based implementation tracks the joint density matrix under the Lindblad master equation, recording time series of relevant observables such as $\langle\sigma_z\rangle$ for the system qubits. The primary quantity analyzed here is $\langle\sigma_z\rangle$ of the system qubit. This choice is intentional for several reasons:

1. Direct Experimental Accessibility :
   - In trapped-ion qubits, $\sigma_z$ corresponds to the population difference between the two internal states, measured via state-dependent fluorescence detection [21]. The bright and dark states are measured experimentally through photon count statistics [22].
   - In neutral-atom qubits, $\sigma_z$ can be read out by fluorescence or absorption imaging after shelving into different hyperfine states [23].
   - In superconducting qubits, $\sigma_z$ dispersive readout of a coupled resonator yields a direct measurement of the basis populations [24].
2. Sensitivity to Coherence and Memory Effects :
   Oscillations in $\langle\sigma_z\rangle$ reflect coherent exchange of information between system and bath, while decay and broadening indicate dissipative or decohering channels [14, 25].
3. Compatibility with FFT Analysis :
   Because the time-domain observable $\langle\sigma_z(t)\rangle$ is real and bounded to [−1,1], a straightforward FFT produces a clean, interpretable frequency-domain signature without the need for additional basis transformations [16].

*Simulation Parameters*

The following parameters were fixed across all cases unless otherwise noted:

- System–Bath coupling: $J_{sb} = 1.0$
- Intra-Layer 1 coupling: $J_{L1}$ varies per case
- Intra-Layer 1 coupling: $J_{L2} = 1.0$
- Inter-Layer ($L_1 \leftrightarrow L_2$) coupling: $J_{L12}$ varies per case
- Bath temperature: T = 0.5 (in $\hbar\omega$/kB units)
- Simulation time: $t_{max}$ = 500 (dimensionless units), 5000-time steps

In this part of our study of the bath's internal structure and its role in memory retention, the system-bath coupling, $J_{sb}$, was intentionally set to a high value of 1.0. This was done to ensure a scenario of significant initial information leakage from the system, effectively "flooding" the bath with information. This approach allows us to better assess how memory is subsequently retained and managed solely due to the bath's engineered structure and the inter-qubit interactions within it. This choice makes it easier to see how bath topology affects non-Markovian dynamics.

*Memory Categories*

By varying the parameters, we identify three distinct bath memory categories that isolate the effect of bath structure on spectral signatures, as shown in Table 1. Figure 3 illustrates the simulation schematic for the 6-qubit setup and the corresponding bath memory categories.

Table 1: Three cases of memory categories based on bath structures and connectivity

| Case | Description | $J_{L1}$ | $J_{L12}$ | $\gamma_{L1}$ | $\gamma_{L2}$ | Expected memory |
|---|---|---|---|---|---|---|
| 1- Fully dissipative | Strong dissipation in both layers; inter-layer coupling enabled | 0.05 | 0.02 | 0.3 | 0.3 | Close to Markovian |
| 2-Retained memory | Intra-layer coupling to Layer 2 suppressed; Layer 1 and 2 dissipate normally. | 0.05 | 0.001 | 0.02 | 0.02 | Non-Markovian |
| 3- Transferred Memory | Enhanced coupling inter-layer enabled; suppressed thermal dissipation in Layer 2 | 0.05 | 0.75 | 0.02 | 0.001 | Partially Markovian |

For each case, we performed a Fast Fourier Transform (FFT) on the time series of the system qubit to extract the dominant frequency components. These spectra serve as fingerprints of the bath structure, revealing:

- The presence or absence of persistent oscillations
- Frequency shifts induced by inter-layer coupling
- Broadening or suppression linked to dissipation strength

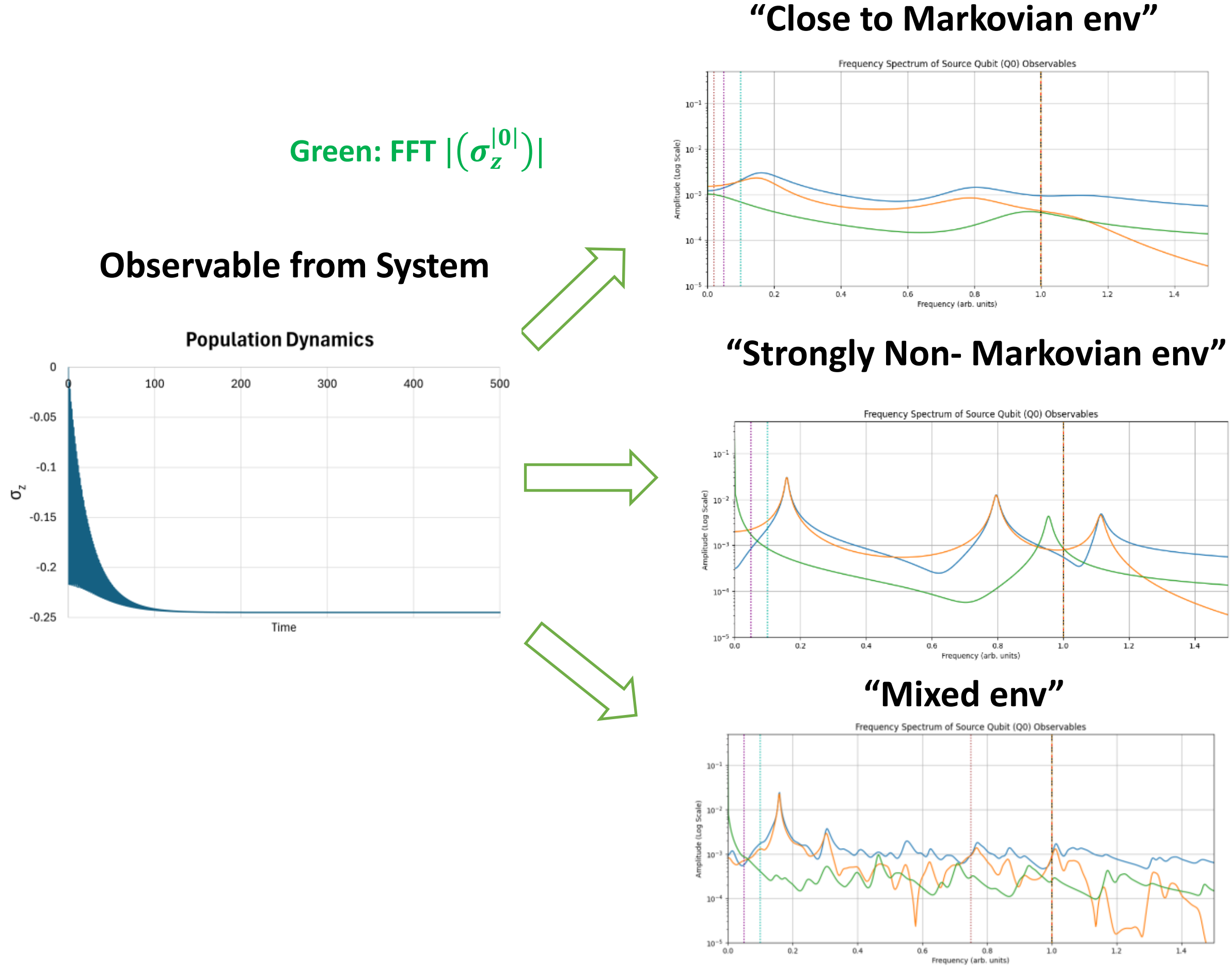

**Figure 3** The workflow of the signal processing for the 6-qubit system-bath generating the normalized FFT spectra of $\left|\sigma_z^{|0|}\right|$ (Green) and the resulting bath memory categories based on the spectra. Extra Blue and Orange curves are from FFTs of $\left|\sigma_x^{|0|}\right|$ and $\left|\sigma_y^{|0|}\right|$ respectively.

In Case 1, the spectrum is broadened with reduced amplitude, consistent with rapid memory loss – close to Markovian. In Case 2, the FFT shows sharp peaks corresponding to well-defined memory oscillations in Layer 1 – Non-Markovian. In Case 3, peaks shift and split, indicating partial transfer of memory into Layer 2. More expansions of these broad memory-bath categories are shown in the Supplementary Section S3.

*General Observations*

- Enabling inter-layer flow and increasing thermal dissipation produces behavior closer to Markovian dynamics. Both inter-layer coupling strength and dissipation rates play critical roles in allowing energy and information to leave the first layer, thereby reducing observable memory effects.
- Conversely, non-Markovianity and trace distance backflow are enhanced when intra-layer coupling between the two bath layers is suppressed. By pinching off energy and information flow to Layer 2, dynamics are confined to the first layer, which reverberates back to the source and exhibits clear memory retention. The relatively modest thermal relaxation in Layer 1 also helps slow down the decoherence rate.
- In practical devices, neither one of these two extreme cases is followed. Certainly, leakage to secondary layers is often unavoidable. However, carefully tuning the coupling strength can still leave identifiable spectral signatures in the FFT signals, especially when accumulated over multiple runs or devices. Thus, these features can then be used as a diagnostic of bath structure.

**5. From Observables to Machine Learning-Based Inference of Bath Structure**

With the generation of the observable signals, which come primarily from the spectral signatures obtained from $\langle \sigma_z \rangle$ and to some extent the coherent strength features of $\langle \sigma_x \rangle, \langle \sigma_y \rangle$, we can move beyond purely qualitative interpretation to a quantitative classification and inference framework.

To ensure comparability between runs and reduce bias from signal amplitude variations, first each FFT spectrum is normalized individually. The normalization process is applied to all three observables ($\sigma_x, \sigma_y, \sigma_z$), producing dimensionless spectra where the maximum amplitude in each channel is scaled to unity. This ensures that differences in spectral shape, rather than raw magnitude, dominate the classification.

Following normalization, the combined FFT features from all three channels are concatenated into a single feature vector per simulation run. We then apply Principal Component Analysis (PCA) to compress the feature space with controllable N dimensions that we can vary depending on the complexity of the spectra, retaining only the leading components that capture the majority of spectral variance. This dimensionality-reduced representation is then

**Simulated normalized FFT's**

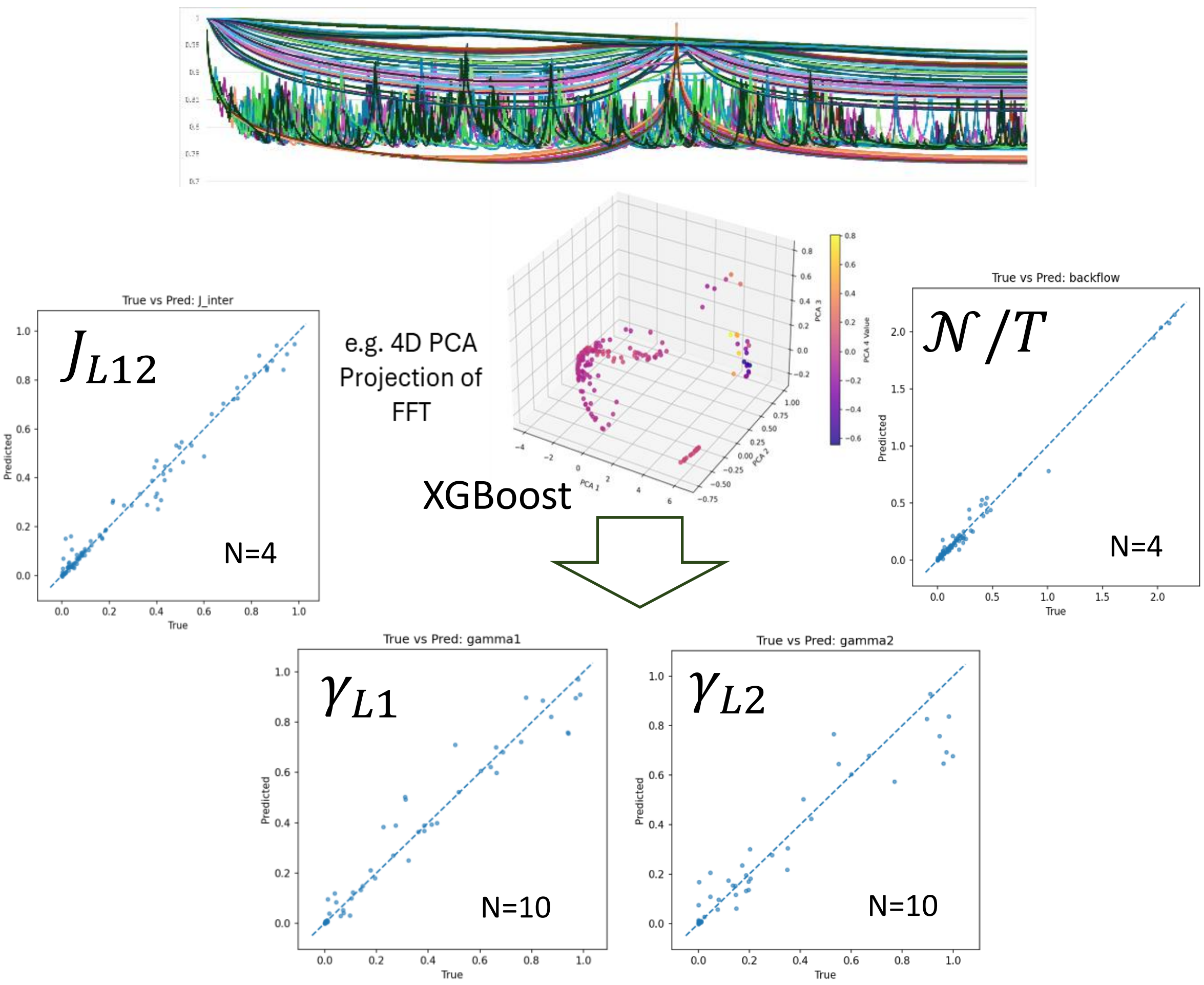

**Figure 4**. Schematics of the workflow after collecting the database of normalized and compressed FFT signals where their PCA projections are then used as training data to map out the bath parameters $(J_{L12}, \gamma_{L1}, \gamma_{L2}, Backflow)$ using n dimensional PCA with N dimension ranges from 4-10.

used as input for supervised learning models, most notably XGBoost regression to perform inverse engineering: predicting underlying bath structure parameters (e.g., inter-layer coupling strengths, thermal dissipation rates) from the measured spectra. An example application is shown in Figure 4, where the regression model successfully recovers key thermal dissipation and inter-layer bath coupling features from the FFT data.

While this is by no means an exhaustive exploration of the methodology, it demonstrates a straightforward yet powerful extension of the simulation analysis pipeline using accessible machine learning tools. This approach can be expanded to other regression or classification architectures and refined by incorporating additional observables or higher-order spectral features. The methodology is also foundational; in the future, more advanced techniques, such as generative AI, could be utilized to infer more complex bath structures from spectral data.

While the full FFT spectra can span hundreds of frequency bins, we find that a PCA projection with a limited dimensionality of four components (see Figure 4) is sufficient to retain accurate inference of key bath parameters. This is especially true for those governing vertical leakage, such as inter-layer coupling and dissipation rates. This dimensionality reduction not only improves computational efficiency but also enhances interpretability, as each principal component reflects a distinct spectral feature tied to bath topology and dissipation. The success of low-dimensional PCA underscores the robustness of spectral fingerprints as a diagnostic tool, enabling a scalable and physics-aware inference. In addition, the FFT analysis can optionally be further enhanced through Digital Signal Processing (DSP) methods, such as multitaper Discrete Prolate Spheroidal Sequences (DPSS ) techniques [26], which reduce variance and sharpen resolution of narrow resonances. More details of the FFT and machine learning procedures are provided in Supplementary Sections S4 and S5.

Lastly, we note that while very simple bath models can sometimes be solved analytically, this possibility disappears quickly once multiple layers, mixed couplings, and structured connectivity are considered. In these more realistic regimes, analytic approaches become impractical or altogether intractable. Machine learning provides a scalable tool that complements, rather than replaces, analytic theory. By capturing correlations and extracting effective parameters directly from high-dimensional dynamics, ML extends the range of problems that can be addressed beyond what closed-form methods allow.

### 6. Connection to the continuum model of non-Markovian

It is instructive to examine briefly the connection between the nodal-based approach that we propose and the continuum model framework advanced in the Nakajima–Zwanzig (NZ) formalism. This comparison not only reveals the underlying commonalities between the two methods but also highlights practical advantages of a nodal representation. This is particularly relevant in the context of present-day hardware engineering, where the logical qubit interacts not with an abstract, infinite bath, but with a finite, structured environment represented by qubit nodes, which may correspond to physical elements or modeling constructs.

In the NZ approach [5,15], the system's evolution takes the form:

$$\frac{d}{dt}\rho_S(t) = -i[H_s, \rho_S(t)] + \int_0^t \mathcal{K}(t-t')\rho_S(t')dt' \quad (36)$$

where the time-nonlocal memory kernel $\mathcal{K}(t-t')$ captures how states influence the present. The kernel can be built up directly from a prescribed spectral density $J(\omega)$ through a bath correlation function $C(t)[14]$**:**

- For zero temperature:

$$C(t-t') = \int_0^\infty d\omega\, J(\omega)e^{-i\omega(t-t')} \quad (37)$$

- For finite temperature:

$$C(t-t') = \int_0^\infty d\omega\, J(\omega) \left[\coth\left(\frac{\beta\omega}{2}\right)\cos\big(\omega(t-t')\big) - i\sin\big(\omega(t-t')\big)\right] \qquad (38)$$

In the idealized Markovian limit, $J(\omega)$ is broad and featureless, leading to a correlation function that decays rapidly and monotonically, producing negligible memory and no significant backflow. Conversely, when $J(\omega)$ is structured or narrowband, $C(t)$ exhibits slowly decayed oscillations, and the resulting dynamics are distinctly non-Markovian.

In its mathematical form, the bath is then modeled as a set of modes with frequencies $\omega_k$ and $g_k$ couplings to the system, which are characterized collectively by the spectral density [17]:

$$J(\omega) = \sum_k |g_k|^2 \delta(\omega - \omega_k) \quad (39)$$

where $\delta(\omega - \omega_k)$ is the Dirac delta function which selects the frequency $\omega_k$for mode $k$.

If the bath contains a very large number, the discrete mode index $k$ can be replaced by a continuous frequency variable $\omega$. We can define the density of states (DOS):

$$\rho(\omega) = \sum_k \delta(\omega - \omega_k) \qquad (40)$$

This function tells us how many modes exist per unit frequency near ω. The sum over $k$ in Eq. (37) can now be rewritten as an integral over frequencies weighted by the density of states. The replacement:

$$\sum_k f(\omega_k) \rightarrow \int_0^\infty \rho(\omega) f(\omega) d\omega \qquad (41)$$

turns the discrete sum in Eq. (37) into an integral over mode frequencies.
Using $f(\omega_k) = |g_k|^2 \delta(\omega - \omega_k)$, we obtain

$$J(\omega) = \int_0^\infty \rho(\omega') |g(\omega')|^2 \delta(\omega - \omega') d\omega' \quad (42)$$

where $|g(\omega')|^2$ is the coarse-grained coupling strength for modes near $\omega$. Using the sifting property of the delta function:

$$\int f(\omega')\delta(\omega - \omega')\, d\omega' = f(\omega) \qquad (43)$$

$$J(\omega) = \rho(\omega)|g(\omega)|^2 \; (44)$$

where:

- $\rho(\omega)$ tells you how many bath modes exist near a given $\omega$ frequency i.e. the density of states.
- $|g(\omega)|^2$ tells you how strongly those modes couple to the system.
- The product is the spectral density $J(\omega)$, which is the key quantity entering the Nakajima–Zwanzig memory kernel and directly controls the system's non-Markovian dynamics.
- Here we adopt the condensed-matter convention in which the spectral density $J(\omega)$ incorporates normalization factors, so no explicit $\frac{1}{2\pi}$ prefactor appears. This choice is consistent with standard treatments in density-functional theory and open-system physics.

*General relation between $J(\omega)$and $C(t)$:*
The bath correlation function $C(t)$ is the Fourier transform of the spectral density.

- *Discrete mode representation:*

$$C(t) = \sum_k J_k e^{-i\omega_k t} \quad (45)$$

  where $J_k = |g(k)|^2$ is the coupling strength of mode $k$ and $\omega_k$ its frequency.

- *Continuum limit:*
  When the bath modes become dense in frequency, the sum turns into an integral weighted by the density of states ρ(ω):

$$C(t) = \int_0^\infty J(\omega)\, e^{-i\omega t} d\omega, \text{ with } J(\omega) = \rho(\omega)|g(\omega)|^2 \quad (46)$$

The *width* and *structure* of $J(\omega)$ directly control the decay and oscillatory behavior of $C(t)$:

- A broad $J(\omega)$ → short correlation time, monotonic decay (Markovian limit).
- A narrow or structured $J(\omega)$ → long-lived oscillations, memory effects (non-Markovian).

In the structured-bath simulations presented here, the narrowband and broadband limits of $J(\omega)$ correspond to the most and least memory-retaining configurations, respectively, with intermediate structures yielding partial-memory behavior consistent with the continuum picture. Thus, our nodal approach can be viewed as a *discretized realization* of this same physics, where the "bath" is constructed from interconnected nodes with tunable couplings, namely intra-layer ($J_{L1}$) and inter-layer ($J_{12}$) and site-specific dissipation rates ($\gamma$). In the limit of *very high connectivity and large node count*, the network's collective modes densely populate frequency space, and $J(\omega)$ approaches a continuous spectrum. In this regime, the nodal model converges with the same effective dynamics predicted by the NZ continuum formalism.
That said, the real strength of the nodal representation emerges actually *when connectivity is finite and structured.* By engineering the pattern and strength of couplings, we can selectively engineer $J(\omega)$ and therefore control the form of $C(t)$, producing tailored degrees of non-Markovianity. This is particularly useful for quantum hardware, where the bath is made of real components like residual couplings, control lines, and nearby elements, not an abstract continuum. In such environments, the ability to model and manipulate a structured bath provides a direct bridge between open-system theory and practical noise engineering.

Here, we illustrate this connection by examining the three cases that demonstrate how structured connectivity manifests in $J(\omega)$ and in the corresponding bath correlation function $C(t)$. For clarity, throughout the figures we show the one-sided magnitude spectrum $|J(\omega)|$ for $\omega \geq 0$ which captures the dominant frequency fingerprints while omitting the phase information present in the full complex. In the Markovian-like case, $C(t)$decays rapidly and monotonically reflecting the short correlation time of a broad $|J(\omega)|$. In contrast, in the non-Markovian-like case, $C(t)$ exhibits long-lived oscillations, arising from a structured or narrowband $|J(\omega)|$ where a few frequency components maintain coherence over extended times. Finally, we also tested an intermediate case in which $|J(\omega)|$ was partially smeared, reflecting spectral characteristics in between the two extremes.

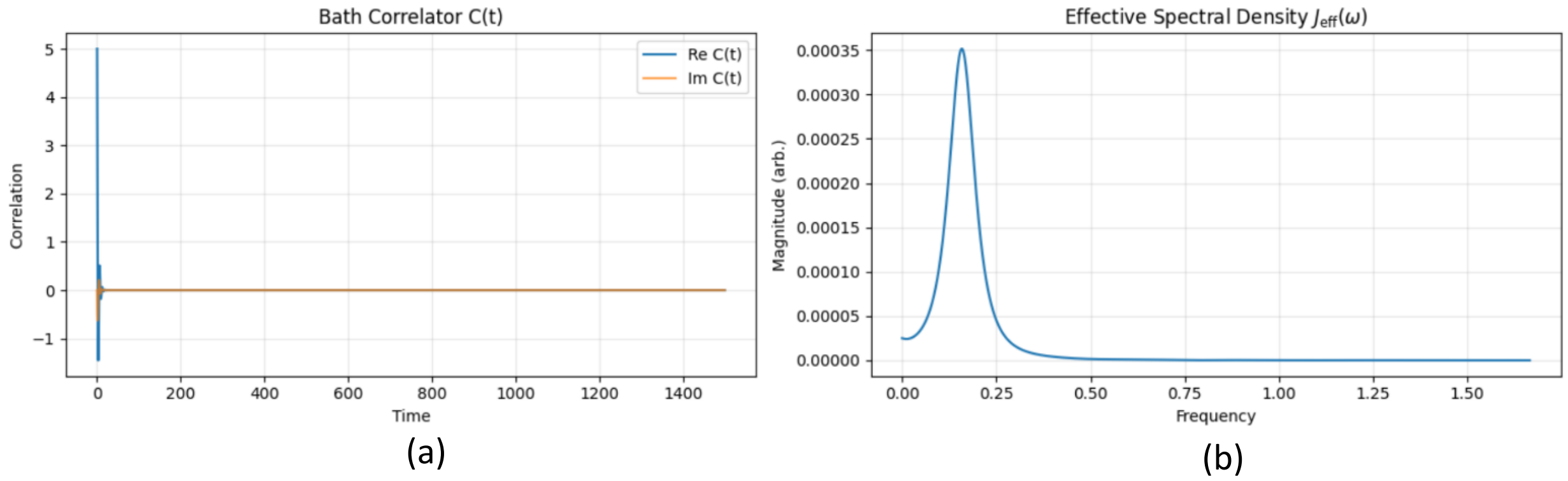

**Figure 5a–b.** Strong Markovian regime obtained with nodal parameters $J_{sb} = 0.8, J_{L1} = 1.0$, $J_{L12} = 0.5$, $J_{L2} = 1.0$, $\gamma_{L1} = 0.2$ , $\gamma_{L2} = 0.2$ (a) Bath correlation function $C(t)$ showing monotonic decay and negligible oscillatory structure, consistent with minimal memory effects. (b) Corresponding spectral density $|J(\omega)|$exhibiting a broad, featureless single-hump profile, indicating a densely populated and strongly mixed bath mode distribution.

<u>Case 1 – Broad, Featureless Spectral Density (Near-Markovian Limit)</u>

In the first limiting case, the nodal network is configured with high connectivity and uniform coupling strengths, such that every node interacts nearly equally with many others. The resulting collective mode structure densely covers frequency space, and the couplings g(ω) vary only weakly with $\omega$. In this situation, the spectral density is broad and essentially flat over the frequency range relevant to the system. From the NZ perspective, a broad, featureless $|J(\omega)|$ yields a bath correlation function that decays rapidly and monotonically, leaving negligible memory. This produces a dynamical regime close to Markovian, with minimal backflow of information from the bath to the system. In the nodal picture, this corresponds to an environment where the structured nature of the bath is “washed out” by high connectivity, effectively simulating a continuum limit.

Figure 5a-b shows a case of a strong Markovian regime with nodal parameters ($J_{L1} = 1.0, J_{L12} = 0.8, J_{L2} = 1.0, \gamma_{L1} = 0.3$ , $\gamma_{L2} = 0.3$) as evidenced by the very small trace-distance backflow and the significant monotonic decay of the correlation function $C(t)$ shown in Fig. 5a, consistent with the absence of significant memory. The corresponding spectral density $|J(\omega)|$ as depicted in Fig 5b is broad and featureless with a single hump, indicating that the bath modes are densely distributed over frequency space and strongly mixed. This form of $|J(\omega)|$ produces a rapidly decaying $C(t)$ in the Nakajima–Zwanzig framework, exactly as observed here using our nodal-based model. The close agreement between the simulation and the NZ continuum expectation confirms that high bath connectivity and large dissipation rates drive the system toward the Markovian limit.

Case 2 – Narrowband or Structured Spectral Density (Strongly Non-Markovian Limit)

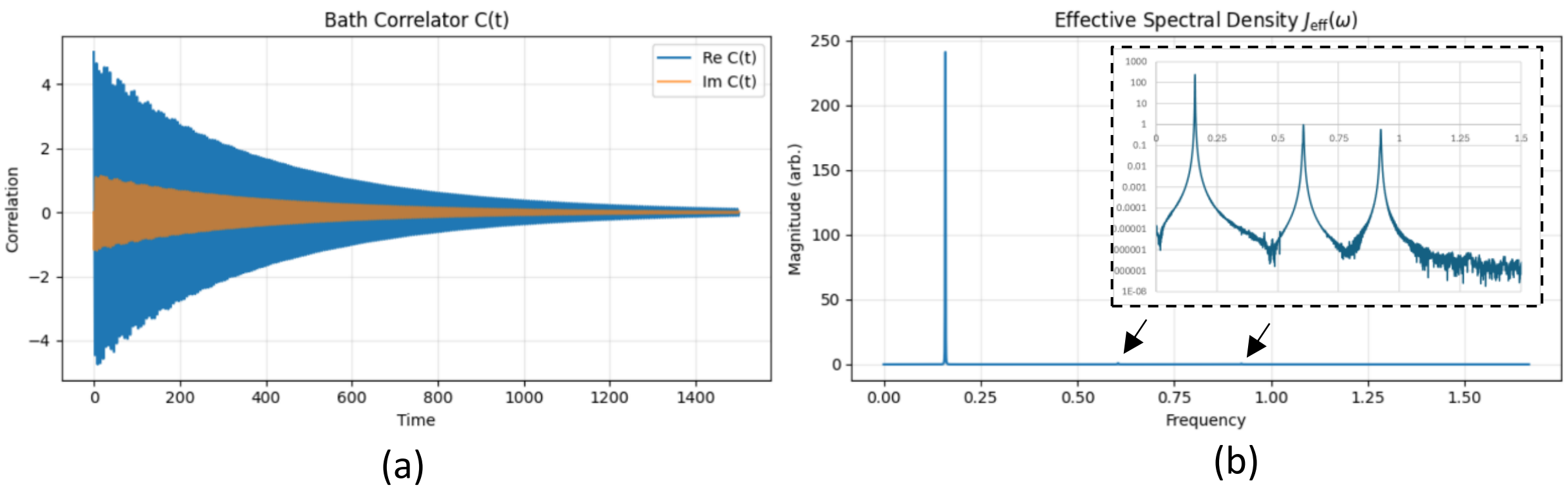

**Figure 6a–b**. Example of a non-Markovian regime (backflow = 1.55), obtained with nodal parameters ($J_{sb} = 0.8, J_{L1} = 0.5,\ J_{L12} = 0.003, J_{L2} = 0.1, \gamma_{L1} = 0.0008, \gamma_{L2} = 0.0008$). The bath correlation function $C(t)$ (a) exhibits long-lived oscillations with a very slow decay envelope, indicative of persistent memory. This behavior originates from the effective spectral density $|J(\omega)|$ which is sharply peaked at a small set of discrete frequencies as shown in (b).

At the opposite extreme, the network connectivity is sparse and highly structured. Coupling strengths are tuned so that only a subset of inter-node connections dominates, leading to discrete frequency bands or sharp peaks in bath density of states $\rho(\omega)$ and/or the coupling envelope $|g(\omega)|^2$. The spectral density J(ω) then develops narrow features, often modeled by Lorentzian or multi-peaked profiles [14]:

$$J(\omega) \propto \sum_n \frac{\Gamma_n}{(\omega-\omega_n)^2+\Gamma_n^2} \qquad (47)$$

Figures 6a-b show a case of a strongly non-Markovian regime, characterized by pronounced oscillations in the correlation function $C(t)$ from Fig. 6a and a relatively large trace-distance backflow , indicating persistent memory effects. The corresponding spectral density $|J(\omega)|$ in Fig. 6b is narrowband and a structured, with distinct peak corresponding to well-resolved bath modes. In the Nakajima–Zwanzig framework, such a structured $|J(\omega)|$ yields long-lived, oscillatory $C(t)$, precisely as reproduced by our nodal-based simulation. This shows that by lowering bath connectivity and dissipation, we can design environments that keep strong non-Markovian behavior.

Case 3 - Intermediate Regime: Partial Memory Behavior

We lastly introduce a third, intermediate case. This case is defined by nodal parameters that lie between the two extremes previously discussed: intra-layer couplings are set to $J_{L1} = J_{L2} = 0.5$, a reduced inter-layer coupling of $J_{L12} = 0.15$, and bath dissipation rates of $\gamma_{L1} = \gamma_{L2} = 0.1$. The system parameters remain unchanged.

The results of this simulation, presented in Figure 7a-b, demonstrate a clear intermediate behavior. The bath correlation function $C(t)$ (Fig. 7a) shows a slower decay than the Markovian case (Fig. 5a), and it retains a more pronounced oscillatory structure than the earlier suggestion, indicating significant but not maximal memory effects. This behavior is a direct consequence of the reduced inter-layer coupling, which limits the rapid flow of information away from the system-proximal bath. Similarly, the spectral density $|J(\omega)|$ as shown in (Fig. 7b) is no longer sharply peaked but instead presents a broader, less structured profile than the non-Markovian case. While it does not yet exhibit the single, broad hump of the Markovian limit, its wider spread of dominant frequencies demonstrates a more densely populated and mixed bath mode distribution, consistent with the partial loss of memory.

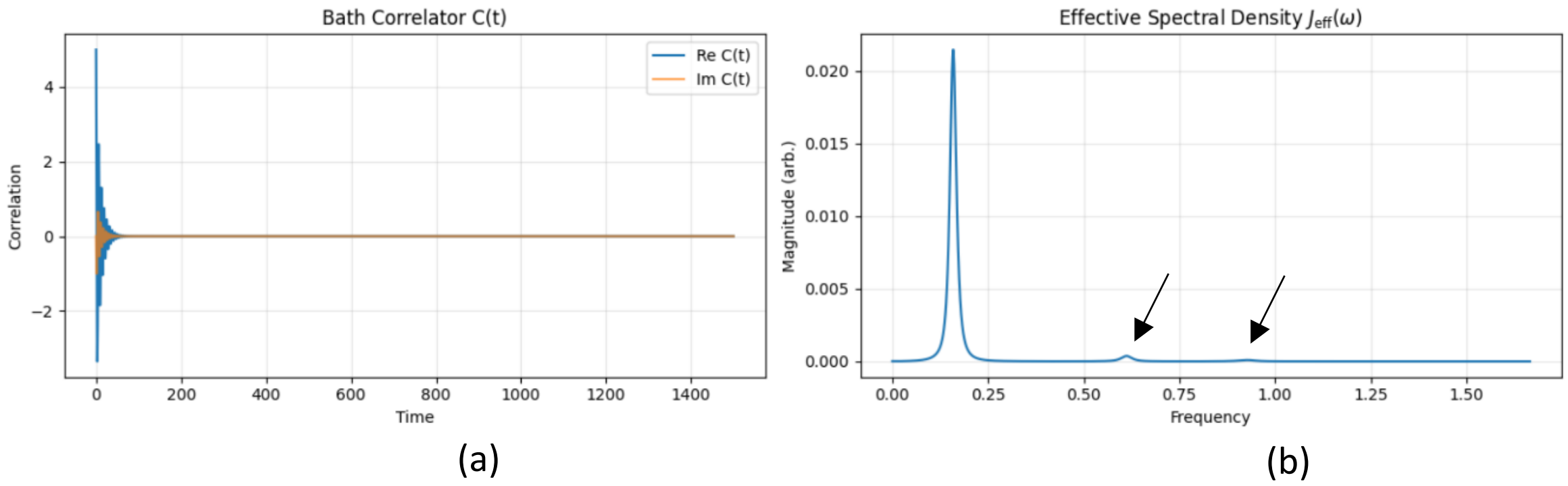

**Figure 7a–b**. Intermediate Regime, obtained with nodal parameters($J_{sb} = 0.8, J_{L1} = 0.5, J_{L12} = 0.05, J_{L2} = 0.5, \gamma_{L1} = 0.05, \gamma_{L2} = 0.05$) (a) The bath correlation function C(t) shows a decay rate and oscillatory behavior that are clearly between the two extreme cases, indicating a partial memory regime. (b) The corresponding spectral density $|J(\omega)|$ exhibits a broader profile than the non-Markovian case, but it retains a structured, multi-peaked shape that reflects the intermediate density and connectivity of the bath modes.

This intermediate case provides additional evidence that our nodal framework allows for a continuous and predictable tuning of the bath's properties. By systematically adjusting the inter- and intra-layer couplings and dissipation rates, we can engineer environments that fall anywhere along the spectrum from strongly non-Markovian to effectively Markovian.

The evolution of the spectral density across the three cases provides a compelling narrative of how bath structure dictates memory effects. In the near-Markovian limit, high connectivity and strong dissipation homogenize the bath to such an extent that all discrete spectral features vanish, leaving only a broad, featureless hump. This demonstrates that the specific structure of the bath is "washed out" in the Markovian regime (see Figure 5b). In contrast, in the strongly non-Markovian regime, sparse connectivity results in a pronounced, sharply defined peak accompanied by other small, sharp peaks (see arrows in Figure 6b). These features correspond to a few dominant bath modes that sustain long-lived oscillations and drive the system's memory. Finally, in the intermediate case, a subtle but important observation is the presence of a more prominent and broadened secondary peak (see arrow in Figure 7b). This feature is a direct manifestation of the bath's discrete structure, now partially "smeared out" by increased connectivity and dissipation. The broadening of the peaks reflects a wider distribution of interacting modes, indicating a partial convergence toward the continuum limit. This shift from a smooth continuum to sharp peaks shows that our nodal model can represent a wide range of memory effects. Overall, the versatility is key, allowing for the simulation of a wide range of realistic quantum systems and providing a direct physical connection to the continuum models described by the Nakajima-Zwanzig formalism.

One major advantage of discretizing the near-field bath into explicit layers and couplings is the ability to diagnose and quantify how energy and information flow through the environment. Rather than treating memory as an abstract kernel, the structured bath makes it possible to measure delay directly, identify pathways of transfer, and distinguish between bottlenecked and oscillatory regimes. This diagnostic power is detailed in the Supplementary Section S6, where cross-correlation analysis and flux traces reveal how transport bottlenecks, oscillatory segments, and synchronous locking emerge depending on the interplay of system–bath injection, inter-layer coupling, and dissipation.

In this context, a highly related perspective comes from the analytically solvable Ising spin-bath model developed by Krovi et al. [27], which connects directly to the Nakajima–Zwanzig framework by providing exact open-system dynamics alongside systematic comparisons to NZ, time-convolutionless, and post-Markovian master equations. Their results show how memory kernels that appear abstractly in NZ can be realized explicitly through discrete spin environments, with spectral features and backflow emerging from the bath's coupling constants. Our approach extends this continuity by embedding the bath explicitly as a finite, structured network with tunable topology, translating the kernel language into spectral fingerprints that can be directly measured, simulated, and analyzed using physics-aware machine learning.

Beyond NZ-based and solvable spin-bath approaches, tensor network (TN) methods such as matrix product states (MPS) and matrix product operators (MPO) have also emerged as powerful tools for simulating non-Markovian dynamics [28, 29]. By encoding time as a chain of tensors, TN methods compress correlations and track long memory effects that would otherwise be too costly to simulate. Our framework, however, takes a different approach. Instead of compressing correlations into an abstract form, we model direct connections whereby energy and information flow between nodes. This makes it possible to see how memory is stored, transferred, or lost along specific pathways.

TN methods are efficient compared to brute-force simulations, but when the bath retains structured or extended memory, the bond dimension grows quickly, and the calculations become heavy. In particular, when TNs are used to model correlated near-field noise, the bond dimension inflates rapidly and the simulation becomes exhausting. The root of this difficulty lies in the faceless-bath assumption built in from the start: all correlations are absorbed into an abstract tensor chain, without distinction between localized, structured interactions and genuinely extended memory. Yet in real devices, what appears as "long-range memory" is usually just a few strong couplings close to the system. For example, in superconducting qubits, two-level system (TLS) defects act as localized near-field baths that generate sharp spectral features of non-Markovianity and backflow, despite involving only a handful of modes. Our structured bath framework starts directly from these minimal motifs, such as a triangle or a layered bath, and reproduces the essential features without the overhead of large bond dimensions.

This difference points to a broader shift. Even in advanced non-Markovian theory, the paradigm remains one of total specification: the belief that to control a system we must first build a complete model of its environment. This equates simulating with managing, committing us to costly calculation rather than adaptive control, especially when correlated effects arise from near-field environments. Our framework avoids this trap by representing only the minimal motifs that generate memory. The near-field is not treated as a faceless band but as a reduced, quantized network whose pathways can be directly tracked. This allows both efficiency and physical transparency: the bath is no longer hidden in abstraction but revealed as a structured network whose role in producing non-Markovianity becomes experimentally meaningful. If a TN description were expanded for such systems, most of its complexity would reduce to the same small set of couplings that we include directly. Our nodal framework therefore provides both simplicity and practical relevance, showing that a small structured bath is not only sufficient but also the more realistic source of memory in quantum devices.

**7. Beyond Bath Topology: The Influence of System-Bath Coupling**

At the beginning of this paper, we set $J_{system-bath}$ ($J_{sb}$) to be a relatively low value of 0.2 to show the thermalization effect as observed by Eisert's group[1]. Later on, in the processing the bath engineering analysis, we purposely set the high value of $J_{sb}$ to discern the systematic role of bath topology. Our analysis of the minimal 3-qubit system revealed that with a weak system-bath coupling ($J_{sb} = 0.2$), memory backflow is significantly amplified, resulting in a high non-Markovianity score. In contrast, our subsequent 6-qubit simulations utilized a much higher $J_{sb}$ of 1.0 to demonstrate how a more complex bath structure can manage memory albeit with the lower degree of retention and even under conditions of strong initial leakage. These two contrasting scenarios underscore the critical role of $J_{sb}$ as the primary gateway for energy and memory to leave the system.

A key point is that weak coupling *does not* equate to memoryless dynamics. In our simulations, the apparent amplification of backflow at $J_{sb} = 0.2$ arises from two complementary effects: (i) limited connectivity at the system–bath interface, which restricts the number of escape channels available, and (ii) slow relaxation within the bath, which prevents rapid dissipation into the far field. Together these factors trap information locally and permit partial return to the system. Thus, memory should be seen not as a monotonic function of coupling strength but as the outcome of both interface exchange rates and bath dissipation.

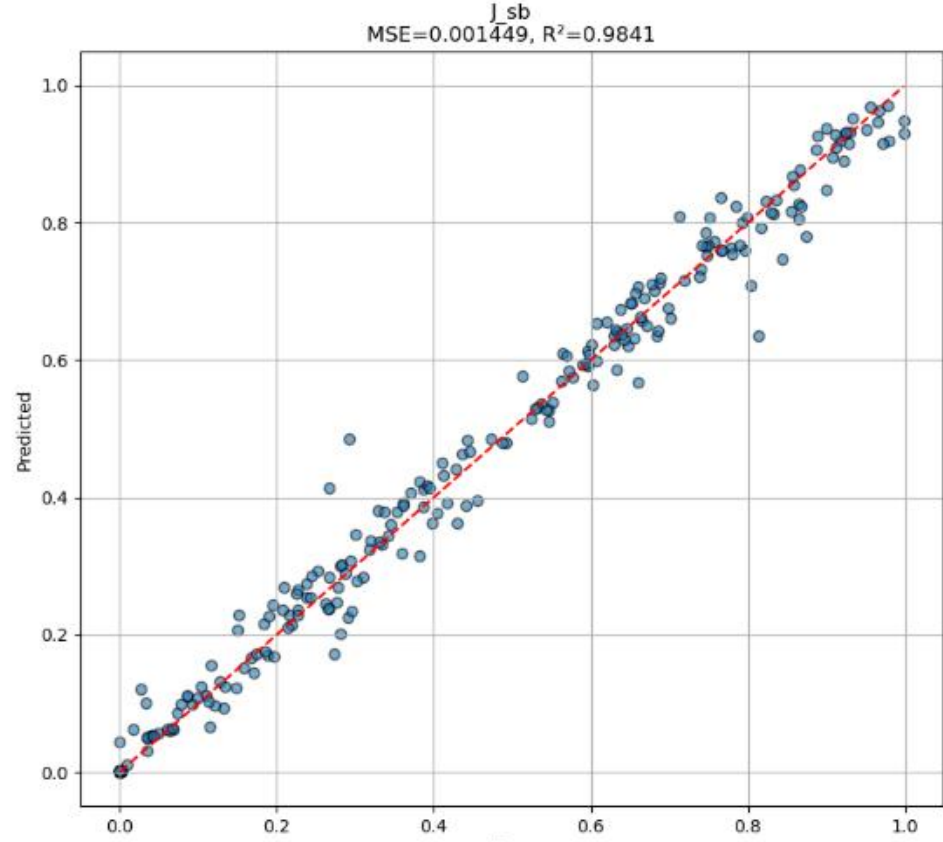

**Figure 8.** Machine learning parity plot for predicted vs. actual $J_{sb}$ values with PCA = 4.

The behavior of the system-bath coupling is effectively a mirror of the role played by the inter-layer coupling $J_{L12}$, which can be predicted by ML tools as shown in Figure 4. Both parameters act as bottlenecks for memory flow, and by tuning them, we can control the rate at which memory dissipates from the system or the proximal bath. This finding underscores a critical point: the ability to resolve and amplify memory effects is not solely dependent on the bath's engineered structure but is also fundamentally tied to the rate of memory exchange at the system-bath interface. We can now use the similar spectral fingerprints along with ML tools to predict the value of $J_{sb}$, even with a smaller set of data, as a testament to the power of spectral fingerprints. Figure 8 shows a plot where the $J_{sb}$ from our spectral fingerprint analysis, using a combined PCA and XGBoost approach, are compared to the original input values. The close proximity of the predicted points to the original points verifies that our strategy is robust and holds practical applications.

## 8.An example of application toward exceptional point

The framework presented can also be used to help identify and characterize the proximity of a quantum system to an Exceptional Point (EP). Exceptional points are spectral degeneracies that occur in non-Hermitian systems, where both the eigenvalues and the corresponding eigenvectors coalesce. They are of significant interest in quantum physics and optics due to their unique properties, such as enhanced sensing and unidirectional energy transfer. In this example, we apply the method to assess one of the markers for the proximity to an EP i.e. using the minimal distance to eigenvalue pairs (henceforth we call it as a Dissipative Exceptional Point Spacing or “DEPS”), which quantifies the minimal spectral separation between eigenvalue pairs of the system’s effective non-Hermitian Hamiltonian. Using QuTiP, we can calculate the instantaneous effective Hamiltonian, including both coherent and dissipative effects as:

$$H_{eff}(t) = H - \frac{i}{2}\sum_k L_k^{\dagger} L_k \qquad (48)$$

where $H$ is the coherent system + bath Hamiltonian and $L_k$’s are the Lindblad collapse operators. The complex eigenvalues of $H_{\text{eff}}(t)$ are denoted $\{\lambda_m(t)\}$, with

$$\lambda_m(t) = E_m(t) - \frac{i}{2}\Gamma_m(t) \qquad (49)$$

where $E_m(t)$ is the instantaneous energy and $\Gamma_m(t)$ is the decay rate. The complex eigenvalues used for the DEPS calculation is also conveniently extracted in QuTiP using the *.eigenenergies()* method applied to the effective non-Hermitian Hamiltonian $H_{eff}(t)$.
The DEPS at time $t$ can then be defined as:

$$DEPS(t) = \min_{m \neq n} |\lambda_m(t) - \lambda_n(t)| \qquad (50)$$

and the minimum over the simulation time is defined as:

$$DEPS_{min} = \min[DEPS(t)] \qquad (51)$$

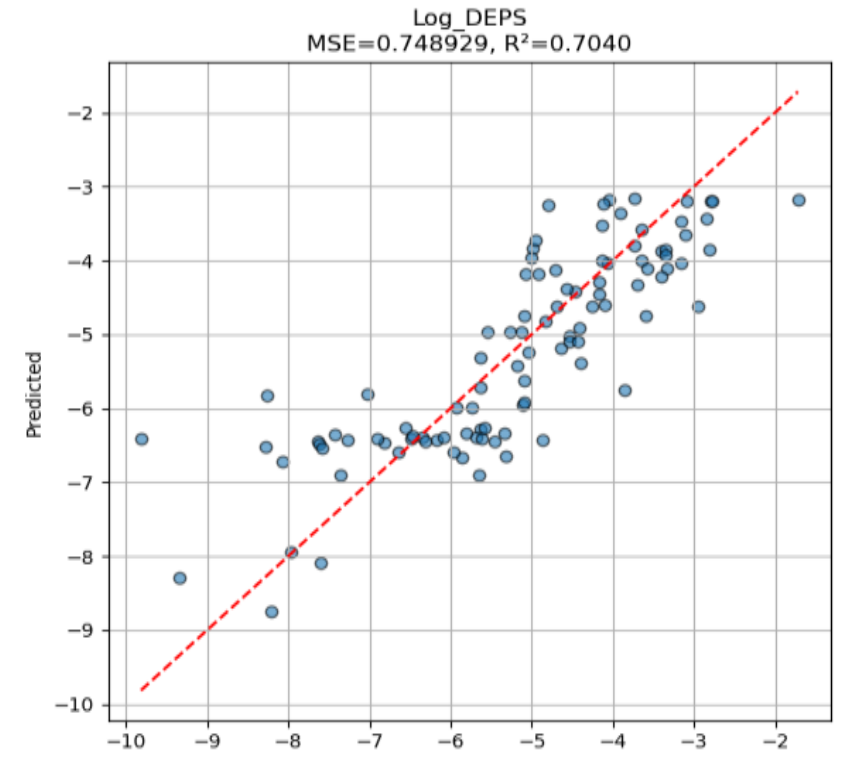

**Figure 9**. Machine learning parity plot for predicted vs. actual log(DEPS) values at PCA = 4.

In our present setup, while the density matrix and physical observables evolve with time, the construction of the effective Hamiltonian makes $H_{eff}$ time independent. Consequently, the eigenvalue spectrum of $H_{eff}$ is fixed in time, and the computed DEPS value is itself time-independent, so a single evaluation already equals $DEPS_{min}$. A more precise approach would indeed allow for explicitly time-dependent Hamiltonians or dissipators, in which case the spectrum and DEPS must be tracked dynamically; however, the present treatment remains grounded in the time-independent $H_{eff}$ defined above.

An EP is approached when $DEPS_{min} \rightarrow 0$, meaning two eigenvalues and their eigenvectors coalesce. Small but finite values indicate near-EP operation. In the structured bath model, EP proximity correlates with the system's ability to retain memory. Figure 9 presents a parity plot comparing ML-predicted versus true values using PCA-compressed FFT features and XGBoost. While the correlation is not as strong as in regression models trained on vertically defined inter-layer couplings such as $J_{\text{system–bath}}$ or $J_{L12}$, it nonetheless provides a valuable route for inferring proximity to exceptional points. In particular, the model still captures spectral trends that guide the search toward the lowest eigenvalue spacings.

Finally, we note that DEPS is just a practical indicator but not a full proof of an exceptional point. A true EP requires eigenvector coalescence in addition to vanishing eigenvalue spacing. As a complementary diagnostic we can compute the Petermann factor[30]:

$$K_k = \frac{\|L_k\|^2 \|R_k\|^2}{|\langle L_k | R_k \rangle|^2} \qquad (52)$$

for each eigenmode of $H_{eff}$ where $R_k$ and $L_k$ are the right/left eigenvectors of Heff (non-Hermitian, biorthogonal). Intuitively, $K_k$ measures the loss of biorthogonality: it stays near 1 when modes are well-separated and grows large as two modes become nearly self-orthogonal**,** which is what happens approaching an EP. Thus, a large $K_{max}$ is a practical, basis-invariant indicator of EP proximity. This is straightforward to evaluate in QuTiP: build $H_{eff}$ from the coherent Hamiltonian H and collapse operators $L_k$, obtain left and right eigenvectors (e.g., via SciPy's `eig` on `H_eff.full()` which returns aligned left/right vectors), and then compute $K_k$ from the formula above. Here, we can extract $K_{max}$ value as a robust one-number summary for scans. We use two indicators: the correlation function C(t) and the FFT spectrum of $\langle\sigma_z(t)\rangle$.

As a proof of concept, we consider two extreme regimes: (i) a highly non-Markovian case (Fig. 10a–b) and (ii) a strongly thermal, weakly coupled case (Fig. 10 c-d). These examples are intended only to illustrate the complementary role of the Petermann factor alongside log(DEPS); a full parameter sweep required for a systematic mapping lies beyond the scope of this work. In the non-Markovian regime (a-b), we observe larger $K_{max}$ and smaller DEPS values, consistent with long-lived, near-resonant modes; accordingly, the bath correlation $C(t)$ shows sustained oscillations. In the strongly dissipative case (c-d), $K_{max}$ is relatively modest and DEPS is larger; $C(t)$decays monotonically. Together, these time- and frequency-domain signatures suggest that sustained information retention in the immediate bath, as identified by non-Markovianity, correlates with proximity to EPs for these extreme cases. We caution, however, that intermediate regimes can exhibit mixed signatures; a more nuanced interpolation across parameter space, including jointly analyzing $C(t)$ and system-weighted spectral metrics, is warranted for a definitive mapping, which we leave for future work.

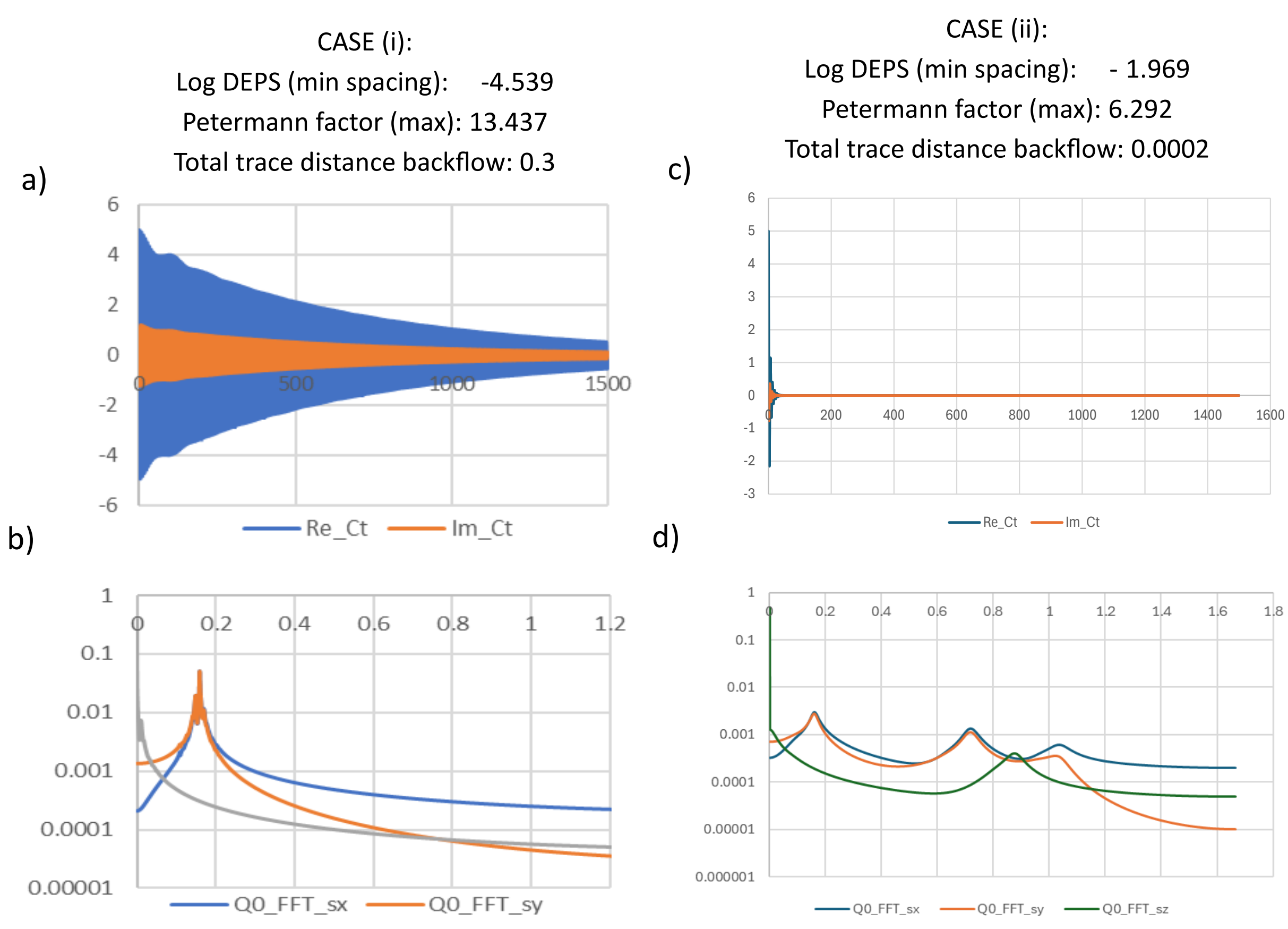

**Figure 10** — EP proximity indicators in two extreme regimes:
(a-b) Non-Markovian (structured bath): correlation $C(t)$ shows long-lived oscillations; FFTs of $\langle\sigma_z(t)\rangle$ exhibit narrow and clustered peaks. The spectrum of $H_{eff}$ yields small DEPS or more negative log (DEPS) and biorthogonal analysis gives a large $K_{max}$. (c-d) Strongly thermal, weak coupling (more Markovian): $C(t)$ decays rapidly; FFTs are broadened; the eigenvalues indicate larger DEPS and a modest $K_{max}$.

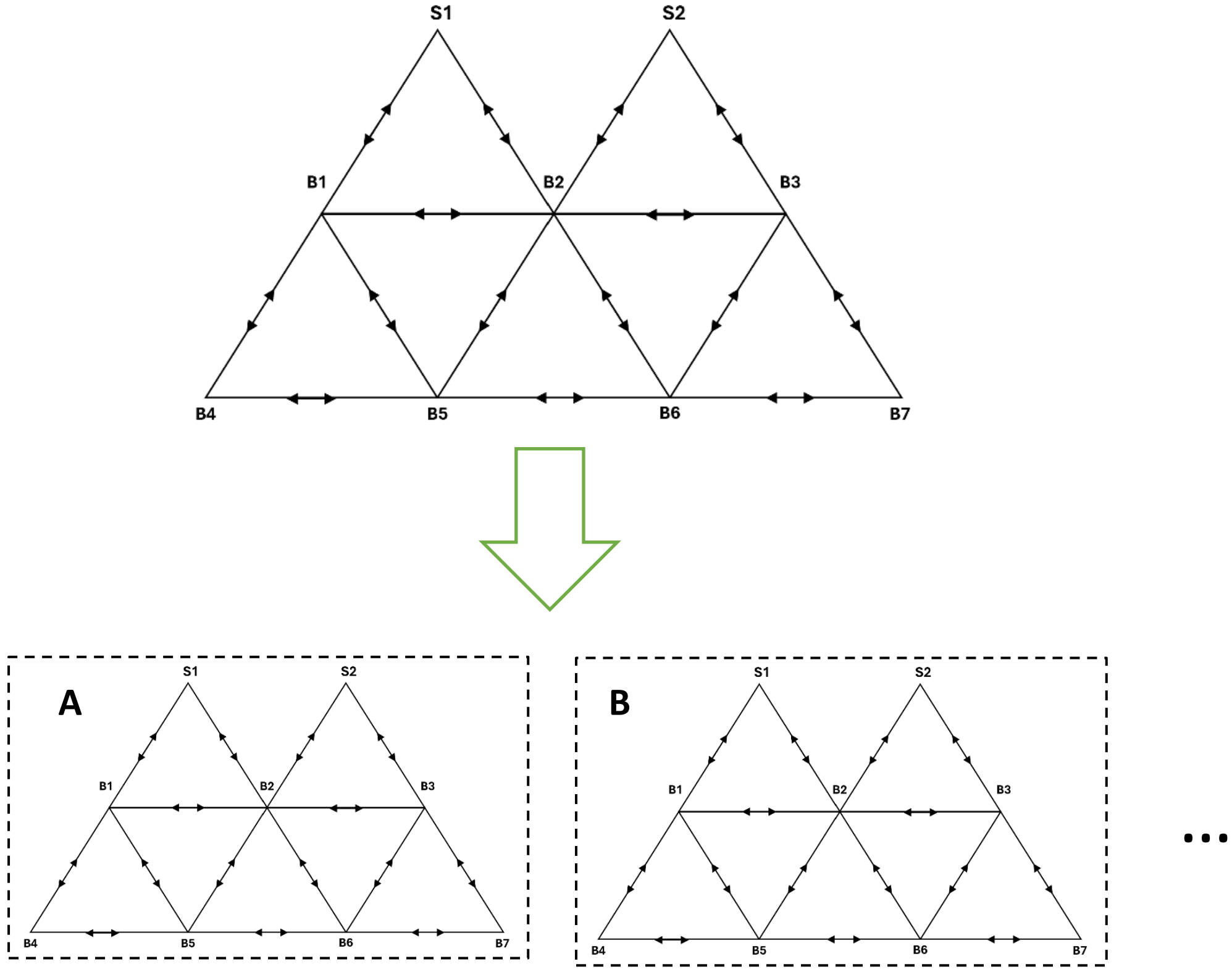

**Figure 11**. Generalization of the structured bath model to a many-body logical qubit system. This modular approach can be used to model the environmental coupling for two logical qubits.

**9. Generalization to many-body logical qubit systems**

The triangle-based bath geometry, initially explored with three and six qubits, can be extended to model more complex, many-body logical (system) qubit systems (Figure 11). This involves building larger, interconnected structures where multiple system qubits (e.g., $S_1$ and $S_2$) are each coupled to their own hierarchical bath layers. A schematic example in Figure 11 shows two system qubits ($S_1$ and $S_2$) with a shared bath qubit ($B_2$) in the first layer ($L_1$) and additional bath qubits in the second layer ($L_2$). This design allows for the investigation of how memory propagates and is distributed across multiple logical qubits and their shared and separate environmental elements. An alternative to this design in Figure 11 is where the connected baths on each unit are instead insulated at the first layer, with the connection occurring only at the second layer, leading to the study of isolated dynamics between the two logical qubits. More details of these formalisms can be found in the Supplementary Section S8.

Going forward, this generalization and modular approach can be quite useful to assess the qubit ecosystem more comprehensively. By compartmentalizing the  bath architecture, memory structures can be distributed and preserved across different environmental regions, protecting useful memory in a controlled manner. The approach follows a finite-element analogy for quantum baths, where the environment is represented as discrete elements with explicit couplings and layers. This perspective makes the model more realistic for physical quantum devices, where system–bath correlation functions are inherently discretized by the hardware's defined structure. Furthermore, this extension provides a clearer diagnostic of how bath–bath connectivity and topology influence the collective dynamics of multiple coupled system qubits, establishing a direct link between the geometry of the environment and observable quantities such as entropy growth and energy flow.

**10. Perspectives**

Using QuTiP, this work formalizes simplified mathematical form practices already implemented across quantum platforms. Techniques such as Rydberg blockade, detuning surrounding qubits, or engineering phonon modes to suppress specific thermal channels can be understood as shaping structured, memory-retaining environments. Here, these ideas are made explicit so their effects can be modeled, measured, and linked to experimentally accessible spectral fingerprints.

Rather than describing the environment as a smooth continuum, we adopt a discretized approach in which the  bath is composed of discrete quantum elements with explicit couplings and layers. Geometry and connectivity are directly tied to observables such as entropy growth, energy flow, and trace distance backflow. Machine learning tools allow not only inverse engineering of bath structures but also active search for optimal bath conditions, such as those near exceptional points.

In analogy with the finite element method [31], the model need not resolve the entire bath at uniform resolution. Refinement is concentrated on the immediate $L_1$ and $L_2$ layers, where the strongest system–bath coupling and memory effects reside. More distant and lesser influential bath layers can be coarse-grained or represented by continuum kernels. Further, these coarse-grained elements can be coupled through higher-level transport models, thus preserving essential non-Markovian features while keeping qubit counts low enough for tractable simulation. The coarse-graining can be done either in real space using adjacency-based graph Laplacians, or in reciprocal/frequency space using spectral-overlap kernels. The latter potentially leverage the dynamical similarity in frequency space. This is particularly advantageous in multi-layer baths where correlated oscillations including horizontal echo and reverberation effects in two-logical-qubit units would involve non-adjacent nodes.

A central insight is that *weak coupling does not imply memoryless dynamics*. The system never passes energy directly to the far field; it always interacts first with the structured near field environment. If dissipation there is sluggish, information becomes trapped, swirling like echoes in a chamber. These echoes seed backflow, reshaping system evolution in bursts of restored coherence. Engineering thermal dissipation channels is therefore as critical as tuning coupling strengths: without this, the near field environment inevitably remembers and eventually feeds back.

From a control perspective, the same spectral diagnostics naturally extend to Dynamical Decoupling (DD). Traditionally, DD has relied on generic pulse sequences to suppress noise. With bath fingerprints in hand, two potential refinements emerge:

- *Targeted DD sequences.* Identified bath frequencies can be directly suppressed, preventing coherent backflow from reinforcing into strong decoherence channels. The goal is not full decoupling—which is unrealistic at typical relaxation rates—but diffusing and confusing memory echoes so they cannot accumulate constructively.
- *Adaptive real-time control.* Because the near field environment evolves with temperature, charge noise, or instabilities, diagnostics can be repeated in a feedback loop. Qubits update their spectral fingerprints, and DD sequences are tuned accordingly. This transforms DD into an adaptive protocol, aligning naturally for autonomous hardware-aware quantum error correction (QEC).

Placed in a broader context, there is a *philosophical* similarity to Majorana zero modes. Both employ delocalization as a strategy to preserve information [32]. In our framework, memory is compartmentalized across bath regions, forming a controlled environment in which useful information is protected and selectively propagated. The analogy is also practically relevant, since Majorana qubits display unconventional decoherence pathways when coupled to environments [33], highlighting the broader role of structure and topology in mitigating information loss.

Additionally, this setup can serve as a teaching module in QuTiP, allowing students to vary couplings and temperatures, run the solver, and observe entropy, fidelity, and trace-distance backflow. This connects open quantum systems theory to device-level intuition. An instructive analogy comes from corrosion in molten salt reactors, where ionic double layers form structured environments at the metal–electrolyte interface [34–36]. These layers, like quantum bath layers, remain in place and influence dynamics, reinforcing the link between quantum dissipation and materials science.

## 11. Limitations and Future Directions

While the present framework offers a compact and pedagogically useful approach to modeling structured quantum baths, there are several limitations that are worth mentioning here. These topics will be part of our future directions.

*1. Limited layer resolution*.

The current nodal construction focuses on resolving only the most immediate and experimentally relevant layers, namely Layer 1 ($L_1$) and Layer 2 ($L_2$). This is justifiable since these immediate baths can potentially dominate coherence and memory effects in current quantum architectures. These layers should express the memory features that traditional continuum models cannot easily capture. However, in some systems, restricting the nodal bath structure to only two layers may not adequately sufficient especially if more subtle or long-range bath correlations are warranted. This limitation can be addressed by extending the nodal network to additional layers or refining inter-node structure through coarse graining, either in real space or reciprocal space. As noted in the finite-element analogy earlier, this is equivalent to locally refining mesh resolution if needed while keeping the rest of the domain remain coarser, preserving efficiency without losing essential physics.

*2. Assumed bath topology.*

In the present proof-of-concept, the bath topology is assumed *a priori* rather than derived from experimental data or first-principles modeling. While this makes the framework computationally efficient, it may limit generalizability. In future work,  we could leverage generative AI and spectral-fingerprint inversion to discern bath structure directly from measured system dynamics. This could allow such a model capturing device-specific connectivity patterns without relying on predefined topologies.

*3. Complementary role to continuum models.*

It is quite clear that the framework we proposed here is not meant to replace continuum methods such as HEOM, Nakajima–Zwanzig, or time-convolutionless master equations. Instead, it serves as a bridge connecting the broad continuum models and detailed nodal models that may exhaustively track every bath element. By keeping key features such as topology, local couplings, and memory effects, the nodal approach can (i) link simulations to measurable signals like full frequency spectra, (ii) create structured data for machine-learning inference or optimization, and (iii) support quantum control strategies depending on observable backflow and non-Markovianity. Because of this middle role, the model can also be further refined as the quantum hardware advances. Certainly, extending such a framework to multi-qubit systems becomes essential to further enhance error-mitigation strategies in real quantum devices.

*4. Signal masking in coherence observables.*

While our framework successfully leverages the population signal to infer bath parameters, it is admittedly more difficult to extract spectral features related to inter-bath couplings ($J_{L1}$ and $J_{L2}$) from the coherence observables ($\langle\sigma_x\rangle$ and $\langle\sigma_y\rangle$). These signals are overwhelmed by the much stronger, dominant coherent oscillation of the system qubit, which acts as a powerful carrier signal. This masking effect means that our analysis must pivot to the $\langle\sigma_z\rangle$ signal to obtain a clean, usable spectral fingerprint. Future work could explore more advanced signal processing techniques, such as filtering or wavelet transforms, to unmask these subtle features. Additionally, it may be possible to mitigate this masking by experimentally lowering the system qubit's on-site energy (ω), which would reduce the frequency of the dominant coherent signal and create more spectral space for the bath-induced features

**12. Conclusion**

In this work, we have introduced a physically motivated framework for modeling non-Markovian dynamics by representing the environment as a structured, finite-element network. This approach provides a complementary alternative to traditional abstract models by allowing us to directly link the bath's explicit geometry and connectivity to observable phenomena. We have demonstrated that this method offers potentially a useful diagnostic tool for quantum engineers, enabling the extraction of the full frequency spectrum of the system, which provides a rich "fingerprint" of the environment.

Beyond its utility as a diagnostic and pedagogical tool, our framework can serve as a new area for future research exploration. The modular nature of our design allows for a scalable approach to establishing and refining more complex system-bath configurations. This may include strategies to couple multiple logical qubits to an optimized structured environment. It can also establish a new path for inverse engineering by employing machine learning tools to extract bath parameters from spectral data and provide a platform to explore and elucidate advanced phenomena like exceptional points in a realistic, experimentally accessible context.

**Outlook.** Our results point out that structured baths with memory can serve as a foundation for both conceptual understanding and the advancement of scalable, hardware-aware quantum technologies. By extending the formalism to multi-system qubit architectures, we establish scalability beyond single-qubit motifs. By linking spectral diagnostics to adaptive control, we outline a path toward hardware-aware dynamical decoupling that erases or confuses bath memory in real time. And by recognizing that error memory is a resource to be mapped rather than ignored, we open opportunities for hardware-aware quantum error correction with reduced overhead. In this sense, structured baths provide not only a window into non-Markovian physics but also a practical toolset for engineering coherence, stability, and scalability in next-generation quantum devices.

**Acknowledgement**

I would like to thank Prof. Eisert and his group for their recent paper, which inspired a reevaluation of the role of weak system–bath coupling in thermalization and motivated a deeper exploration of bath architectures. Additional perspective came from strong analogies to materials science, particularly corrosion in molten salt reactors, the focus of my Summer 2025 Visiting Faculty Program (VFP) at Berkeley Lab. supported by the Dept. of Energy (DOE).

The machine learning implementations to extract normalized FFT signals were performed at NERSC at Lawrence Berkeley National Laboratory (LBNL), supported by the Quantum Information Science (QIS) at Perlmutter Award. Conceptual and code development also benefited from the assistance of large language models (ChatGPT-5, Co-Pilot, Gemini, among others).

To make this work accessible, a Jupyter notebook (executable in Google Colab) that implements the models and simulations is available at https://github.com/sakidja/quantum_bath.

**References**

1. Bertoni, C., Wassner, C., Guarnieri, G., Eisert, J.: Typical thermalization of low-entanglement states. Communications Physics. 8, 301 (2025). https://doi.org/10.1038/s42005-025-02161-7
2. Breuer, H.-P., Laine, E.-M., Piilo, J.: Measure for the Degree of Non-Markovian Behavior of Quantum Processes in Open Systems. Phys. Rev. Lett. 103, 210401 (2009). https://doi.org/10.1103/PhysRevLett.103.210401
3. Heiss, W.D.: The physics of exceptional points. Journal of Physics A: Mathematical and Theoretical. 45, 444016 (2012). https://doi.org/10.1088/1751-8113/45/44/444016
4. Nakajima, S.: On Quantum Theory of Transport Phenomena: Steady Diffusion. Progress of Theoretical Physics. 20, 948–959 (1958). https://doi.org/10.1143/PTP.20.948
5. Zwanzig, R.: Ensemble Method in the Theory of Irreversibility. The Journal of Chemical Physics. 33, 1338–1341 (1960). https://doi.org/10.1063/1.1731409
6. Breuer, H.-P., Kappler, B., Petruccione, F.: Stochastic wave-function method for non-Markovian quantum master equations. Phys. Rev. A. 59, 1633–1643 (1999). https://doi.org/10.1103/PhysRevA.59.1633
7. Tanimura, Y., Kubo, R.: Time Evolution of a Quantum System in Contact with a Nearly Gaussian-Markoffian Noise Bath. J. Phys. Soc. Jpn. 58, 101–114 (1989). https://doi.org/10.1143/JPSJ.58.101
8. Tanimura, Y.: Numerically “exact” approach to open quantum dynamics: The hierarchical equations of motion (HEOM). The Journal of Chemical Physics. 153, 020901 (2020). https://doi.org/10.1063/5.0011599
9. Brand, D., Sinayskiy, I., Petruccione, F.: Markovian noise modelling and parameter extraction framework for quantum devices. Scientific Reports. 14, 4769 (2024). https://doi.org/10.1038/s41598-024-54598-5

10. Johansson, J.R., Nation, P.D., Nori, F.: QuTiP: An open-source Python framework for the dynamics of open quantum systems. Computer Physics Communications. 183, 1760–1772 (2012). https://doi.org/10.1016/j.cpc.2012.02.021
11. Johansson, J.R., Nation, P.D., Nori, F.: QuTiP 2: A Python framework for the dynamics of open quantum systems. Computer Physics Communications. 184, 1234–1240 (2013). https://doi.org/10.1016/j.cpc.2012.11.019
12. Gorini, V., Kossakowski, A., Sudarshan, E.C.G.: Completely positive dynamical semigroups of N-level systems. Journal of Mathematical Physics. 17, 821–825 (1976). https://doi.org/10.1063/1.522979
13. Lindblad, G.: On the generators of quantum dynamical semigroups. Communications in Mathematical Physics. 48, 119–130 (1976). https://doi.org/10.1007/BF01608499
14. Breuer, H.P., Petruccione, F.: The Theory of Open Quantum Systems. Oxford University Press (2002)
15. Gardiner, C., Zoller, P.: Quantum Noise: A Handbook of Markovian and Non-Markovian Quantum Stochastic Methods with Applications to Quantum Optics. Springer (2004)
16. Nielsen, M.A., Chuang, I.L.: Quantum Computation and Quantum Information: 10th Anniversary Edition. Cambridge University Press, Cambridge (2010)
17. Uhlmann, A.: The “transition probability” in the state space of a $*$-algebra. Reports on Mathematical Physics. 9, 273–279 (1976). https://doi.org/10.1016/0034-4877(76)90060-4
18. Jozsa, R.: Fidelity for Mixed Quantum States. Journal of Modern Optics. 41, 2315–2323 (1994). https://doi.org/10.1080/09500349414552171
19. Leggett, A.J., Chakravarty, S., Dorsey, A.T., Fisher, M.P.A., Garg, A., Zwerger, W.: Dynamics of the dissipative two-state system. Rev. Mod. Phys. 59, 1–85 (1987). https://doi.org/10.1103/RevModPhys.59.1
20. Spohn, H.: Entropy production for quantum dynamical semigroups. Journal of Mathematical Physics. 19, 1227–1230 (1978). https://doi.org/10.1063/1.523789
21. Leibfried, D., Blatt, R., Monroe, C., Wineland, D.: Quantum dynamics of single trapped ions. Rev. Mod. Phys. 75, 281–324 (2003). https://doi.org/10.1103/RevModPhys.75.281
22. Home, J.P., Hanneke, D., Jost, J.D., Amini, J.M., Leibfried, D., Wineland, D.J.: Complete Methods Set for Scalable Ion Trap Quantum Information Processing. Science. 325, 1227–1230 (2009). https://doi.org/10.1126/science.1177077
23. Saffman, M., Walker, T.G.: Analysis of a quantum logic device based on dipole-dipole interactions of optically trapped Rydberg atoms. Phys. Rev. A. 72, 022347 (2005). https://doi.org/10.1103/PhysRevA.72.022347
24. Bianchetti, R., Filipp, S., Baur, M., Fink, J.M., Göppl, M., Leek, P.J., Steffen, L., Blais, A., Wallraff, A.: Dynamics of dispersive single-qubit readout in circuit quantum electrodynamics. Phys. Rev. A. 80, 043840 (2009). https://doi.org/10.1103/PhysRevA.80.043840
25. Scully, M.O., Zubairy, M.S.: Quantum Optics. Cambridge University Press, Cambridge (1997)
26. D. J. Thomson: Spectrum estimation and harmonic analysis. Proceedings of the IEEE. 70, 1055–1096 (1982). https://doi.org/10.1109/PROC.1982.12433
27. Krovi, H., Oreshkov, O., Ryazanov, M., Lidar, D.A.: Non-Markovian dynamics of a qubit coupled to an Ising spin bath. Phys. Rev. A. 76, 052117 (2007). https://doi.org/10.1103/PhysRevA.76.052117

28. Orús, R.: A practical introduction to tensor networks: Matrix product states and projected entangled pair states. Annals of Physics. 349, 117–158 (2014). https://doi.org/10.1016/j.aop.2014.06.013
29. Strathearn, A., Kirton, P., Kilda, D., Keeling, J., Lovett, B.W.: Efficient non-Markovian quantum dynamics using time-evolving matrix product operators. Nature Communications. 9, 3322 (2018). https://doi.org/10.1038/s41467-018-05617-3
30. K. Petermann: Calculated spontaneous emission factor for double-heterostructure injection lasers with gain-induced waveguiding. IEEE Journal of Quantum Electronics. 15, 566–570 (1979). https://doi.org/10.1109/JQE.1979.1070064
31. Zienkiewicz, O.C., Taylor, R.L., Zhu, J.Z.: The Finite Element Method: Its Basis and Fundamentals. Butterworth-Heinemann (2005)
32. Nayak, C., Simon, S.H., Stern, A., Freedman, M., Das Sarma, S.: Non-Abelian anyons and topological quantum computation. Rev. Mod. Phys. 80, 1083–1159 (2008). https://doi.org/10.1103/RevModPhys.80.1083
33. Budich, J.C., Walter, S., Trauzettel, B.: Failure of protection of Majorana based qubits against decoherence. Phys. Rev. B. 85, 121405 (2012). https://doi.org/10.1103/PhysRevB.85.121405
34. Graves, A.D.: The electrical double layer in molten salts: Part 1. The potential of zero charge. Journal of Electroanalytical Chemistry and Interfacial Electrochemistry. 25, 349–356 (1970). https://doi.org/10.1016/S0022-0728(70)80097-3
35. Frischknecht, A.L., Halligan, D.O., Parks, M.L.: Electrical double layers and differential capacitance in molten salts from density functional theory. The Journal of Chemical Physics. 141, 054708 (2014). https://doi.org/10.1063/1.4891368
36. Kłos, J., Lamperski, S.: Analysis of electrical double layer structure in molten salts. The Journal of Chemical Physics. 150, 064704 (2019). https://doi.org/10.1063/1.5082561

**Structured Quantum Baths with Memory: A QuTiP Framework for Spectral Diagnostics and Machine Learning Inference**

Ridwan Sakidja
Dept. of Physics, Astronomy and Materials Science
Missouri State University

SUPPLEMENTARY MATERIALS

## S1 QuTiP Simulation Framework

- We simulate a six-qubit open quantum system with one central qubit coupled to a two-layer bath (Figure 1). The Hamiltonian includes on-site energies and anisotropic Heisenberg couplings: system–bath, intra-layer, and inter-layer. Dissipation is modeled with Lindblad operators—pure dephasing on the system qubit and layer-dependent thermal channels on the bath qubits. The system starts in a coherent superposition, while the bath qubits begin in thermal states.
- The simulation tracks spectral and information-theoretic measures. Bath correlators $C(t)$ are Fourier transformed to extract the spectral density $J(\omega)$ for comparison with continuum models. For clarity, we report the one-sided spectral density) $|J_{eff}(\omega)|$ with $\omega \geq 0$, obtained from the Fourier transform of $C(t)$. A non-Hermitian effective Hamiltonian is analyzed, with minimum eigenvalue spacing indicating proximity to exceptional points. Diagnostics include entropy, fidelity, entanglement negativity, mutual information, and energy-flow checks. Non-Markovianity is measured using the Breuer–Laine–Piilo backflow from trace-distance dynamics between orthogonal system states.
- To connect simulation and data-driven analysis, time-domain observables are Fourier-analyzed with log-enhanced normalization, producing machine-learning feature vectors. These are paired with ground-truth targets (coupling parameters, dissipation rates, distance to eigenvalue pair spacing, dominant spectral frequency, and backflow rate measure) to form datasets written to CSV. Automated routines allow single or batch simulations with randomized parameters, with all diagnostics, plots, and datasets saved to disk and archived for reproducibility.
- This script implements a structured open quantum system with a system qubit coupled to a two-layer bath of five qubits. It integrates Hamiltonian dynamics, Lindblad dissipation, bath spectral analysis, and quantum information diagnostics, and produces datasets suitable for machine learning regression/classification.
- In this study, we employ two time-windows for the six-qubit simulations. The primary configuration uses a uniform grid of 5,000 points spanning t=0 to t=1500, giving a temporal step of $\Delta t = \frac{1500-0}{5000-1} \approx 0.3001$ and a Nyquist frequency of approximately $f_N = \frac{1}{2.\Delta t} \approx 1.666$. A shorter time configuration is also considered, consisting of only 500 window points but with a total of 1500 grid points with $\Delta t = \frac{500-0}{1500-1} \approx 0.3336$ and a Nyquist limit of $f_N = \frac{1}{2.\Delta t} \approx 1.499$. The FFT analysis of the bath correlation function C(t) shows that dominant spectral components cluster near 0.16 cycles per time unit, with secondary peaks below 0.20. Thus, these signals are well within both Nyquist limits, eliminating concerns of aliasing.

- The spectral resolution depends on the sampled duration: For the longer window ($N$=5000, $\Delta t$=0.3001), this gives $\Delta f = \frac{1}{N.\Delta t} \approx 6.67x10^{-4}$, whereas for the shorter window (N=1500,Δt=0.3336), $\Delta f = \frac{1}{N.\Delta t} \approx 2.00x10^{-3}$. Both provide *sufficient spectral resolution* to capture oscillatory features that define bath-memory behavior, with the extended window offering nearly an order of magnitude finer detail for frequency-domain classification.
- Below we also include a few specific points of clarification where further explanation was considered useful. Some of the relations provided here have already appeared in the main text. To avoid redundancy, the equations are presented without numbering in this Supplement. The intent is to give compact definitions and technical clarifications that support the results discussed in the main text, rather than introduce new derivations.

1. System Definition and Hamiltonian
   - Hilbert space**:** 6 qubits (dimension $2^6$=64) - see Figure 1.
   - Local Hamiltonian:

$$H_{local} = \sum_{i=0}^{5} \frac{\omega}{2} \sigma_z^{(i)}, \omega = 1.0$$

   - Interaction Hamiltonian (anisotropic Heisenberg form):

$$H_{int} = \sum_{\langle i,j \rangle} J_{ij} \left( \sigma_x^{(i)} \sigma_x^{(j)} + \sigma_y^{(i)} \sigma_y^{(j)} + \sigma_z^{(i)} \sigma_z^{(j)} \right)$$

     where couplings include:
     - System-Layer1: $J_{01}$ $and$ $J_{02}$.
     - Intra-Layer1: $J_{12}$.
     - Inter-Layer1: Source ⬄ Bath1, Bath2
     - Inter-Layer2: Bath1 ⬄ Bath3, Bath1 ⬄ Bath4 and Bath2 ⬄ Bath4, Bath2 ⬄ Bath5.

     Thus, the total Hamiltonian is: $H = H_{local} + H_{int}$
2. Dissipation and Lindblad Operators
   - The open dynamics are modeled via the Lindblad master equation:

$$\dot{\rho} = -i[H, \rho] + \sum_i \left( L_i \rho L_i^\dagger \right) - \frac{1}{2} \{ L_i^\dagger L_i, \rho \}$$

   - System dephasing:

$$L_{deph} = \sqrt{\gamma_{sys}} \sigma_z^{(0)}$$

   - Thermal bath dissipators: for each bath qubit $j$

$$L_{\downarrow,j} = \sqrt{\gamma(1 + n_{th})} \sigma_-^{(j)}, L_{\uparrow,j} = \sqrt{\gamma n_{th}} \sigma_+^{(j)}$$

with $n_{th} = \frac{1}{e^{\beta\omega - 1}}$. Layer 1 and Layer 2 are assigned to different dissipation rates $\gamma_1$ and $\gamma_2$ respectively.

3. Bath Correlator and Spectral Density
   - We define the bath operator as:

$$B = \sum_{i=1}^{5} \sigma_x^{(i)}$$

     which captures the collective transverse bath fluctuations responsible for decoherence in the system qubit. Technically, this choice is not unique to $\sigma_x^{(i)}$. One may equally define:

$$B = \sum_{i=1}^{5} \sigma_y^{(i)}$$

     since both $\sigma_x$ and $\sigma_y$ components represent transverse fluctuations relative to the system qubit's energy basis.
   - The bath correlation function is then:

$$C(t) = Tr[B(t)B(0)\rho_0]$$

     The bath correlation function $C(t)$ serves as the time-domain fingerprint of environmental fluctuations. It measures how strongly the collective bath operator $B$ at time $t$ remains correlated with its value at time 0, averaged over the initial bath state $\rho_0$. A rapid decay of $C(t)$ corresponds to short-lived or memory-less correlations and Markovian behavior, whereas long-lived oscillatory tails signal memory effects and non-Markovianity. The Fourier transform of $C(t)$ yields the spectral density $J(\omega)$, providing the frequency-domain representation of the bath and connecting the discrete qubit environment to a continuum-like spectral picture.
   - And the Fourier transform yields the spectral density:

$$J(\omega) \propto |\mathcal{F}[C(t)]|$$

   - This construction directly bridges our discrete qubit bath to the continuum-like spectral description used in Nakajima–Zwanzig approaches. In this picture, smaller DEPS values correspond to narrower $J(\omega)$, which manifest as long-lived correlations in $C(t)$.
4. Non-Hermitian Proximity
   - A simplified measure of distance to eigenvalue pair spacing is defined using the non-Hermitian effective Hamiltonian: $H_{eff} = H - \frac{i}{2}\sum_i L_i^\dagger L_i$
   - DEPS = $\min_{k \neq l} |\lambda_k - \lambda_l|$ where $\lambda_k$ are eigenvalues of $H_{eff}$. This measures closeness to coalescence of eigenvalues, i.e. exceptional points.
   - Alongside DEPS, we quantify non-Hermitian proximity using the Petermann factor, which measures the degree of non-orthogonality between left and right eigenvectors of the effective Hamiltonian. For each eigenmode k:

$$K_k = \frac{\|L_k\|^2 \|R_k\|^2}{|\langle L_k | R_k \rangle|^2}$$

where $R_k$ and $L_k$ are the right and left eigenvectors of $H_{eff}$. A normalization-invariant form is used to avoid dependence on eigenvector scaling.

Large values of $K_k$ indicate high sensitivity to perturbations and enhanced noise amplification near exceptional points. In practice, we track both the minimum DEPS and the maximum Petermann factor across eigenmodes to capture complementary signatures of non-Hermitian proximity.

5. Diagnostics from Time-Evolved States

   From the density matrices $\rho(t)$ obtained via QuTiP's mesolve, we computed

   - Entropy: Von Neumann entropy was computed for the system, individual baths, and averaged over all baths using QuTiP's *entropy_vn*.
   - Negativity: Bipartite entanglement between the system and Bath 1 was quantified via *qutip.negativity*.
   - Mutual Information: Total correlations computed as $S(A) + S(B) - S(AB)$ using von Neumann entropies of the reduced density matrices, computed via *entropy_vn*.
   - Energy Decomposition: Local subsystem energies, total energy, and energy flow contributions (coherent vs. dissipative) were tracked using operator expectation values combined with the Hamiltonian and Lindblad terms.
   - Fidelity of Mixed States: The fidelity measure used in this work follows Eq. (29) of the main text. All fidelity calculations were implemented in QuTiP via *qutip.fidelity*, which evaluates the Uhlmann fidelity for general mixed states. This provides a robust measure of state overlap even under dissipative evolution.

6. Initial States

   - System qubit: initialized in a pure state with zero entropy.
   - Bath qubits $(Q_1 - Q_5)$: initialized in thermal states at $\beta$=1.00, each with entropy 0.582.

**S2 Summary of the Numerical Procedures for Non-Markovianity: Trace Distance Backflow**

We quantify non-Markovianity using the Breuer–Laine–Piilo (BLP) trace distance measure. The trace distance between two reduced system states $\rho_1(t)$ and $\rho_2(t)$ is computed in practice with QuTiP's *tracedi*st. To obtain a representative signal, we evolve two orthogonal initial states of the system, $|0\rangle$ and $|1\rangle$, together with identical bath states. While the formal BLP measure is defined via maximization over all pairs:

$$\mathcal{N}_{BLP} = \max_{\rho_{1,2}(0)} \int_{\dot{D}(t)>0} \dot{D}(t)\, dt$$

this choice provides a practical and widely used lower bound that reliably detects information backflow across both weak and stronger coupling regimes studied here.

In the script, the function *blp_backflow_dt_invariant* is used to extract the backflow score from the trace distance time series. This routine first smooths $D(t)$ (using *scipy.signal.savgol_filter* or a moving average) to reduce numerical jitter, then computes the derivative D(t) on the actual tlist grid with *numpy.gradient*. A noise-adaptive threshold is estimated from the tail of the derivative so that only significant positive slopes are counted. Finally, numpy.trapz integrates these positive regions to give the total cumulative backflow N, and the script also reports the backflow rate $\frac{\mathcal{N}}{T}$ i.e $\mathcal{N}$ divided by the simulation time span T. This combination of smoothing, thresholding, and trapezoidal integration makes the result robust and step-agnostic, preventing artificial inflation of backflow due to discretization or small oscillations in the data.

Here we show two examples:

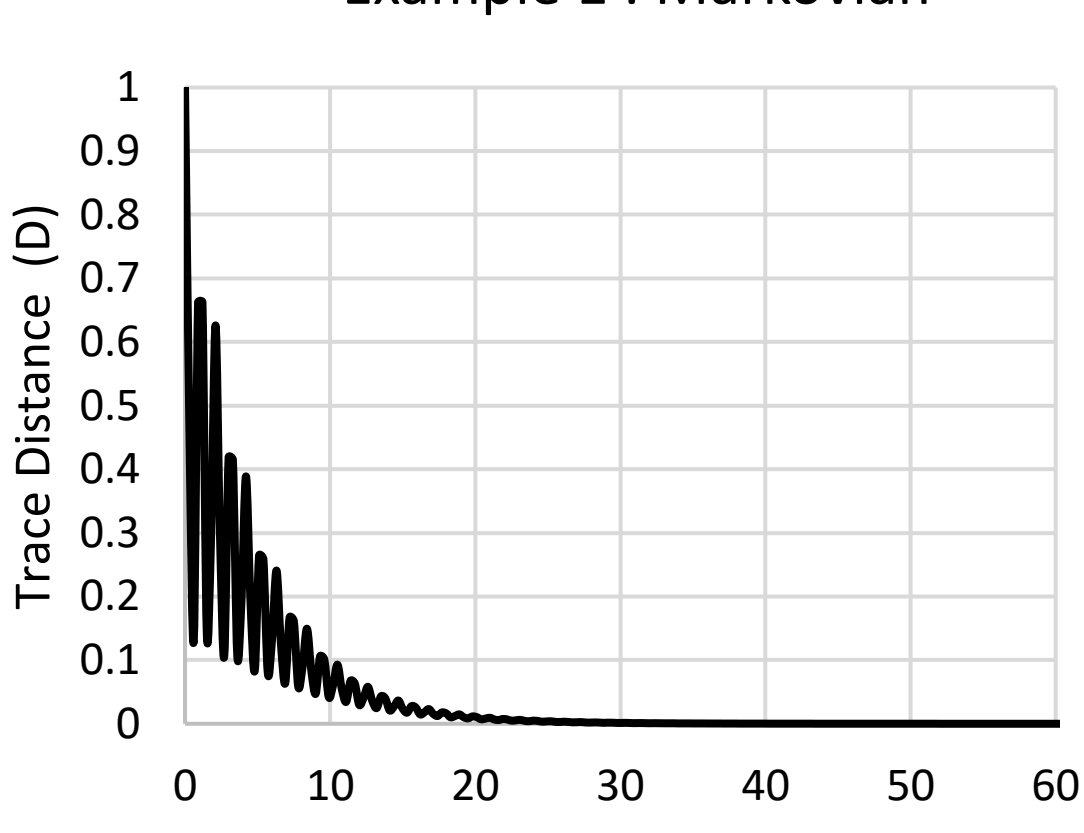

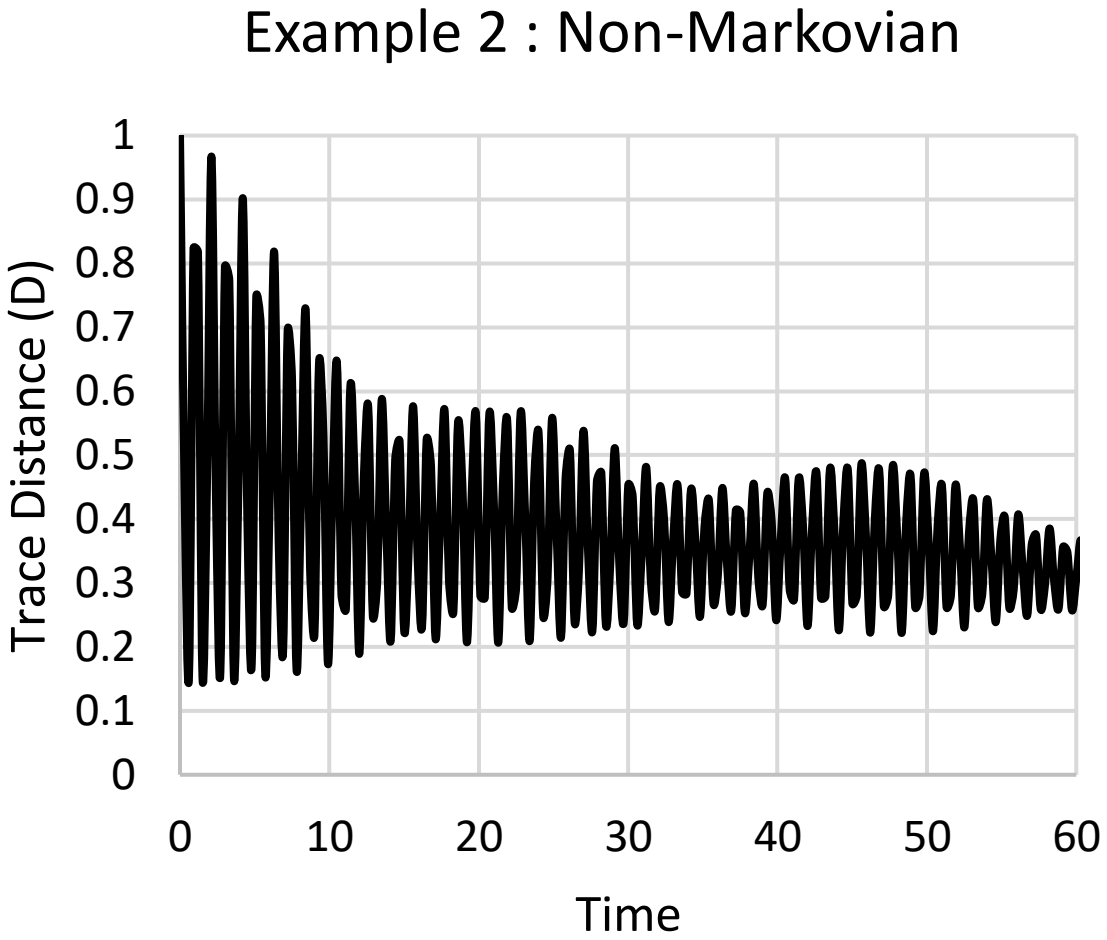

<u>Example 1: Near-Markovian Dynamics</u>

Trace Distance Backflow (Non-Markovianity Measure):$N$=0.0009560 (≈0)
Simulation Parameters:
Total Qubits: 6
System/Bath Energy Gap (ω): 1.00
Inverse Temperature (β): 1.00 (T = 1.00)
Thermal Occupation Number (n_th): 0.582
System-Bath Couplings (Jx, Jy, Jz): 0.100, 0.100, 0.050
J_source_to_bath: 1.000
J_layer1_intra_bath: 0.365
J_layer1_to_layer2_inter: 0.041
J_layer2_intra_bath: 0.844
System Dephasing Rate (γ_sys): 0.0050
Bath Thermalization Rates: Layer 1 = 0.1961, Layer 2 = 0.1858

Simulation Time: 0–1500 (5000 points) – simulation time of only up to 60 shown for clarity.
Interpretation: The extremely low backflow ($N \approx 0$) indicates Markovian dynamics, meaning the system loses information to the bath almost irreversibly. The relatively high thermalization rates ($\gamma \approx 0.19$) ensure that both bath layers quickly relax to equilibrium, suppressing any feedback. Strong intra-layer coupling in Layer 2 ($J_{L2} \approx 0.844$) accelerates this relaxation, making memory effects negligible.

Example 2: Non-Markovian Dynamics

Trace Distance Backflow (Non-Markovianity Measure): $N$= 0.18723979.

Simulation Parameters:

Total Qubits: 6

System/Bath Energy Gap (ω): 1.00

Inverse Temperature (β): 1.00 (T = 1.00)

Thermal Occupation Number (n_th): 0.582

System-Bath Couplings (Jx, Jy, Jz): 0.100, 0.100, 0.050

J_source_to_bath: 1.000

J_layer1_intra_bath: 0.407

J_layer1_to_layer2_inter: 0.043

J_layer2_intra_bath: 0.010

System Dephasing Rate (γ_sys): 0.0050

Bath Thermalization Rates: Layer 1 = 0.0030, Layer 2 = 0.0063

Simulation Time: 0–1500 (5000 points) - simulation time of only up to 60 shown for clarity.

Interpretation: The large backflow ($N \approx 0.844$) implies strong non-Markovianity, meaning information lost by the system partially returns over time. Here, the thermalization rates are two orders of magnitude smaller than in the first case, and Layer 2 coupling is extremely weak ($J_{L2} \approx 0.01$). This combination creates a "memory reservoir" where excitations linger, enabling coherent feedback to the system.

**S3. Results of QuTiP Simulations in the 6-Qubit System–Bath Model**

The following cases expand on the bath memory categories discussed in the main text. Each case is presented in the format: Setup → Purpose → Results → FFT Observations.

We begin with the simulation results for the source qubit and bath qubit population dynamics across five representative cases. These runs illustrate the different bath memory categories, with each case summarizing its setup, purpose, key outcomes, and FFT observations. This structure enables direct comparison of how variations in bath connectivity and dissipation influence relaxation behavior and memory effects.

*Note on dominant frequencies:*

Across all simulations, FFT analysis consistently reveals a robust component near 0.16 cycles per time unit, corresponding to the coherent phase oscillation of the system qubit when initialized in a superposition state. This mode is most pronounced in phase-sensitive observables such as $C(t)$, $\sigma_x$, and $\sigma_y$. In these transverse spectra, secondary peaks around ~ 0.8 cycles per time unit emerge, interpretable as higher harmonics or nonlinear bath-driven modulations of the 0.16 coherent oscillation.

By contrast, $\sigma_z$ spectra—which directly track population dynamics—show weaker sensitivity to the 0.16 coherent mode but instead emphasize bath-dependent structures. In particular, a dominant peak near ~1 cycle per time unit emerges under non-Markovian conditions, where inter-bath coupling $(J_{L12})$ and system-bath exchange $(J_{sb})$ drive memory backflow. This makes $\sigma_z$ the most informative channel for diagnosing bath connectivity, since its complex signatures are experimentally accessible and can be directly leveraged as machine-learning features for inference.

Case 1: Memory Retention in Layer 1

Setup:

- Layer 1 $(Q_1 - Q_2)$ is strongly connected. $J_{L1} = 1.0$,
- Layer 2 has negligible coupling. $J_{L2} = 0.001$,
- The inter-layer coupling between Layer 1 and Layer 2 is kept weak ($J_{12} = 0.001$),
- Low thermal dissipation.

Purpose: Localize memory feedback in Layer 1; Layer 2 remains passive.

FFT Observations:

- Strong, sharp peaks in $Q_1/Q_2$ (Layer 1)
- Signifies internal memory cycling
- Minimal contribution from Layer 2

Case 2a: Fully Dissipative Markovian Limit

Setup:

- All qubit couplings are fully enabled:
  - Intra-Layer 1 $(Q_1 - Q_2)$: $J_{L1} = 1.0$
  - Inter-Layer $(Q_1 - Q_3, Q_1 - Q_4, Q_2 - Q_4, Q_2 - Q_5)$: $J_{12} = 1.0$
  - Intra-Layer 2 $(Q_3 - Q_4, Q_4 - Q_5)$: $J_{L2} = 1.0$
- A stronger thermal dissipation is applied throughout the bath:
  - $\gamma_{thermal} = \gamma_{L1} = \gamma_{L2} = 0.01$ for $Q_1 - Q_5$

Purpose:

To simulate a fully Markovian environment, in which all memory effects are erased through strong dissipation and high connectivity. This can serve as a reference point for a regime dominated by irreversible thermalization and information loss.

FFT Observations:

- The frequency spectrum is relatively flat, low-intensity, and largely featureless.
- This indicates a lack of coherent oscillations, characteristic of rapid thermalization and no memory cycling.
- There is no significant spectral signature of non-Markovian feedback from any bath layer

Case 2b: Strong Dissipation with Quenched Intra-Layer Coupling

Setup:

- Weak inter-layer couplings:
  - $(Q_1 - Q_3, Q_1 - Q_4, Q_2 - Q_4, Q_2 - Q_5)$: $J = 0.001$ – very weak
- Intra-layer couplings suppressed:
  - $Q_1 - Q_2$: J = 0.0001 (Layer 1) - weak
  - $Q_3 - Q_4, Q_4 - Q_5$: J = 0.001 (Layer 2) - weak
- Very strong thermal dissipation: $\gamma_{thermal} = \gamma_{L1} = \gamma_{L2} = 0.3$ for $Q_1 - Q_5$:

Purpose:

To test the extreme limit of coherence suppression, where strong thermal dissipation is combined with severely quenched intra-layer connectivity. This simulates a highly dissipative environment with minimal internal bath structure, expected to eliminate almost all backflow.

FFT Observations:

- Frequency spectrum is further flattened compared to Case 3a, with extremely low spectral density.
- Residual coherence peaks are minimal, possibly reflecting only initial transients or edge effects.
- This confirms that strong dissipation combined with weak intra-bath connectivity yields near-complete Markovian behavior.

Case 3a: Transferred Memory to Layer 2

Setup:

- Full inter-layer coupling: $J_{12} = 1.0$,
- Layer 2 has strong couplings $J_{L2} = 1.0$,
- Low dissipation in both layers: $\gamma_{thermal} = \gamma_{L1} = \gamma_{L2} = 0.0001$

Purpose: Explore memory flow into deeper bath layers

FFT Observations:

- Broader frequency spread
- Spectral weight shifts into $Q_3 - Q_5$ : (Layer 2)
- Non-Markovian memory retained but *reduced*.

Case 3b: Suppressed Intra-Layer-2 Coupling

Setup:

- Same as 3a, but intra-layer-2 coupling almost turned off: $J_{L2} = 0.001$

Purpose: Test whether coherence can persist via inter-layer coupling alone

FFT Observations:

- Surprisingly rich frequency content emerges in Layer 2, despite its internal coupling being nearly eliminated.
- Indicates that coherence backflow *does not* require internal bath coherence within Layer 2. Rather, it is *mediated* through Layer 1.
- Confirms that triangle "3-qubit nucleus" connectivity (System ⇔ Bath 1 ⇔ Bath 2) alone is sufficient to broadcast and retrieve quantum information, even when deeper layers are structurally decohered.
- This resilience of spectral activity in Layer 2 reinforces the idea of *structured bath engineering as a way to control memory dynamics*

CASE 1

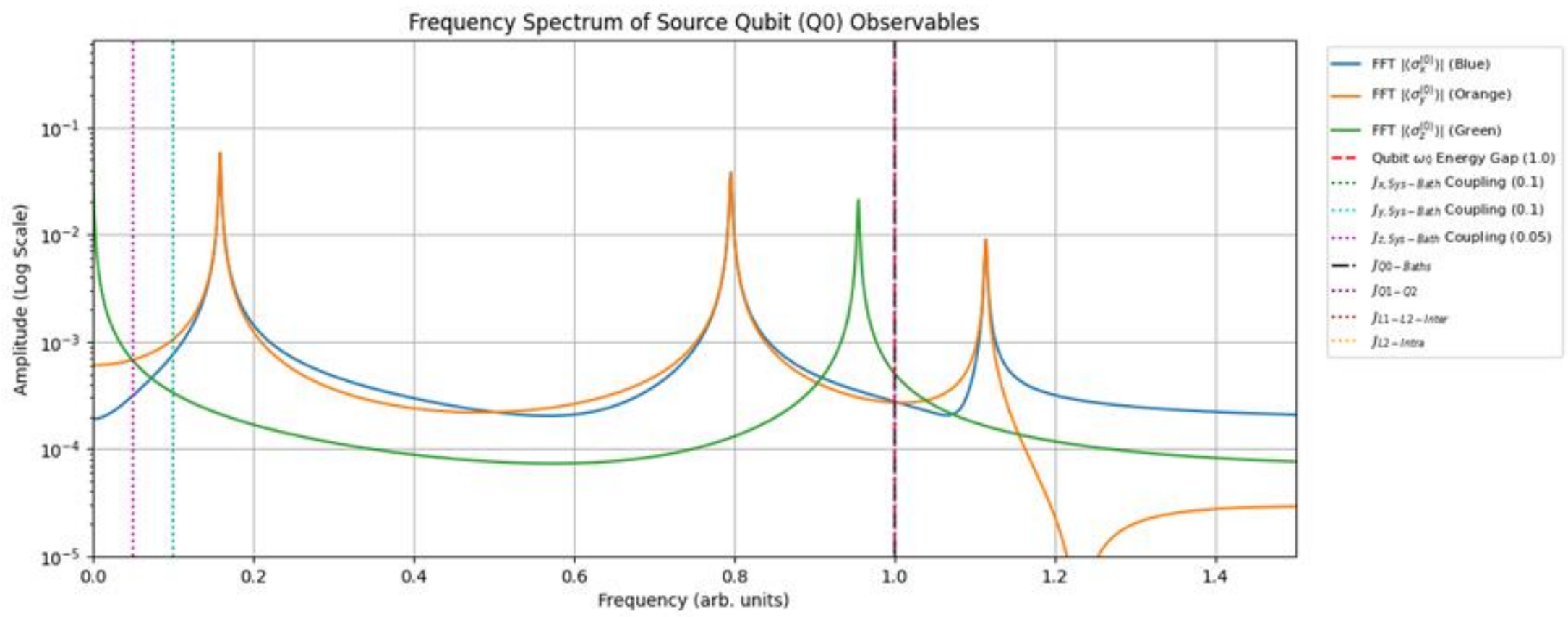

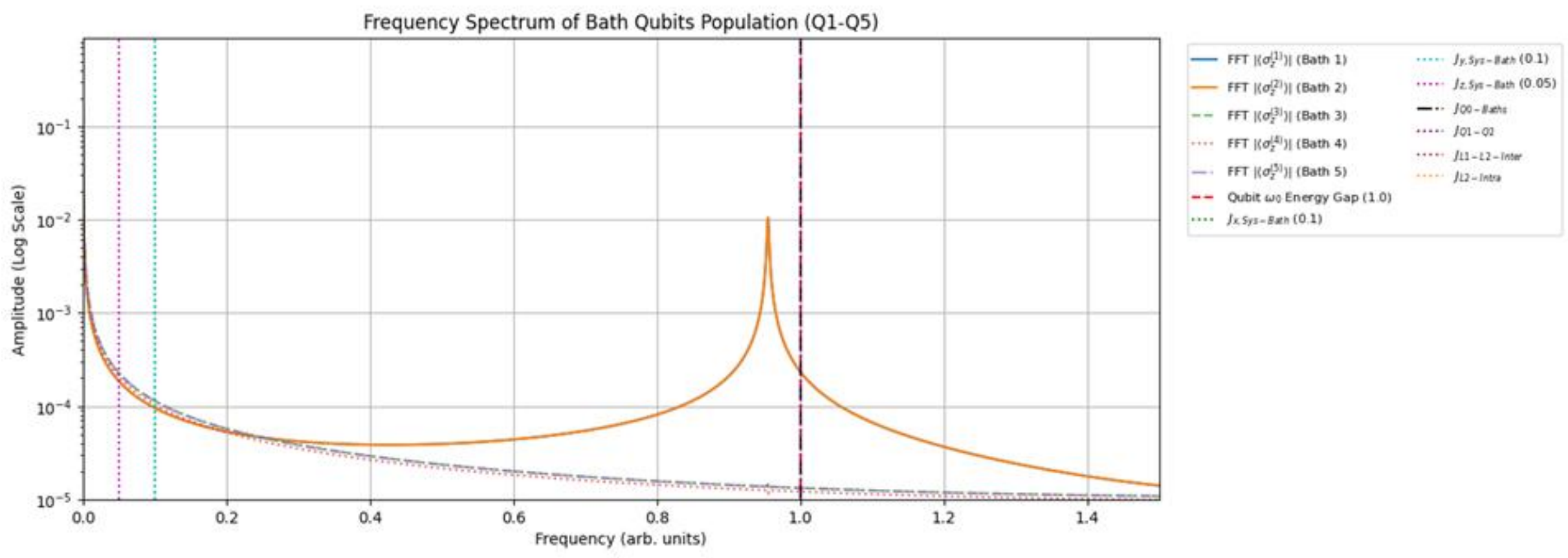

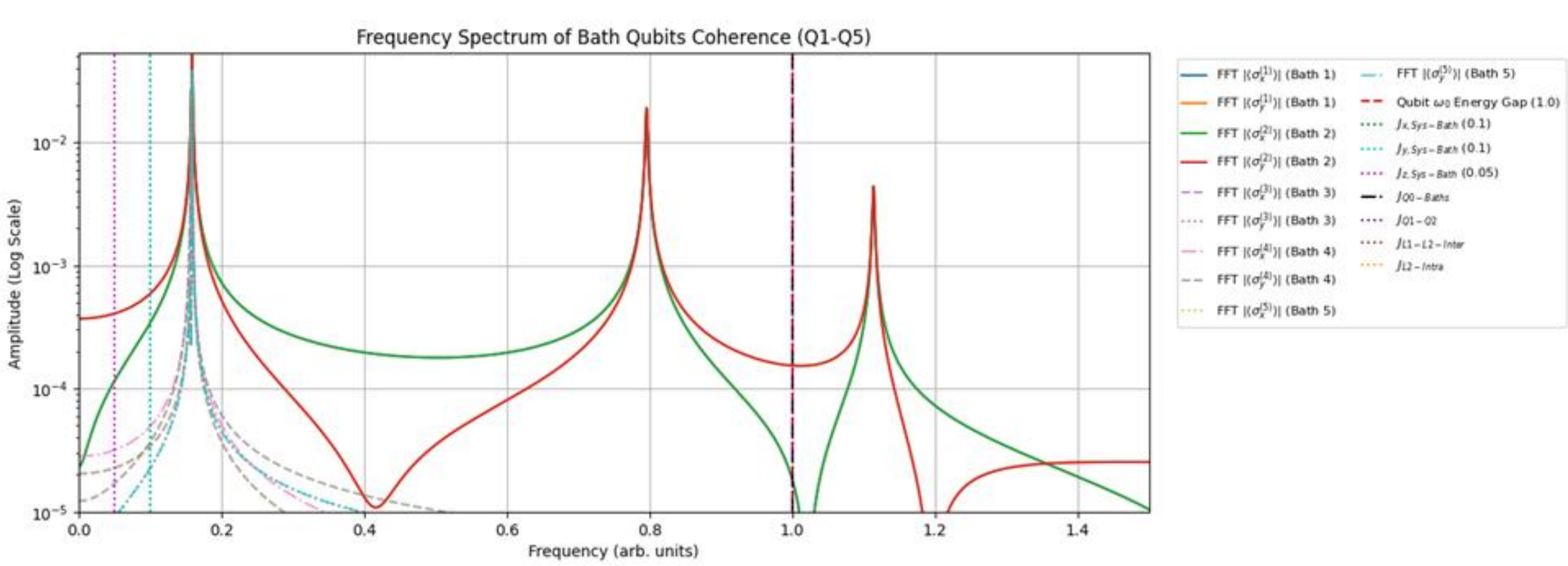

CASE2a:

Frequency Analysis (FFT) of 6-Qubit System Dynamics

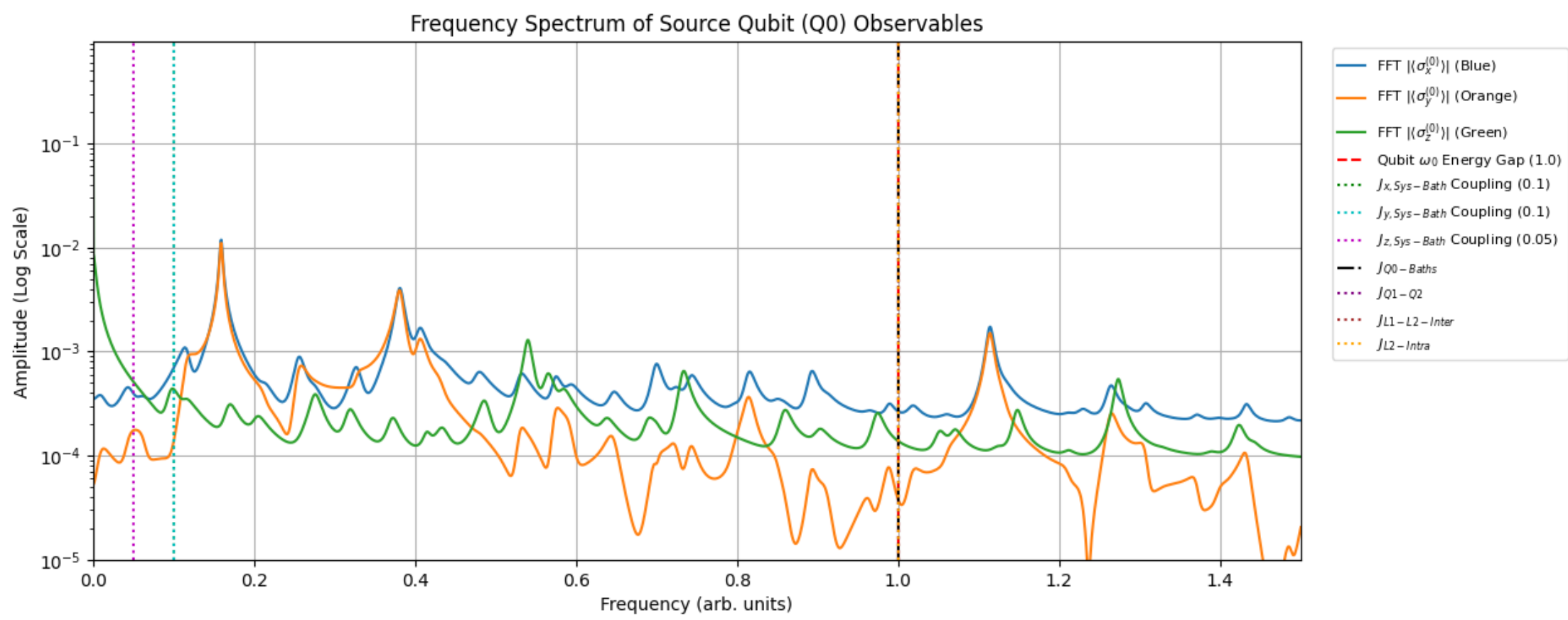

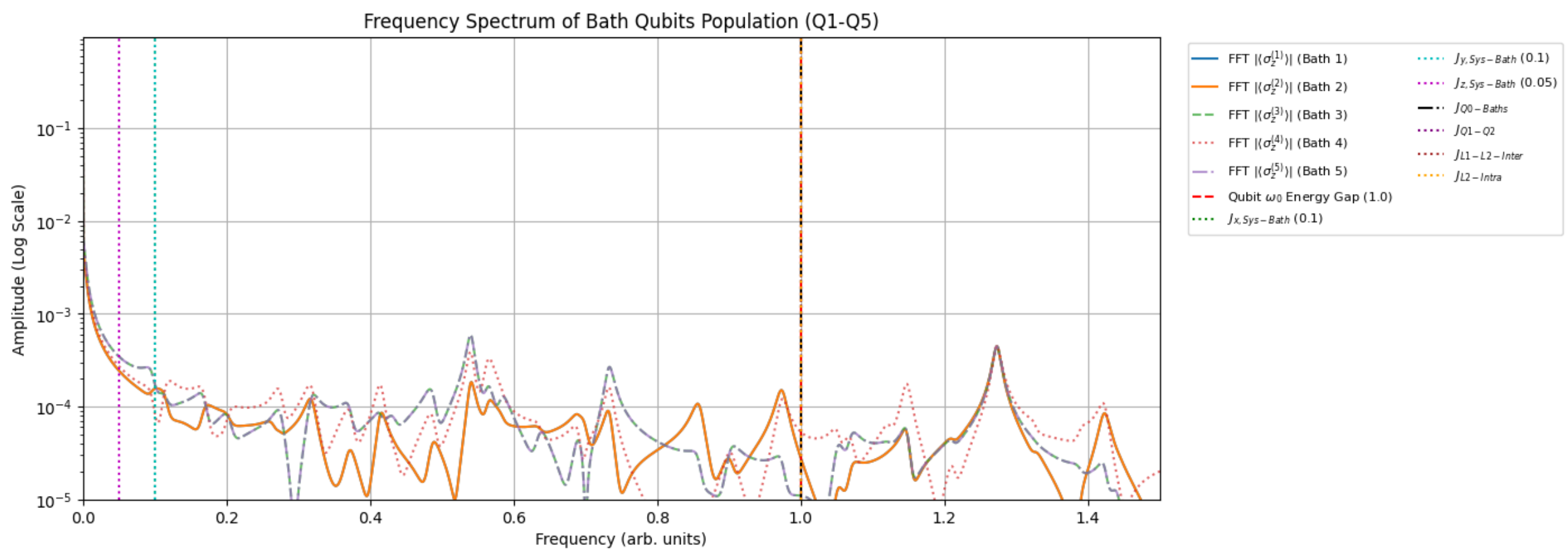

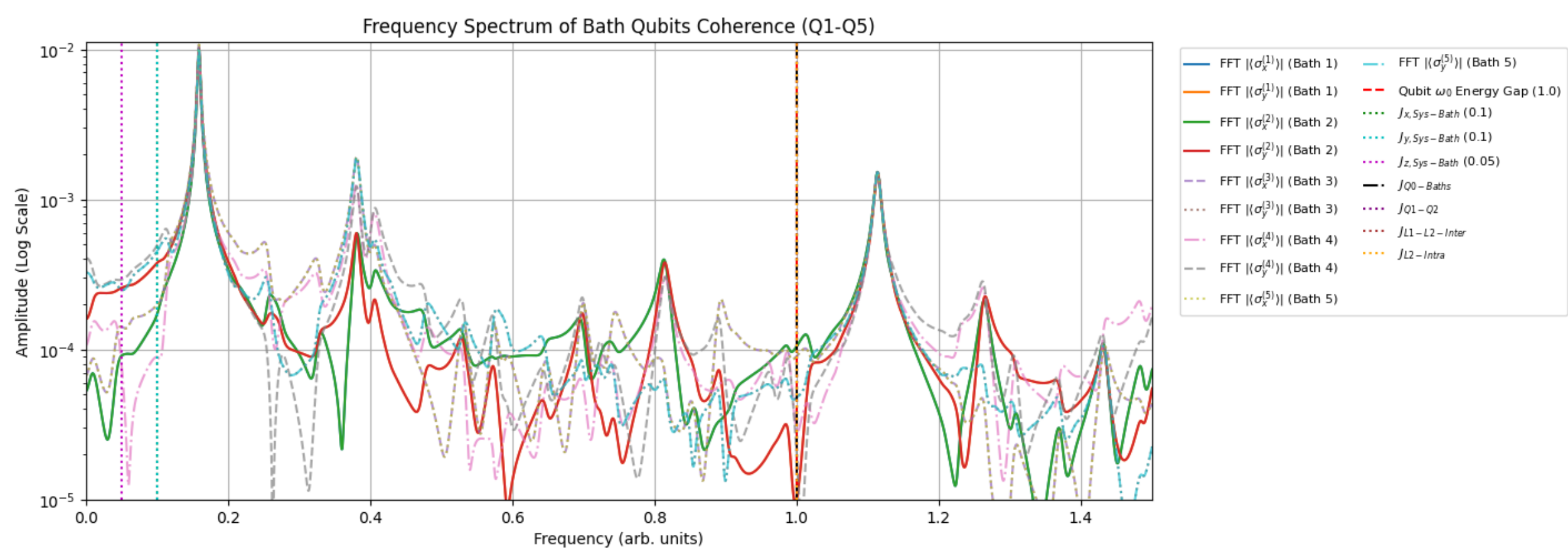

## CASE 2b:

Frequency Analysis (FFT) of 6-Qubit System Dynamics

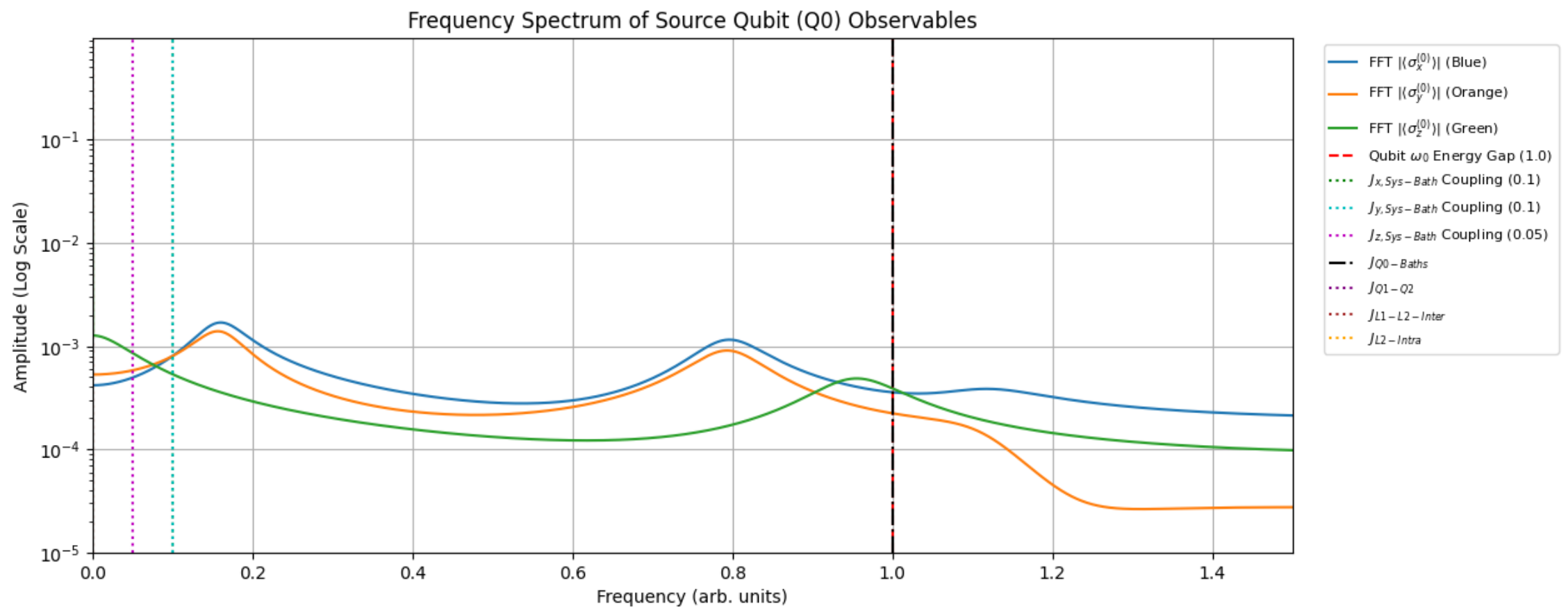

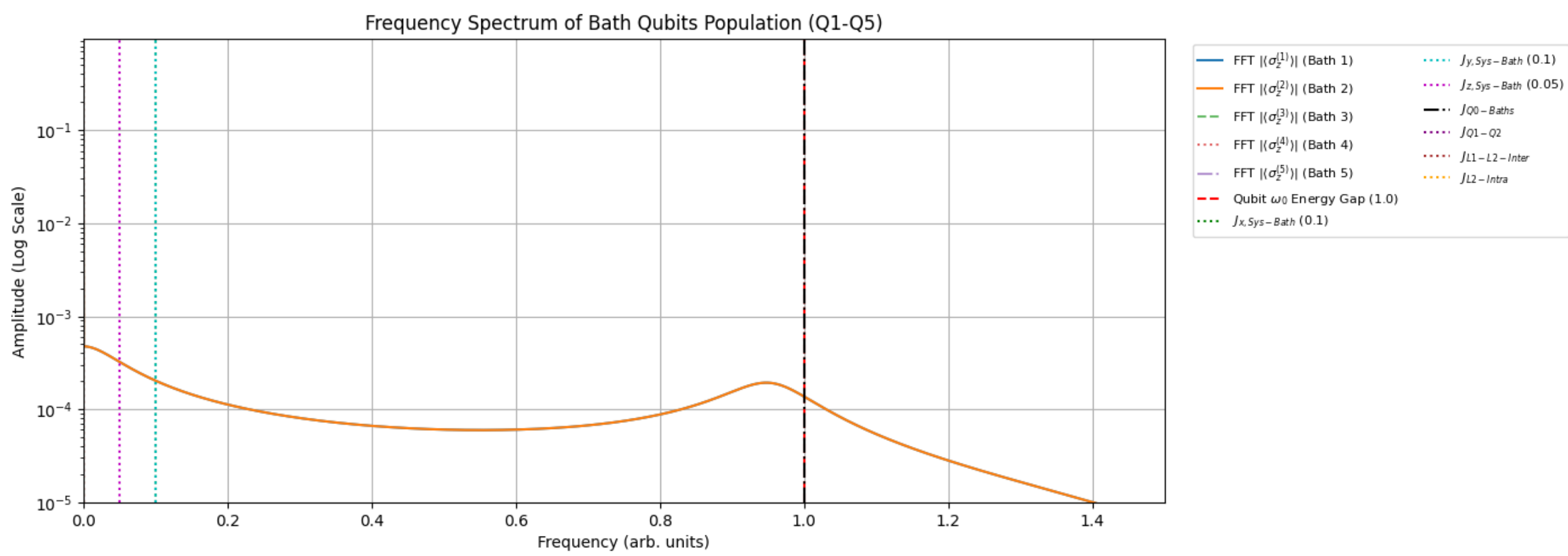

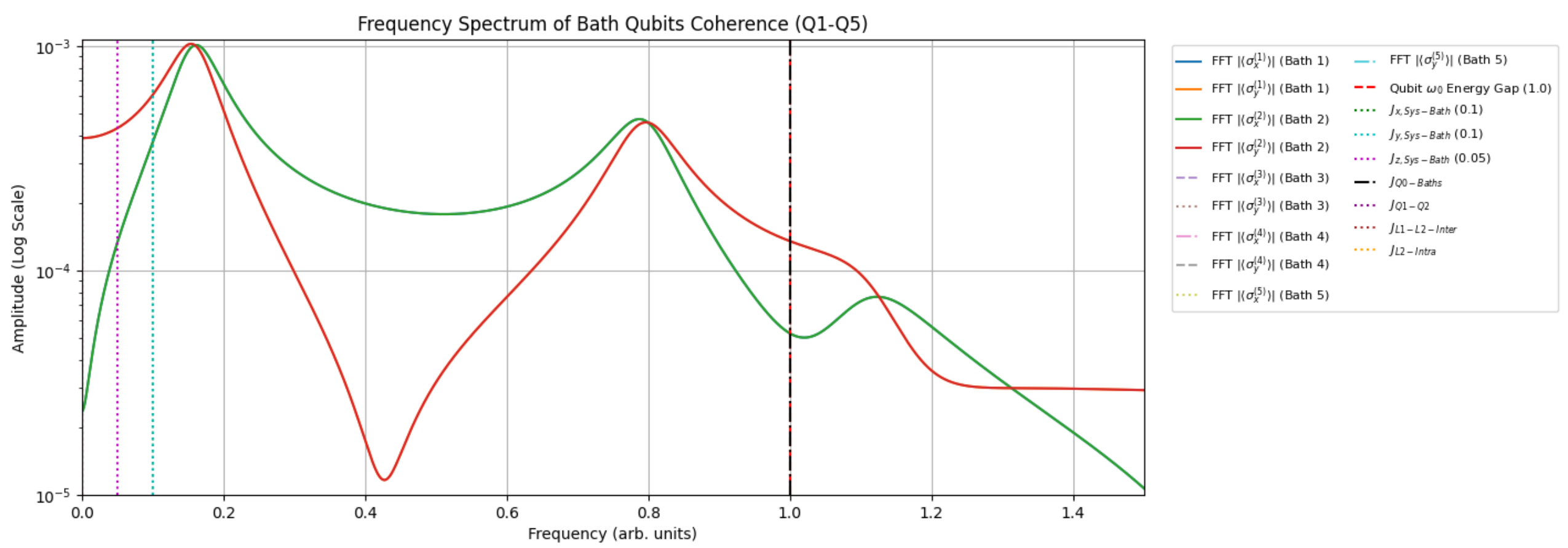

CASE 3a:

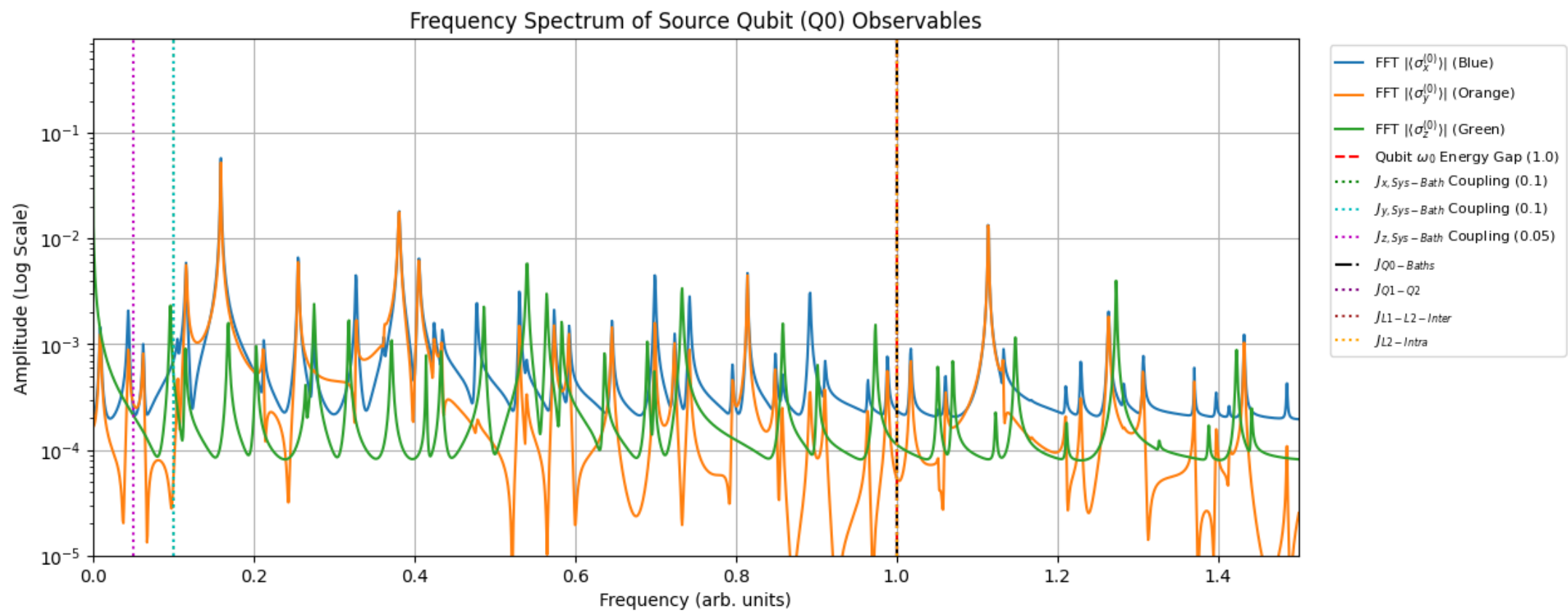

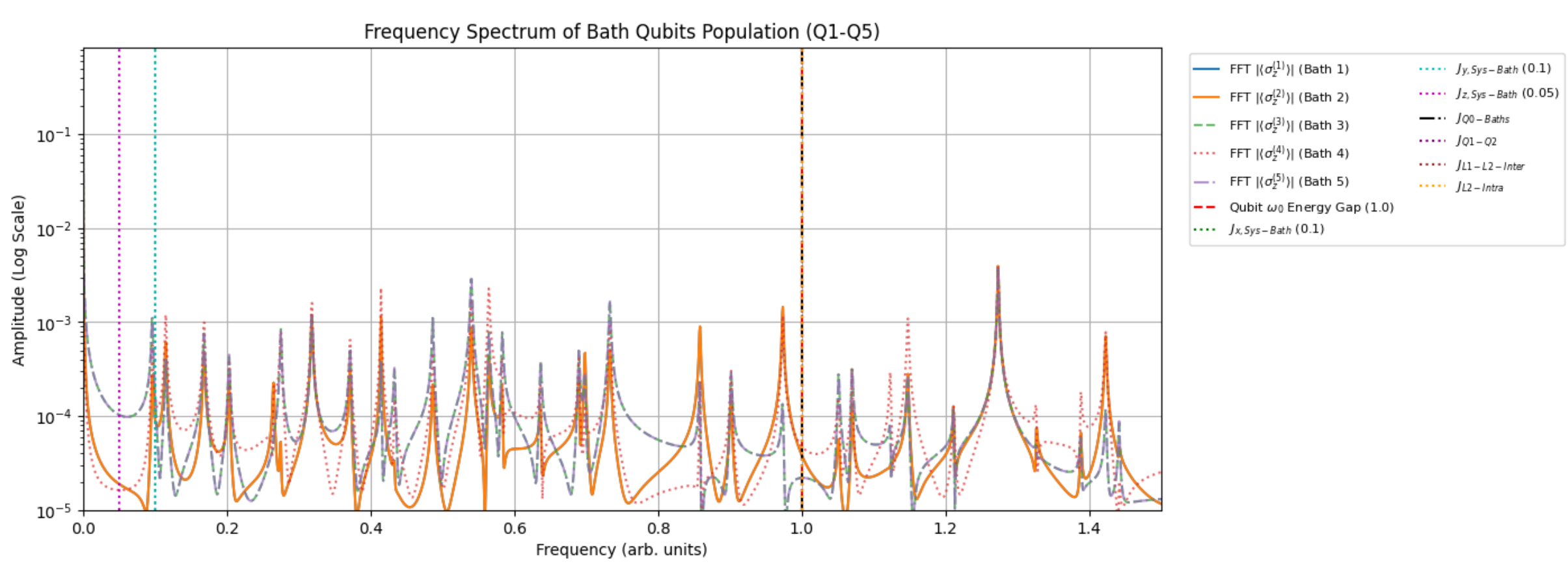

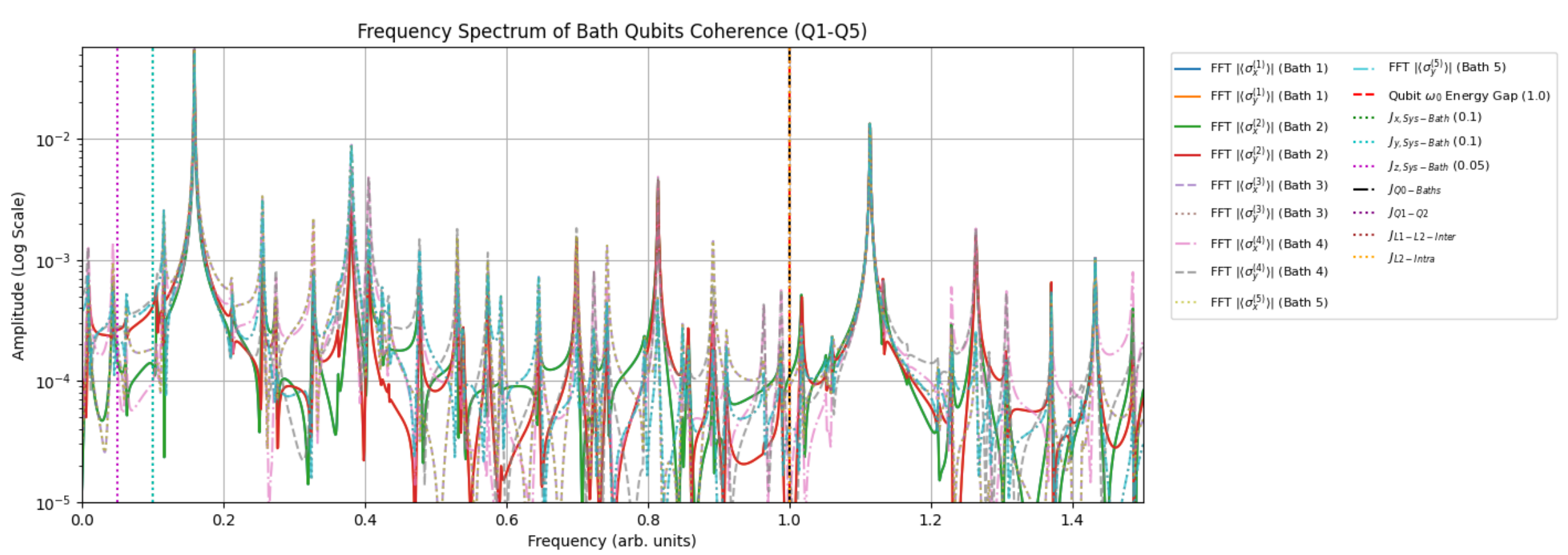

CASE 3b:

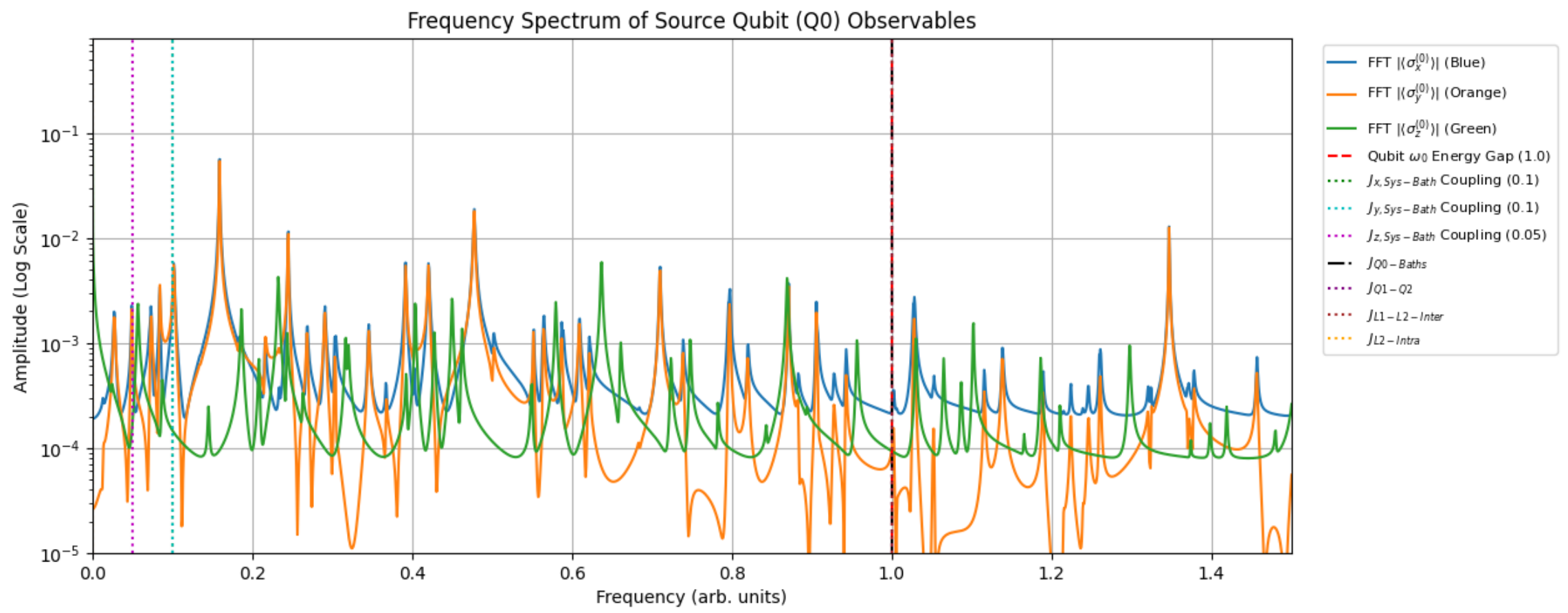

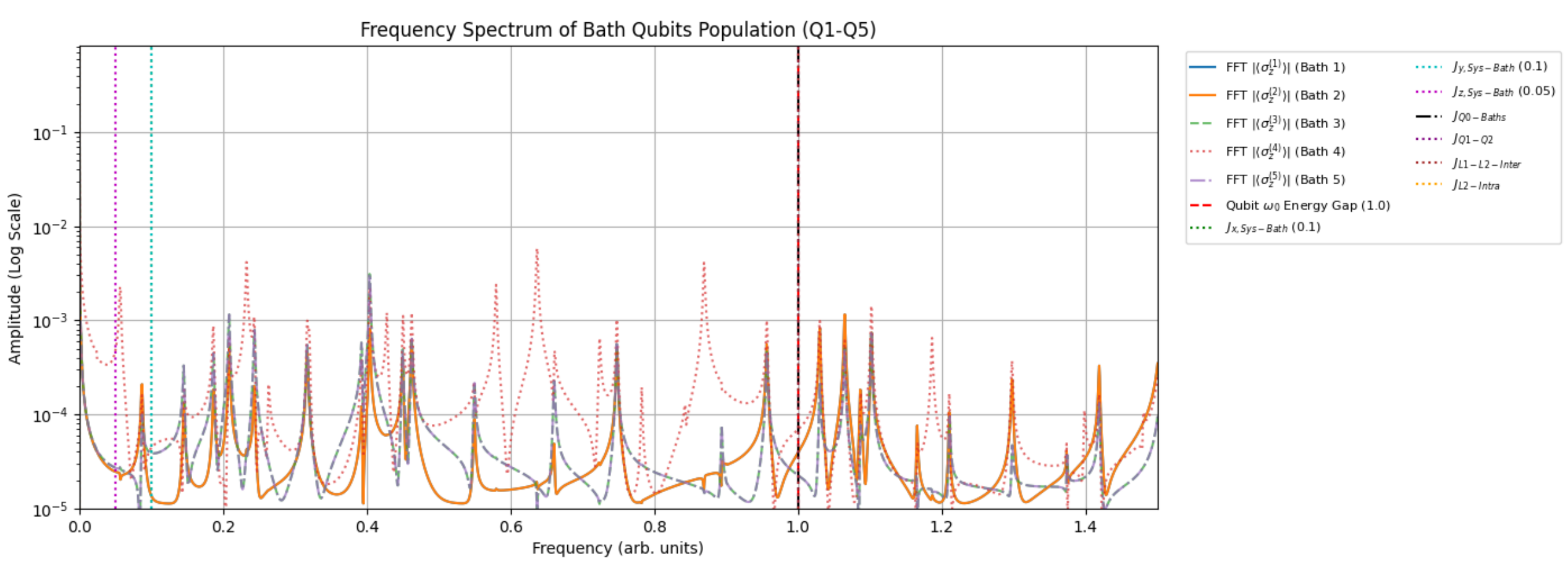

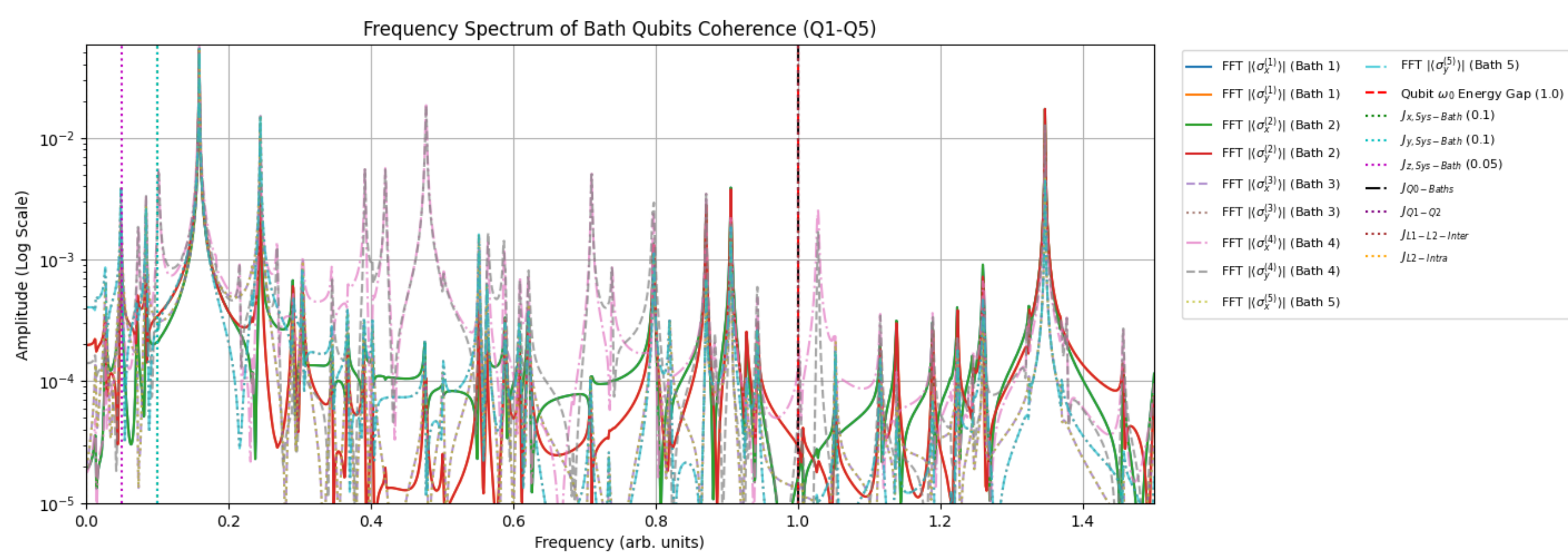

## S4. FFT Normalization and Feature Generation

The FFT processing in this work follows a reproducible multi-step procedure designed for robust comparison across simulation cases and hardware platforms. The implementation is shown in the provided script block and is summarized here.

<u>Step 1 – Observable Collection</u>

For the source qubit $(Q_0)$ we collect the time-domain expectation values for $\langle\sigma_x\rangle, \langle\sigma_y\rangle$ *and* $\langle\sigma_z\rangle$:

$$observables = \{x{:}\,\langle\sigma_x\rangle, y{:}\,\langle\sigma_y\rangle, z{:}\,\langle\sigma_z\rangle\}$$

Note that the $\sigma_z$ component is a directly measurable population signal in most qubit platforms (trapped ions, neutral atoms, superconducting qubits), enabling both simulation–experiment comparisons and practical inverse-engineering workflows.

<u>Step 2 – Mean Removal</u>

Each signal is then mean centered to suppress the DC offset (the constant baseline of the signal):

$$s'(t) = s(t) - \bar{s}$$

This avoids artificial low-frequency peaks dominating the spectrum.

<u>Step 3 – Spectral Computation (Multitaper FFT)</u>

We compute the Fourier transform:

$$FFT[s'(t)] = \sum_{n=0}^{N-1} s'(t_n) e^{-\frac{i2\pi kn}{N}}$$

The output is then converted to a magnitude spectrum and restricted to positive frequencies.

<u>Step 4 – Log-Enhanced Normalization</u>

To ensure visibility of both dominant and weak spectral components, we apply:

$$A_{log} = log_{10}(A + \epsilon)$$

with $\varepsilon = 10^{-12}$ to avoid log (0).

We then normalize:

$$A_{norm} = \frac{A_{log} - \min(A_{log})}{\max(A_{log}) - \min(A_{log})}$$

This confines values to $[0,1]$, allowing fair comparison between runs with different absolute scales.

For visualization, however, the figures display raw one-sided FFT amplitudes (linear magnitude), whereas the ML feature vectors are computed from the mean-removed, log-enhanced, normalized spectra described above.

<u>Additional note on the signal processing:</u>

Although not implemented in the current paper's baseline, one possible enhancement to a single-window FFT is the use of Digital Signal Processing (DSP) techniques such as a multitaper spectral estimator based on *discrete prolate spheroidal sequences* (DPSS). DPSS tapers are orthogonal weighting functions applied to the time-domain signal before Fourier transformation. They are designed to minimize spectral leakage within a specified bandwidth, and by applying multiple orthogonal tapers, one obtains independent spectral estimates whose average reduces variance.

For a mean-centered signal $s'(t_n)$, each taper $v_k(t)$ is applied as

$$S_k(f) = \sum_{n=0}^{N-1} [s'(t_n) v_k(t_n)] e^{-i2\pi f t_n}$$

and the multitaper spectrum is obtained by averaging over the K orthogonal tapers:

$$\tilde{S}(f) = \frac{1}{K} \sum_{k=1}^{K} |S_k(f)|,$$

Here:

- $s'(t_n) = mean\ centered\ signal\ value\ at\ time\ sample\ t_n,$
- $N = number\ of\ time\ sampled,$
- $f = frequnacy\ variables,$
- $v_k(t_n) = k^{th}\ DPSS\ taper,$
- $K = number\ of\ orthogonalized\ taper\ used,$
- $S_k(f) = Fourier\ transform\ of\ the\ tapered\ signal\ for\ taper\ k,$
- $\tilde{S}(f) = mutitaper\ spectrum, obtained\ by\ averaging\ the\ magnitides\ acrossall\ K\ tapes.$

Finally, the one-sided spectrum is scaled as in the baseline FFT (2|S|/N) so that results remain directly comparable across simulation cases. This multitaper procedure reduces variance roughly as $\frac{1}{K}$, suppresses spectral leakage, and makes narrow, structured bath resonances more clearly visible than with a single FFT.

The figure illustrates the effect of different spectral processing steps. The standard FFT (light blue) shows variance and spectral leakage, while the multitaper FFT (orange) produces a smoother and more stable estimate. Applying log-based normalization (dark blue, right axis) rescales the multitaper spectrum to the [0–1] range, highlighting both weak and strong bath resonances on a common scale. Finally, the red dashed line indicates a threshold (0.5) used to systematically identify dominant frequency components.

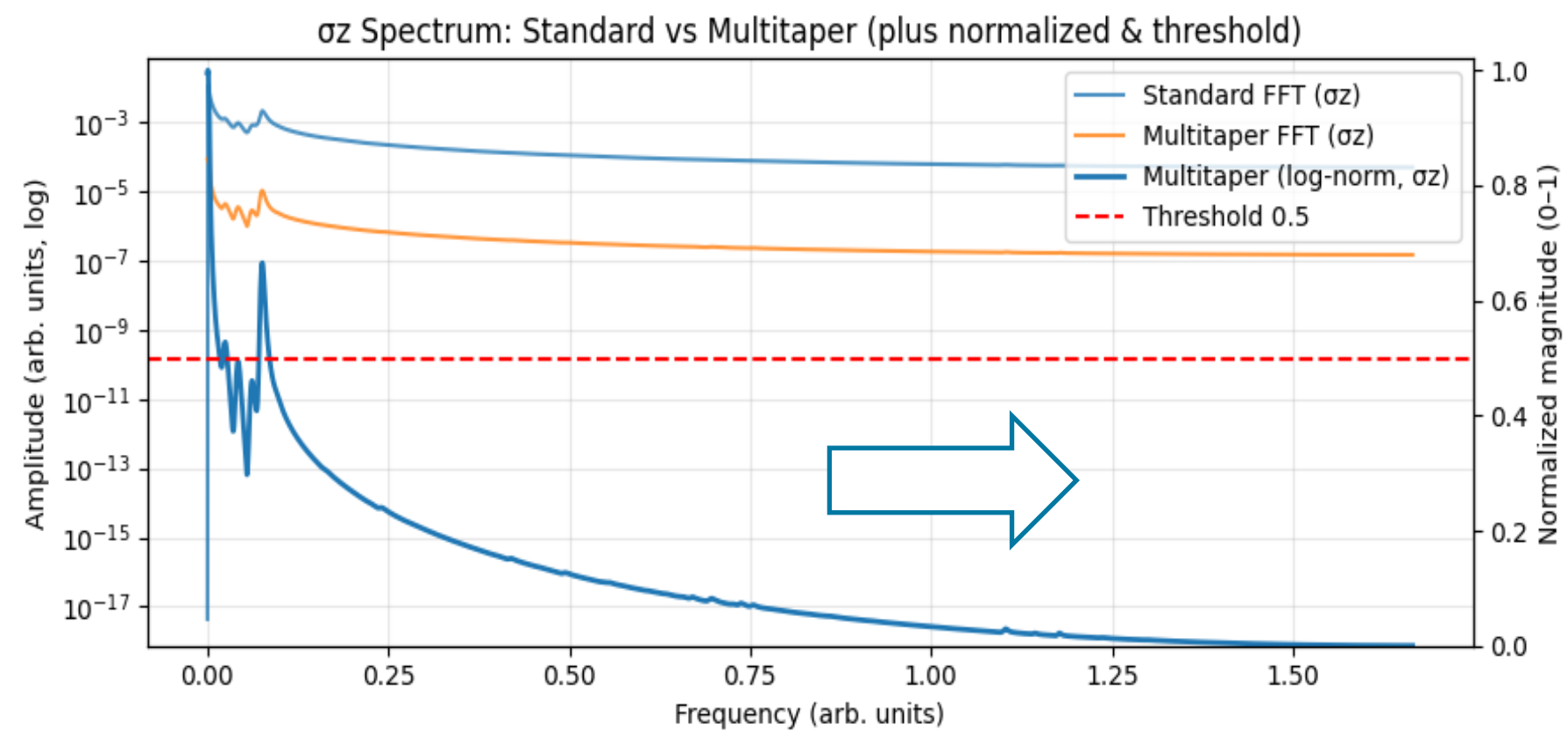

Step 5 – Multi-Axis Feature Merging

Normalized FFT vectors from all three axes are concatenated:

$$F = \left[A_{norm}^{(x)}, A_{norm}^{(y)}, A_{norm}^{(z)}\right]$$

This multi-axis vector becomes the high-dimensional feature input for later PCA compression.

Step 6 – Data Output

CSV outputs are generated for:

- Source qubit observables
- Bath qubit $\sigma_z$ populations
- Bath qubit $\sigma_x/\sigma_y$ coherences

plus, the merged FFT featured vectors and target parameters (couplings, dissipations, backflow rate and DEPS) for machine learning.

## S5. Machine Learning Framework

Our machine learning workflow combines Principal Component Analysis (PCA) for dimensionality reduction and XGBoost for supervised regression. This pipeline is implemented using the scikit-learn and XGBoost libraries and is applied directly to the spectral fingerprint data generated from our simulations.

Principal Component Analysis (PCA)

We employ Principal Component Analysis (PCA) as a pre-processing step to reduce the dimensionality of our high-dimensional, multi-axis feature vectors. The goal is to find a set of orthogonal vectors, known as principal components, that best represent the variance in the data. The first principal component captures the most variance, the second the next most, and so on.

The PCA algorithm performs a linear transformation on the data. For a dataset matrix $X \in \mathbb{R}^{mxn}$ where m is the number of samples and n is the number of features, PCA finds the eigenvectors of the covariance matrix $\sum = \frac{1}{m} X^T X$. These eigenvectors are the principal components, and they are ordered by their corresponding eigenvalues, which indicate the amount of variance captured. We project the data onto a reduced number of n-dimensional principal components, effectively compressing the feature vector while retaining the most relevant information.

Gradient Boosting (XGBoost)

After PCA, we use a gradient boosting algorithm, specifically XGBoost, for our supervised learning tasks. XGBoost is a powerful and robust ensemble method that combines a series of weak prediction models (decision trees) to create a single, highly accurate model. The model is built sequentially, with each new tree correcting the errors of the previous ones. The objective is to minimize a loss function, $\mathcal{L}(\theta)$, by adding a new function, $f_t$ , at each step $t$ to the ensemble:

$$\hat{y}^{(t)} = \hat{y}^{(t-1)} + f_t(x)$$

We chose this algorithm for its effectiveness in handling complex, non-linear relationships between spectral fingerprint features and the physical parameters of interest. Using a multi-output regressor, our model is trained to predict multiple targets simultaneously:

- inter-layer coupling ($J_{L12}$),
- system–bath coupling ($J_{sb}$),
- non-Markovian backflow rate,
- and the logarithm of the minimal distance to eigenvalue pair spacing.

The model takes the principal components of the spectral data as input and outputs predictions for these physical quantities.

Typical model hyperparameters: n estimators=300, max depth = 6, learning rate = 0.02, subsample = 0.9. Evaluation metrics: Mean Squared Error (MSE) and coefficient of determination ($R^2$) for each target.

Data partitioning

We employ an 80/20 train–test split to ensure balanced evaluation of model performance.

## S6. Structured Bath Delay Analysis

S.6.A. From global kernels to layer-resolved memory

Section 6 connected the finite-bath framework to continuum descriptions such as the Nakajima–Zwanzig (NZ) formalism. NZ provides a global kernel: it captures how the system as a whole exchanges information with its environment, and through that kernel one can diagnose the existence of memory and even quantify overall backflow. What it does not reveal is where inside the bath the memory resides, or how it propagates between different parts of the environment.

The structured bath construction fills this gap. By discretizing the bath into explicit layers, the same logic used in NZ can be applied internally: correlation functions no longer describe only system–environment exchange, but also layer-by-layer transfer within the bath. In this way, the familiar idea of a memory kernel becomes a spatially resolved diagnostic, letting us watch memory being injected into $L_1$, delayed, and then passed into $L_2$.

This extension is not a replacement for NZ but a refinement of its diagnostic power. NZ gives the global view; the structured bath exposes the circulation of memory inside the environment.

S.6.B. Correlation diagnostics

To carry out this analysis, we measure correlations between the layer signals.

- Autocorrelation of $L_1$ or $L_2$ shows how long that layer "remembers itself."
- Cross-correlation *between* $L_1$ and $L_1$ shows how memory propagates forward.

The general expression:

$$C_{1\to 2}(\tau) = \langle \delta_{\sigma_z}^{(L1)}(t)\delta_{\sigma_z}^{(L2)}(t+\tau)\rangle_t \ ,\tau^* = \arg\max C_{1\to 2}(\tau),$$
$$\text{with } \delta_{\sigma_z}^{(Lk)} = \delta_{\sigma_z}^{(Lk)} - \langle \delta_{\sigma_z}^{(Lk)}\rangle_t.$$

the interpretation of the time of peak-lag:

- $\tau^* < 0$: $L_2$ trails $L_1$by $\tau^*$
- $\tau^* = 0$: $L_1$ and $L_2$ are synchronous
- The height of the peak = strength of the correlation
- The width and oscillations = how long and in what form the memory persists

Thus, the very same correlation logic used in NZ kernels now serves as a microscope inside the bath, layer by layer.

S.6.C. The 36-case dataset

We scanned system–bath injection ($J_{sb}$), inter-layer coupling ($J_{L12}$), and thermal dissipation ($\gamma_{L1}, \gamma_{L2}$) across 36 cases.

Table S1. Complete diagnostics of the 6-qubit system.

6-Qubit System Diagnostics : $\beta = 0.5, \omega = 1.0, J_{x,y,z} = [0.1,0.1,0.05]$

$J_{sb}$: 0.1-0.4-1.0, $J_{L12}$: 0.0001-0.001-0.1-1.0

$\gamma_{L1}$: 0.02 vs. 0.005 vs 0.002 , $\gamma_{L2}$: 0.04 vs. 0.01 vs 0.004

| No. | CASE | $J_{sb}$ | $J_{L12}$ | $J_{L1}$ | $J_{L2}$ | $\gamma_{L1}$ | $\gamma_{L2}$ | Peak lag |
|---|---|---|---|---|---|---|---|---|
| 1 | 1a-1 | 0.1 | 0.0001 | 0.04 | 0.06 | 0.02 | 0.04 | -4.8 |
| 2 | 1b-1 | 0.1 | 0.01 | 0.04 | 0.06 | 0.02 | 0.04 | -5.1 |
| 3 | 1c-1 | 0.1 | 0.1 | 0.04 | 0.06 | 0.02 | 0.04 | -6.3 |
| 4 | 1d-1 | 0.1 | 1.0 | 0.04 | 0.06 | 0.02 | 0.04 | 0.0 |
| 5 | 2a-1 | 0.4 | 0.0001 | 0.04 | 0.06 | 0.02 | 0.04 | -10.8 |
| 6 | 2b-1 | 0.4 | 0.01 | 0.04 | 0.06 | 0.02 | 0.04 | -11.4 |
| 7 | 2c-1 | 0.4 | 0.1 | 0.04 | 0.06 | 0.02 | 0.04 | -6.0 |
| 8 | 2d-1 | 0.4 | 1.0 | 0.04 | 0.06 | 0.02 | 0.04 | 0.0 |
| 9 | 3a-1 | 1.0 | 0.0001 | 0.04 | 0.06 | 0.02 | 0.04 | -4.8 |
| 10 | 3b-1 | 1.0 | 0.01 | 0.04 | 0.06 | 0.02 | 0.04 | -11.4 |

| 11 | 3c-1 | 1.0 | 0.1 | 0.04 | 0.06 | 0.02 | 0.04 | -6.0 |
|---|---|---|---|---|---|---|---|---:|
| 12 | 3d-1 | 1.0 | 1.0 | 0.04 | 0.06 | 0.02 | 0.04 | -0.3 |
| 13 | 1a-2 | 0.1 | 0.0001 | 0.04 | 0.06 | 0.005 | 0.01 | -23.7 |
| 14 | 1b-2 | 0.1 | 0.01 | 0.04 | 0.06 | 0.005 | 0.01 | -24.6 |
| 15 | 1c-2 | 0.1 | 0.1 | 0.04 | 0.06 | 0.005 | 0.01 | -6.3 |
| 16 | 1d-2 | 0.1 | 1.0 | 0.04 | 0.06 | 0.005 | 0.01 | 0.0 |
| 17 | 2a-2 | 0.4 | 0.0001 | 0.04 | 0.06 | 0.005 | 0.01 | -29.7 |
| 18 | 2b-2 | 0.4 | 0.01 | 0.04 | 0.06 | 0.005 | 0.01 | -30.9 |
| 19 | 2c-2 | 0.4 | 0.1 | 0.04 | 0.06 | 0.005 | 0.01 | -6.3 |
| 20 | 2d-2 | 0.4 | 1.0 | 0.04 | 0.06 | 0.005 | 0.01 | 0.0 |
| 21 | 3a-2 | 1.0 | 0.0001 | 0.04 | 0.06 | 0.005 | 0.01 | -29.7 |
| 22 | 3b-2 | 1.0 | 0.01 | 0.04 | 0.06 | 0.005 | 0.01 | -30.9 |
| 23 | 3c-2 | 1.0 | 0.1 | 0.04 | 0.06 | 0.005 | 0.01 | -6.3 |
| 24 | 3d-2 | 1.0 | 1.0 | 0.04 | 0.06 | 0.005 | 0.01 | 0.0 |
| 25 | 1a-3 | 0.1 | 0.0001 | 0.04 | 0.06 | 0.002 | 0.004 | -73.8 |
| 26 | 1b-3 | 0.1 | 0.01 | 0.04 | 0.06 | 0.002 | 0.004 | -43.8 |
| 27 | 1c-3 | 0.1 | 0.1 | 0.04 | 0.06 | 0.002 | 0.004 | -6.3 |
| 28 | 1d-3 | 0.1 | 1.0 | 0.04 | 0.06 | 0.002 | 0.004 | 0.0 |
| 29 | 2a-3 | 0.4 | 0.0001 | 0.04 | 0.06 | 0.002 | 0.004 | -73.8 |
| 30 | 2b-3 | 0.4 | 0.01 | 0.04 | 0.06 | 0.002 | 0.004 | -37.5 |
| 31 | 2c-3 | 0.4 | 0.1 | 0.04 | 0.06 | 0.002 | 0.004 | -6.3 |
| 32 | 2d-3 | 0.4 | 1.0 | 0.04 | 0.06 | 0.002 | 0.004 | 0.0 |
| 33 | 3a-3 | 1.0 | 0.0001 | 0.04 | 0.06 | 0.002 | 0.004 | -73.8 |
| 34 | 3b-3 | 1.0 | 0.01 | 0.04 | 0.06 | 0.002 | 0.004 | -37.5 |
| 35 | 3c-3 | 1.0 | 0.1 | 0.04 | 0.06 | 0.002 | 0.004 | -6.3 |
| 36 | 3d-3 | 1.0 | 1.0 | 0.04 | 0.06 | 0.002 | 0.004 | 0.0 |

To provide a more detailed analysis, the lag values from the 36-case dataset are condensed into three heatmaps, one for each thermal dissipation set. These heatmaps provide a global map of memory transfer by visualizing how the peak lag time ($\tau^*$) changes as a function of the system-bath injection strength ($J_{sb}$) and the inter-layer coupling ($J_{L12}$). The color scale of the maps shows the peak lag, with darker colors representing longer delays.

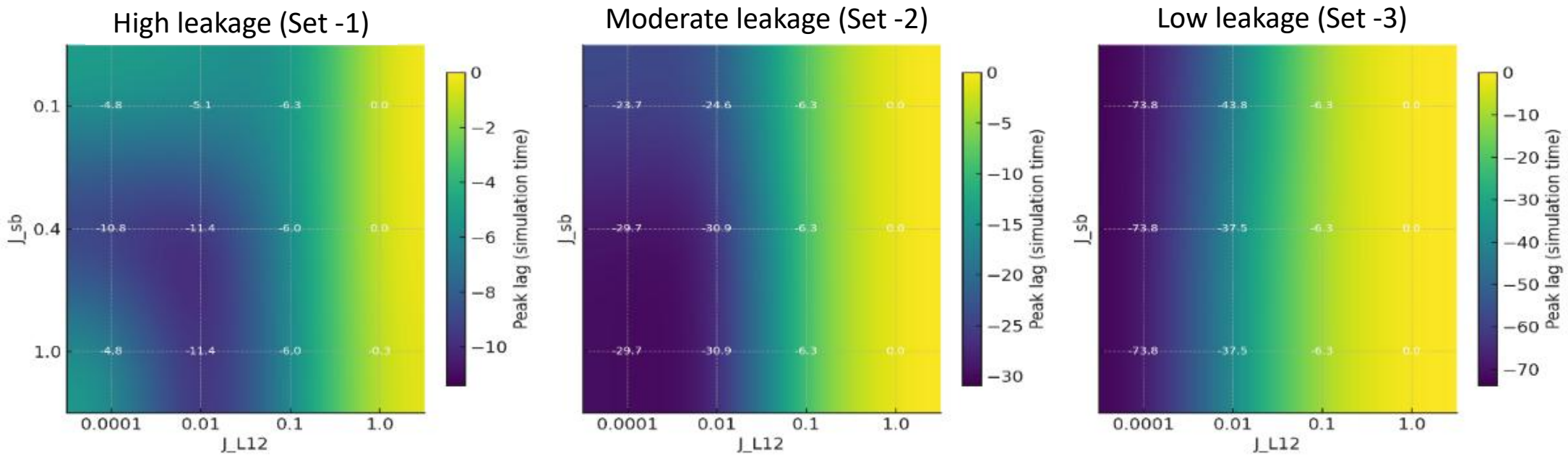

A few key observations can be made from this heat map:

**High Leakage** (Set –1, $\gamma_{L1}, \gamma_{L2} = 0.02, 0.04$): In this regime, the system is strongly coupled to a thermal bath, and information dissipates rapidly. The heatmaps show only modest delays, ranging from –4.8 to –11.4. The dark colors (indicating delays) are confined to a narrow region of low inter-layer coupling, and the delays collapse quickly at wide bridges where the coupling is high. This behavior is characteristic of a bath that quickly erases memory, making it appear more Markovian.
**Moderate Leakage** (Set –2, $\gamma_{L1}, \gamma_{L2} = 0.005, 0.01$): As the thermal leakage rate is reduced, the system-bath injection and inter-layer coupling have a more pronounced effect on the memory. The heatmaps show significantly amplified delays ranging from –23 to –31, but these delays remain bounded. This is a transitional regime where the structured bath begins to retain memory, but the leakage is still strong enough to prevent the extreme delays seen in the low-leakage scenario.
**Low Leakage** (Set –3, $\gamma_{L1}, \gamma_{L2} = 0.002, 0.004$): This is the most revealing set, demonstrating maximal amplification of memory effects. The heatmaps show significant regions of deep purple, representing a peak lag of –73.8 at narrow bridges (low inter-layer coupling). As the inter-layer coupling increases, the delays shrink stepwise and only fully collapse at unity where the layers are perfectly coupled. This demonstrates that over-decoupling the bath layers can create a "bottleneck" that traps memory, leading to *very long-lived non-Markovian effects.*
Together, the table and heatmaps give the global map of memory transfer, providing both numerical diagnostics and physical interpretation of the bath's behavior. They visually demonstrate that thermal dissipation ($\gamma$) is a dual control knob that presents a trade-off between protecting coherence and amplifying memory.

S.5.D. Representative Case Studies: A Visualized Spectrum of Memory

To interpret the patterns revealed by the 36-case dataset, we will now analyze four representative cases. For each case, we will examine two key figures: the normalized layer signals and the $L_1 \leftrightarrow L_2$ cross-correlation function. The normalized layer signals show the collective, time-dependent behavior of each bath layer, while the cross-correlation function diagnoses the flow of memory between them.

1.Case 1 (1a-1): High Thermal Leakage, Short-Lived Memory
$(J_{sb} = 0.1, J_{L12} = 0.0001, \gamma_{L1} = 0.02, \gamma_{L2} = 0.04$
*Layer Signals (normalized)*: As seen in the plot below, the normalized layer signals for $L_1$ and $L_2$ decay quickly, showing that information dissipates rapidly into the environment. This quick decay is characteristic of a high-leakage bath.

*Cross-Correlation* $L_1 \leftrightarrow L_2$: The corresponding cross-correlation plot shows a sharp peak with a modest negative lag of approximately -4.8. The narrow width of the peak indicates that while there is a memory effect, it is very short-lived. This case shows that even with weak injection and high thermal leakage, some short-lived memory is still retained.

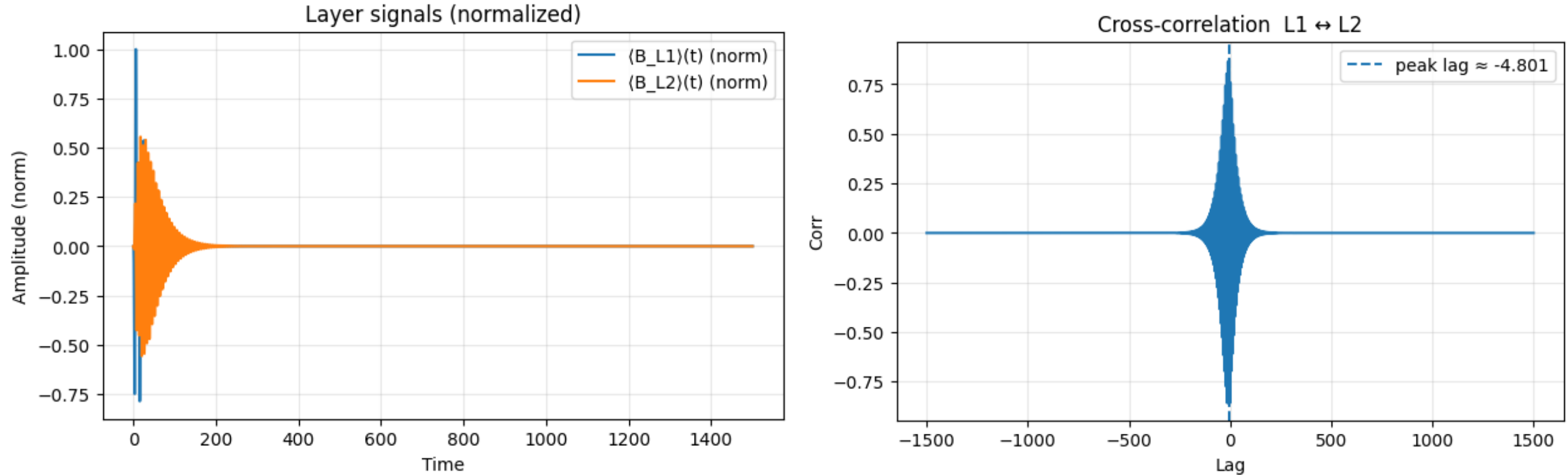

2. Case 25 (1a-3): The Bottleneck Regime
(1a-3, squeezed source and bridge, $J_{sb} = 0.1$, $J_{L12} = 0.0001$).
*Layer Signals (normalized)*: In this low-leakage scenario, the layer signals show a significantly slower decay. The oscillations persist for a much longer time, demonstrating that information remains trapped within the bath layers.

*Cross-Correlation* $L_1 \leftrightarrow L_2$: The cross-correlation plot shows a very broad, pronounced peak at a large negative lag of -73.8. The plot highlights the presence of dense, high-frequency oscillations contained within this large, long-lived correlation function. This pronounced delay is a clear signature of a "bottleneck" effect, where information is squeezed and takes a long time to propagate from $L_1$ to $L_2$. This illustrates a key finding: over-decoupling the bath layers can trap memory rather than erasing it, leading to long-lived memory effects.

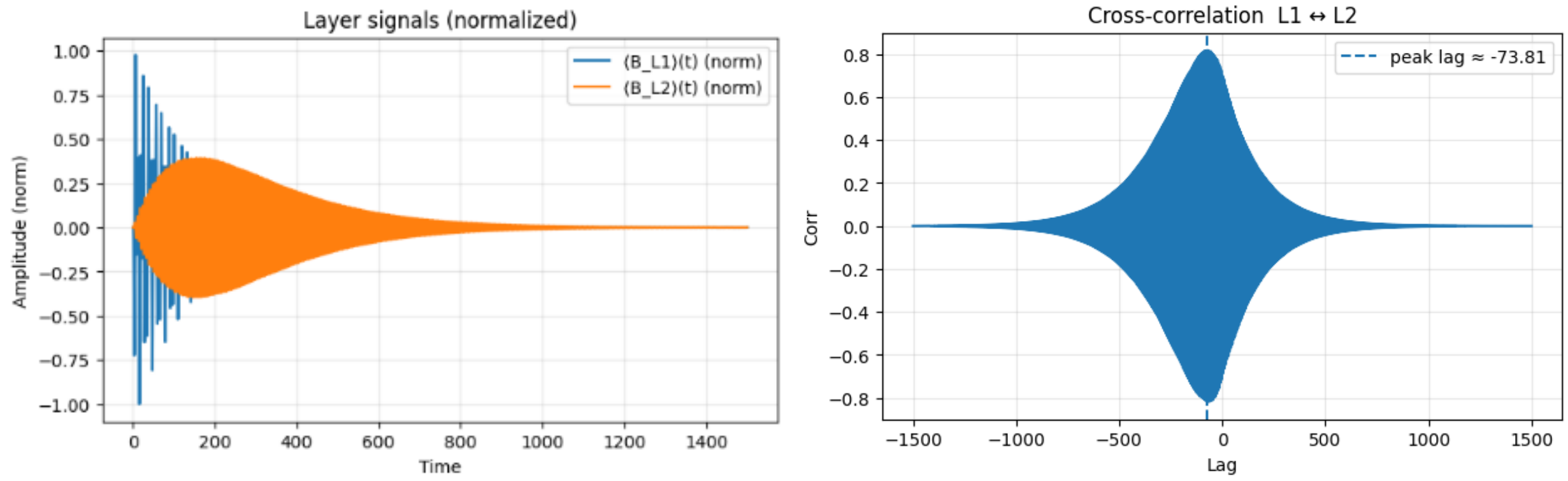

3. Case 30 (2b-3): The Intermediate Bridge, Oscillatory Fragmentation
*Layer Signals (normalized)*: The layer signals for this case show a decay that is faster than Case 25 but slower than Case 1. The signals exhibit persistent oscillations that gradually fade out.
*Cross-Correlation* $L_1 \leftrightarrow L_2$: The cross-correlation plot for this case shows a peak at a lag of -37.5, which is smaller than in the bottleneck case. More importantly, the central peak is flanked by distinct oscillatory "shoulders". This indicates that as memory propagates, it fragments into multiple timescales, a signature of the "intermediate bridge" regime.

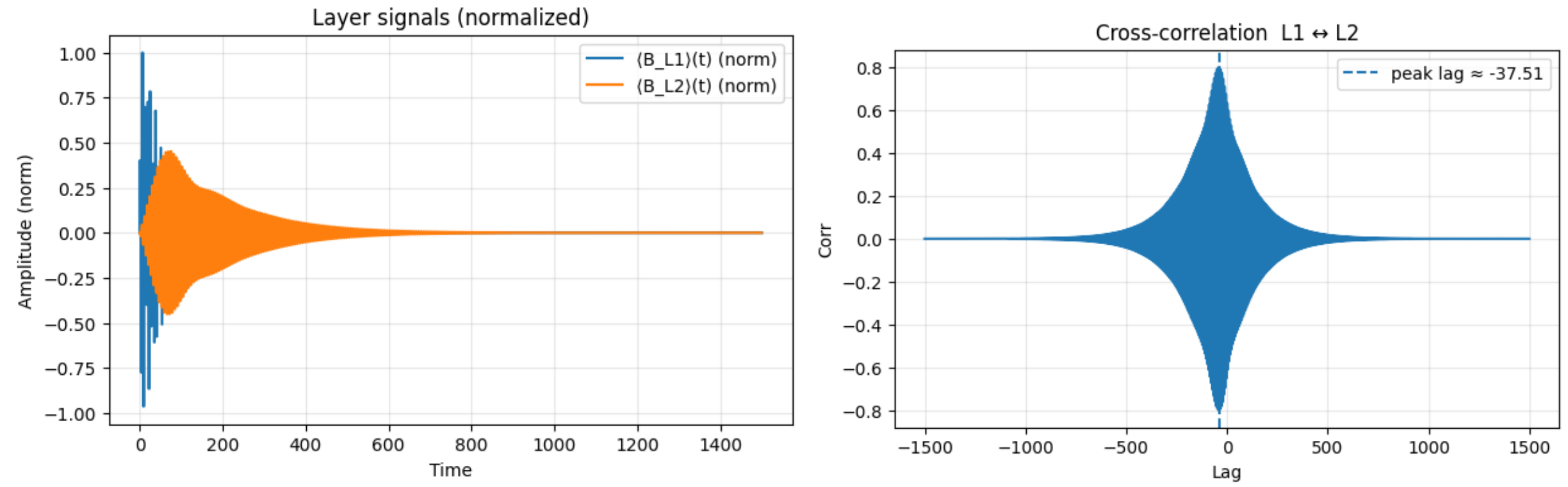

4. Case 32 (2d-3): The Wide Bridge, Synchronous Locking

*Layer Signals (normalized)*:: The layer signals in this case are nearly perfectly synchronized. Both $L_1$and $L_2$ signals decay at a similar rate, and their oscillations are aligned.

*Cross-Correlation* $L_1 \leftrightarrow L_2$: The cross-correlation plot confirms this synchrony with a peak centered at a lag of approximately 0. However, even with this synchrony, the cross-correlation function shows persistent oscillatory ringing. This demonstrates a crucial point: a peak lag of zero does not necessarily mean the absence of memory. Instead, it indicates that the memory is shared and propagates synchronously across the layers, which can still lead to non-Markovian behavior.

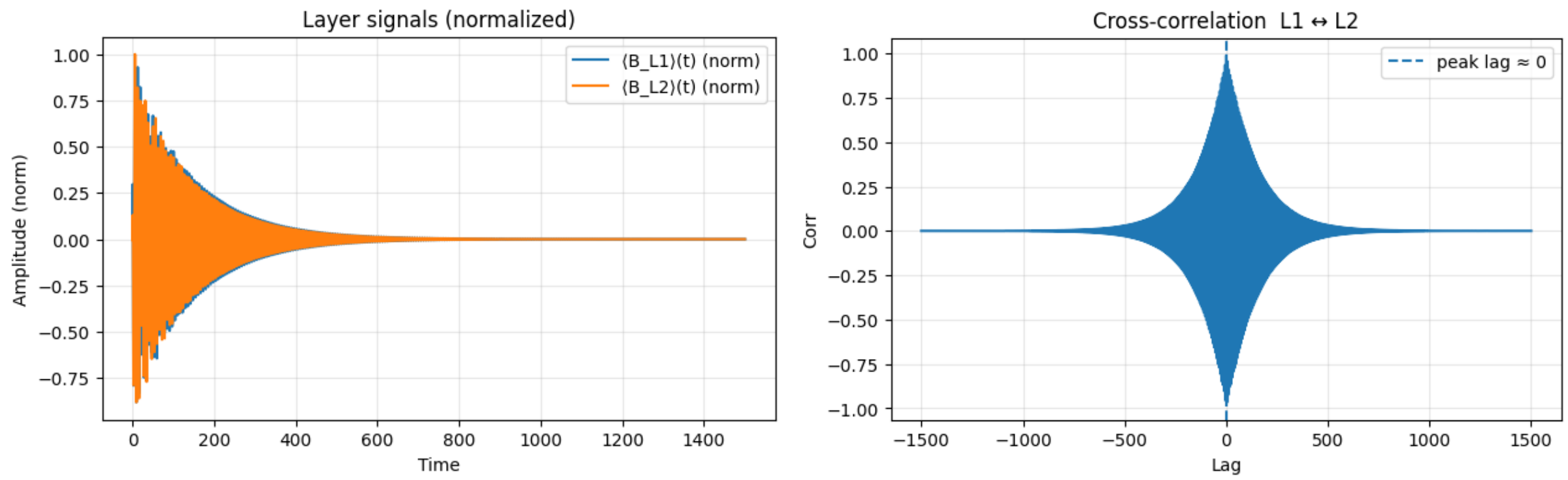

These illustrate the spectrum: bottleneck → oscillatory fragmentation → synchronous locking.

6.6.E. Role of Thermal Dissipation

The thermal leakage rate, $\gamma$, acts as a dual control knob, governing whether memory is suppressed or amplified within the bath. It is not simply a "good" or "bad" parameter but rather represents a trade-off between fragile Markovianity and structured long-memory feedback.

- Large $\gamma$: A large thermal leakage rate collapses delays and narrows oscillations, making the bath appear more Markovian. However, this comes at the cost of accelerating dephasing in the system.

- Small $\gamma$: A small thermal leakage rate preserves coherence but as demonstrated in our case studies, produces long delays and oscillatory kernels, meaning the bath retains memory.

Where NZ provides a global view of memory through kernels, structured baths deliver a layer-resolved view, showing how memory is injected, delayed, and transferred across the bath. The correlation functions that underpin NZ become more powerful when applied inside the bath, providing both numerical diagnostics (lags, heatmaps) and physical interpretation (bottlenecks, oscillations, synchrony).

The 36-case scan demonstrates this systematically, the case studies illustrate the regimes, and the dissipation analysis highlights the design principle: optimal environments balance coherence protection against memory amplification.

S.6.F. Conclusion

Where the Nakajima–Zwanzig (NZ) formalism provides a global view of memory through kernels, our structured bath framework delivers a layer-resolved view, showing precisely how memory is injected, delayed, and transferred across the bath. The same correlation functions that underpin NZ become more powerful when applied inside the bath, providing both numerical diagnostics (lags, heatmaps) and a clear physical interpretation (bottlenecks, oscillations, synchrony).

The 36-case scan demonstrates this systematic analysis, the case studies illustrate the various regimes of memory transfer, and the dissipation analysis highlights a fundamental design principle: *optimal environments balance coherence protection against memory amplification.* This approach views structured baths not as a source of noise to be eliminated, but as a potential collaborative partner in the design of quantum systems.

**S7. Extension of structured bath for two logical qubits**

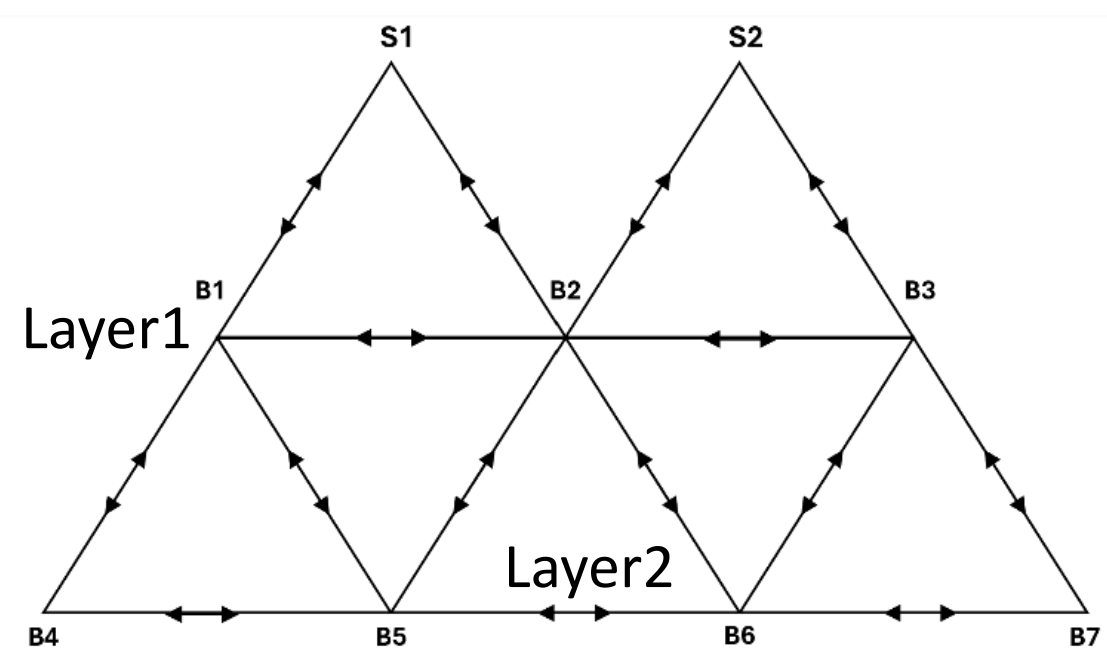

Here, we extend the formulism toward two logical (system) qubits with shared structured baths as pictured in Figure 11. No direct $S_1 \leftrightarrow S_2$ coupling is included, so that any correlations or entanglement arise solely through the structured bath. The logical qubits are labeled $S_1$ and $S_2$. Layer 1 would comprise of $B_1$, $B_2$, $B_3$ with $B_2$ shared by both systems. Layer 2 is made of $B_4$, $B_5$, both attached to the $S_1$ side, and $B_6$, $B_7$, both attached to the $S_2$ side. Note that $B_5$ and $B_6$ are also attached to the shared bath $B_2$. We define the global system as:

$$Q = \{S_1, S_2, B_1, \dots, B_7\}$$

The Pauli vectors are defined as:

$$\vec{\sigma}^{(i)} = \left(\sigma_x{}^{(i)}, \sigma_y{}^{(i)}, \sigma_z{}^{(i)}\right)$$

and

$$\vec{\sigma}^{(i)}.\vec{\sigma}^{(j)} = {\sigma_x}^{(i)}{\sigma_x}^{(j)} + {\sigma_y}^{(i)}{\sigma_y}^{(j)} + {\sigma_z}^{(i)}{\sigma_z}^{(j)}$$

The local Hamiltonian is defined as:

$$H_{loc} = \sum_{q\in Q} \frac{\omega_q}{2}{\sigma_z}^{(q)} = \frac{\omega_{S_1}}{2}{\sigma_z}^{(S_1)} + \frac{\omega_{S_2}}{2}{\sigma_z}^{(S_2)} + \sum_{k=1}^{7} \frac{\omega_{B_k}}{2}{\sigma_z}^{(B_k)}$$

**Couplings:**

System to layer 1:

$$H_{SB} = J_{S_1B_1}\vec{\sigma}^{(S_1)}.\vec{\sigma}^{(B_1)} + J_{S_1B_2}\vec{\sigma}^{(S_1)}.\vec{\sigma}^{(B_2)} + J_{S_2B_2}\vec{\sigma}^{(S_2)}.\vec{\sigma}^{(B_2)} + J_{S_2B_3}\vec{\sigma}^{(S_2)}.\vec{\sigma}^{(B_3)}$$

Within layer 1:

$$H_{L1} = J^{(1)}_{B_1B_2}\vec{\sigma}^{(B_1)}.\vec{\sigma}^{(B_2)} + J^{(1)}_{B_2B_3}\vec{\sigma}^{(B_2)}.\vec{\sigma}^{(B_3)}$$

Within layer 2:

$$H_{L2} = J^{(2)}_{B_4B_5}\vec{\sigma}^{(B_4)}.\vec{\sigma}^{(B_5)} + J^{(2)}_{B_5B_6}\vec{\sigma}^{(B_5)}.\vec{\sigma}^{(B_6)} + J^{(2)}_{B_6B_7}\vec{\sigma}^{(B_6)}.\vec{\sigma}^{(B_7)}$$

Between layer 1 and layer 2:

$$H_{L12} = J_{B_1B_4}\vec{\sigma}^{(B_1)}.\vec{\sigma}^{(B_4)} + J_{B_1B_5}\vec{\sigma}^{(B_1)}.\vec{\sigma}^{(B_5)} + J_{B_2B_5}\vec{\sigma}^{(B_2)}.\vec{\sigma}^{(B_5)} + J_{B_2B_6}\vec{\sigma}^{(B_2)}.\vec{\sigma}^{(B_6)} + J_{B_3B_6}\vec{\sigma}^{(B_3)}.\vec{\sigma}^{(B_6)} + J_{B_3B_7}\vec{\sigma}^{(B_3)}.\vec{\sigma}^{(B_7)}$$

An optional extension toward a direct system edge:

$$H_{SS} = K_{S_1S_2}\vec{\sigma}^{(S_1)}.\vec{\sigma}^{(S_2)}$$

<u>Total coherent Hamiltonian:</u>

$$H = H_{loc} + H_{SB} + H_{L1} + H_{L2} + H_{L12} + H_{SS}$$

Dissipators

$$\mathcal{D}[L]\rho = L\rho L^\dagger - \frac{1}{2}\{L^\dagger L, \rho\}$$

1. System pure dephasing for $S_1$ and $S_2$ following the same approach as in section 2.4:

$$L^{(S_j)}_{deph} = \sqrt{\gamma_{S_j}}{\sigma_z}^{(S_j)}$$

$$\mathcal{D}^{(S_j)}_{deph}[\rho] = \gamma_{S_j}\left({\sigma_z}^{(S_j)}\rho{\sigma_z}^{(S_j)} - \rho\right)$$

2. Bath thermal relaxation and excitation:
   For simplicity, we employ one global thermal occupation $n_{th} = \frac{1}{e^{\beta\omega}-1}$ (the same as Section 2.4).

$$L^{(B_k)}_{\downarrow} = \sqrt{\gamma_{B_k}(1+n_{th})}\,\sigma^{(B_k)}_{-}(emision) \qquad L^{(B_k)}_{\uparrow} = \sqrt{\gamma_{B_k}n_{th}}\sigma^{(B_k)}_{+}\,(absorption)$$

$$\mathcal{D}^{(B_k)}_{thermal}[\rho] = L^{(B_k)}_{\downarrow}\rho L^{(B_k)\dagger}_{\downarrow} - \frac{1}{2}\left\{L^{(B_k)\dagger}_{\downarrow}L^{(B_k)}_{\downarrow}, \rho\right\} + L^{(B_k)}_{\uparrow}\rho L^{(B_k)\dagger}_{\uparrow} - \frac{1}{2}\left\{L^{(B_k)\dagger}_{\uparrow}L^{(B_k)}_{\uparrow}, \rho\right\}$$

Expanded for each bath $B_k$ :

$$\mathcal{D}_{thermal}^{(B_k)}[\rho] = \gamma_{B_k}(1+n_{th})\left(\sigma_-^{(B_k)}\rho\sigma_+^{(B_k)} - \frac{1}{2}\left\{\sigma_+^{(B_k)}\sigma_-^{(B_k)},\rho\right\}\right) + \gamma_{B_k}n_{th}\left(\sigma_+^{(B_k)}\rho\sigma_-^{(B_k)} - \frac{1}{2}\left\{\sigma_-^{(B_k)}\sigma_+^{(B_k)},\rho\right\}\right)$$

Thus, the Master equation for the two-logical-qubit structured bath:

$$\frac{d\rho}{dt} = -i[H,\rho] + \sum_{j=1}^{2}\gamma_{S_j}\left(\sigma_z{}^{(S_j)}\rho\sigma_z{}^{(S_j)} - \rho\right) + \sum_{k=1}^{7}\gamma_{B_k}(1+n_{th})\left(\sigma_-^{(B_k)}\rho\sigma_+^{(B_k)} - \frac{1}{2}\left\{\sigma_+^{(B_k)}\sigma_-^{(B_k)},\rho\right\}\right) + \gamma_{B_k}n_{th}\left(\sigma_+^{(B_k)}\rho\sigma_-^{(B_k)} - \frac{1}{2}\left\{\sigma_-^{(B_k)}\sigma_+^{(B_k)},\rho\right\}\right)$$

$H$ is defined from the total coherent Hamiltonian.

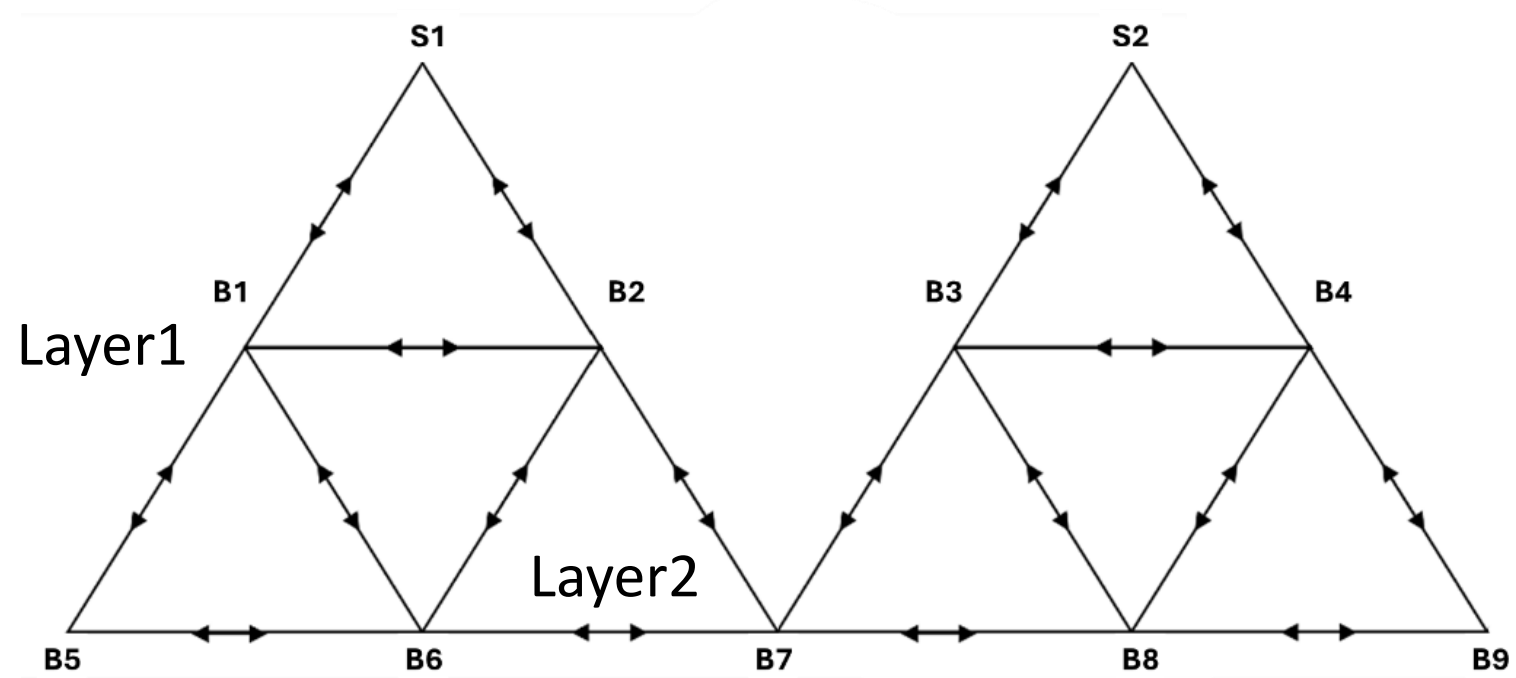

As an alternative, we can also consider two logical (system) qubits $S_1$, $S_2$ whose near-field baths are not shared at Layer-1 but are coupled through a five-site Layer-2 chain $B_5 - B_6 - B_7 - B_8 - B_9$ (see the figure above). Layer-1 nodes are ($B_1$, $B_2$) on the $S_1$ side and ($B_3$, $B_4$) on the $S_2$ side.

Nodes: $Q = \{S_1, S_2, B_1, \ldots, B_9\}$, with similarly $\vec{\sigma}^{(i)}.\vec{\sigma}^{(j)} = \sigma_x{}^{(i)}\sigma_x{}^{(j)} + \sigma_y{}^{(i)}\sigma_y{}^{(j)} + \sigma_z{}^{(i)}\sigma_z{}^{(j)}$

Local term:

$$H_{loc} = \sum_{q\in Q}\frac{\omega_q}{2}\sigma_z{}^{(q)} = \frac{\omega_{S_1}}{2}\sigma_z{}^{(S_1)} + \frac{\omega_{S_2}}{2}\sigma_z{}^{(S_2)} + \sum_{k=1}^{9}\frac{\omega_{B_k}}{2}\sigma_z{}^{(B_k)}$$

System– $L_1$ couplings:

$$H_{SB} = J_{S_1B_1}\vec{\sigma}^{(S_1)}.\vec{\sigma}^{(B_1)} + J_{S_1B_2}\vec{\sigma}^{(S_1)}.\vec{\sigma}^{(B_2)} + J_{S_2B_3}\vec{\sigma}^{(S_2)}.\vec{\sigma}^{(B_3)} + J_{S_2B_4}\vec{\sigma}^{(S_2)}.\vec{\sigma}^{(B_4)}$$

Within $L_1$:

$$H_{L1} = J_{B_1B_2}^{(1)}\vec{\sigma}^{(B_1)}.\vec{\sigma}^{(B_2)} + J_{B_3B_4}^{(2)}\vec{\sigma}^{(B_3)}.\vec{\sigma}^{(B_4)}$$

Within $L_2$:

$$H_{L2} = J_{B_5B_6}^{(2)}\vec{\sigma}^{(B_5)}.\vec{\sigma}^{(B_6)} + J_{B_6B_7}^{(2)}\vec{\sigma}^{(B_6)}.\vec{\sigma}^{(B_7)} + J_{B_7B_8}^{(2)}\vec{\sigma}^{(B_7)}.\vec{\sigma}^{(B_8)} + J_{B_8B_9}^{(2)}\vec{\sigma}^{(B_8)}.\vec{\sigma}^{(B_9)}$$

Between layer 1 and layer 2:

$$H_{L12} = J_{B_1B_5}\vec{\sigma}^{(B_1)}.\vec{\sigma}^{(B_5)} + J_{B_1B_6}\vec{\sigma}^{(B_1)}.\vec{\sigma}^{(B_6)} + J_{B_2B_6}\vec{\sigma}^{(B_2)}.\vec{\sigma}^{(B_6)} + J_{B_2B_7}\vec{\sigma}^{(B_2)}.\vec{\sigma}^{(B_7)} + J_{B_3B_7}\vec{\sigma}^{(B_3)}.\vec{\sigma}^{(B_7)} + J_{B_3B_8}\vec{\sigma}^{(B_3)}.\vec{\sigma}^{(B_8)} + J_{B_4B_8}\vec{\sigma}^{(B_4)}.\vec{\sigma}^{(B_8)} + J_{B_4B_9}\vec{\sigma}^{(B_4)}.\vec{\sigma}^{(B_9)}$$

Optional direct system edge:

$$H_{SS} = K_{S_1S_2}\vec{\sigma}^{(S_1)}.\vec{\sigma}^{(S_2)}$$

Total Hamiltonian:

$$H = H_{loc} + H_{SB} + H_{L1} + H_{L2} + H_{L12} + H_{SS}$$

Dissipators remain exactly as in Section 2.4 and the previous segment above: system qubits have pure dephasing only; all bath qubits $B_k$ use the thermal emission/absorption channels with a single global .